\documentclass[fleqn,usenatbib,usenames,dvipsnames]{mnras}

\usepackage[T1]{fontenc}

\usepackage[utf8]{inputenc}
\usepackage{amssymb}
\usepackage{tabularx}
\usepackage{amsmath}
\usepackage{tensor}
\usepackage{mathrsfs}
\usepackage{enumitem}
\usepackage{lipsum}
\usepackage{comment}
\usepackage{orcidlink} 
\usepackage[]{xcolor} 

\usepackage{newtxtext,newtxmath} 
\usepackage[normalem]{ulem} 

\newcommand{\figref}[1]{Fig.~\ref{#1}}
\newcommand{\tabref}[1]{Tab.~\ref{#1}}
\renewcommand{\eqref}[1]{Eq.~(\ref{#1})}

\newcommand{\e}[1]{\times 10^{#1}} 

\title[Signatures of Exploding Supermassive Stars]{Signatures of Exploding Supermassive PopIII Stars at High Redshift in JWST, EUCLID and Roman Space Telescope}

\author[C. Jockel et al.]{
Cédric Jockel\,\orcidlink{0009-0007-7617-7178},$^{1}$\thanks{E-mail: cedric.jockel@aei.mpg.de}
Kyohei Kawaguchi\,\orcidlink{0000-0003-4443-6984},$^{1,2,3}$
Sho Fujibayashi\,\orcidlink{0000-0001-6467-4969},$^{1,4,5}$
Masaru Shibata\,\orcidlink{0000-0002-4979-5671}$^{1,3}$
\\
$^{1}$Max Planck Institute for Gravitational Physics (Albert Einstein Institute), Am Mühlenberg 1, 14476 Potsdam, Germany\\
$^{2}$Institute for Cosmic Ray Research, The University of Tokyo, 5-1-5 Kashiwanoha, Kashiwa, Chiba 277-8582, Japan\\
$^{3}$Center for Gravitational Physics and Quantum Information, Yukawa Institute for Theoretical Physics, Kyoto University, Kyoto, 606-8502, Japan\\
$^{4}$Frontier Research Institute for Interdisciplinary Sciences, Tohoku University, Sendai 980-8578, Japan\\
$^{5}$Astronomical Institute, Graduate School of Science, Tohoku University, Sendai 980-8578, Japan
}

\date{Accepted XXX. Received YYY; in original form ZZZ}

\pubyear{\the\year{}}

\begin{document}
\label{firstpage}
\pagerange{\pageref{firstpage}--\pageref{lastpage}}
\maketitle

\begin{abstract}
Recently discovered supermassive black holes with masses of $\sim10^8\,M_\odot$ at redshifts $z\sim9$--$11$ in active galactic nuclei (AGN) pose severe challenges to our understanding of supermassive black hole formation. One proposed channel are rapidly accreting supermassive PopIII stars (SMSs) that form in large primordial gas halos and grow up to $<10^6\,M_\odot$. They eventually collapse due to the general relativistic instability and could lead to supernova-like explosions. This releases massive and energetic ejecta that then interact with the halo medium via an optically thick shock. 
We develop a semi-analytic model to compute the shock properties, bolometric luminosity, emission spectrum and photometry over time. The initial data is informed by stellar evolution and general relativistic SMS collapse simulations.
We find that SMS explosion light curves reach a brightness $\sim10^{45\mathrm{-}47}\,\mathrm{erg/s}$ and last $10$--$200$ years in the source frame -- up to $250$--$3000$ years with cosmic time dilation. This makes them quasi-persistent sources which vary indistinguishably to little red dots and AGN within $0.5$--$9\,(1+z)$ yrs. Bright SMS explosions are observable in long-wavelength JWST filters up to $z\leq20$ ($24$--$26$ mag) and pulsating SMSs up to $z\leq15$. EUCLID and the Roman space telescope (RST) can detect SMS explosions at $z<11$--$12$. Their deep fields could constrain the SMS rate down to $10^{-11}$Mpc$^{-3}$yr$^{-1}$, which is much deeper than JWST bounds.
Based on cosmological simulations and observed star formation rates, we expect to image up to several hundred SMS explosions with EUCLID and dozens with RST deep fields.
\end{abstract}



\section{Introduction} \label{sec:intro}
The James Webb Space Telescope (JWST) has opened the gate to the high redshift universe. During its first years of observations, it has already uncovered a number of mysteries about the early universe. One is the presence of a population of a large number of very red and spatially unresolved sources called little red dots (LRDs) \citep{Akins:2024:2024arXiv240610341A,Matthee:2023utn}. LRDs feature unusually high stellar densities and a v-shaped spectrum. It is suggested that this could be the result of old stellar populations following a bursty star formation period \citep{Wang:2024,weibel2024rubiesrevealsmassivequiescent,Baggen:2024arXiv240807745B} or obscured/dust-reddened active galactic nuclei (AGN) \citep{Li:2024littlereddotsrapidly,Inayoshi:2024arXiv240907805I} -- or a combination of both \citep{Ma:2024jkd,Akins:2024arXiv241000949A}. Another mystery are observed galaxies with disproportionally massive central supermassive black holes (SMBHs) compared to their host galaxies \citep{Mezcua:2024overmassive,Pacucci:2024abl,Junyao:2024arXiv240300074L,Pacucci:2023oci}. Some of these SMBHs compose large fractions of their host galaxies' mass \citep{Furtak:2024,Napolitano:2024arXiv241018763N}, far exceeding the modern universe’s typical $\sim0.1\%$ SMBH-to-galaxy mass ratio \citep{Mezcua:2024overmassive,Castellano:2024jwst,Jeon:2024iml,Pacucci:2023oci}. This can have implications on the co-evolution and formation of SMBHs and their host galaxies \citep{Bhowmick:2024jwc,Harikane:2024ApJ...960...56H,Silk:2024rsf,Husko:2024doa,Li:2024arXiv240906208L,Frosst:2024arXiv240906783F}. Additionally, SMBHs with inferred masses of over $10^{7-8}\,M_\odot$ only at $\sim 500\,$Mry after the Big Bang have been observed by JWST \citep{Kovacs:2024zfh,Bogdan:2023ilu,Larson:2023ApJ...953L..29L,Maiolino:2023zdu}. The earliest and most massive candidates to date being GN-z11 with $M\sim 1.5 \times 10^6\,M_\odot$ at $z=10.6$ ($460\,$Myr) \citep{Maiolino:2023zdu} and GHZ9 with $M\sim 8\times 10^7\,M_\odot$ at $z=10.4$ \citep{Kovacs:2024zfh}. \\

The discovery of those early SMBHs especially poses challenges to our understanding of SMBH formation. The standard picture of black hole (BH) growth is that stellar mass BHs are formed after collapse of massive stars and then accrete surrounding gas. The first stars in the early universe -- population III (PopIII) stars -- are expected to form roughly until $200\,$Myr after the Big Bang and have masses of up to several hundred solar masses (see \citealt{Klessen:2023qmc} for a review). They collapse to stellar mass BHs of $\sim 10^{1\mathrm{-}2}\,M_\odot$ after their evolution. The resulting BHs would then need to grow by accretion to the observed SMBH sizes. This scenario however has come under increased pressure with recent JWST observations (also see \citealt{Regan:2024wsu} for a review of various seeding and growth scenarios). This is because the initial seed BHs would need to grow with super-Eddington accretion rates (see \citealt{Mayer:2018vrr} for a review) over several $100\,$Myr with high duty-cycles \citep{Lupi:2023oji}, which might be difficult to sustain \citep{Alvarez:2008vw}. Frequent BH mergers might also be a path to rapid growth \citep{Ryu:2016dou,Tagawa:2016mwr}. However, for these scenarios, they need to be located in ideal environments which can feed sufficient amounts of gas and merger partners for the whole growth period \citep{Jeon:2024iml}. Gas accretion in particular might also be difficult due to BH feedback \citep{Su:2024aim}. \\

Another scenario is that SMBHs grow from initial heavy seeds with masses of a few $10^{4{\rm -}5}\,M_\odot$. These seeds could form through the collapse of large, extremely metal poor primordial clouds in the early universe. The resulting seeds could then grow at sub-Eddington accretion rates to reach the observed AGN masses by $z=10$. 
This scenario is called the direct collapse black hole (DCBH) scenario and typically involves the formation of rapidly accreting supermassive PopIII stars. This has attracted a growing amount of attention in the last years (see \citealt{Inayoshi:2019fun,Hosokawa:2018ugt,Woods:2018lty} for reviews). It usually necessitates that the DCBH forms inside a large ($\sim 10^{7-8}\,M_\odot$, $\sim 1000\,$pc, see \citealt{Patrick2023:10.1093/mnras/stad1179}) atomically cooled halo (ACH) of primordial gas. Cooling in these ACHs would be very inefficient due to the lack of metals and molecular hydrogen. H$_2$ formation in those ACHs would be suppressed in the presence of a strong dissociating background Lyman-Werner radiation \citep{Omukai:2000ic,Oh:2001ex,Bromm:2002hb}, collisional dissociation \citep{Fernandez:2014wia} or large relative baryon-dark matter streaming velocities \citep{Tanaka:2013boa,Hirano:2017wbu}. Significant turbulence \citep{Levine:2007zc,Begelman:2009di,Choi:2013kia} or primordial magnetic fields \citep{Ralegankar:2024ekl,Diaz:2024poo} may additionally support the halo against gravitational fragmentation and collapse. This would lead to high virial temperatures inside the ACH of around $8000\,$K \citep{Omukai:2000ic}, because emission lines of atomic hydrogen become an efficient coolant above this temperature. The ACH would subsequently collapse nearly isothermally and a star-forming disk would form in the centre \citep{Patrick2023:10.1093/mnras/stad1179,Regan:2008jr,Suazo:2019per}. The collapse of ACHs would happen at around $z= 10-20$ \citep{Chon:2016nmh,Patrick2023:10.1093/mnras/stad1179,Regan:2008jr}. \\

This monolithic collapse leads to large infall rates of up to $\mathcal{O}(\,M_\odot/\mathrm{yr})$ on these protostars \citep{Hosokawa:2015ena,Patrick2023:10.1093/mnras/stad1179,Regan:2024wsu,Kimura:2023ApJ...950..184K}. Stellar evolution calculations by \citealt{Hosokawa:2013mba,Omukai:2000ic,Herrington:2022dbu} indicate that these stars would then evolve as rapidly accreting red hypergiants with a large bloated low-density atmosphere and a denser core region with convective and radiative layers. These supermassive stars (SMSs) can grow up to $10^6\,M_\odot$ depending on the accretion rate and angular momentum of the system \citep{Shibata:2024xsl}. This behaviour is also retained in the case of episodic accretion \citep{Sakurai:2015MNRAS.452..755S} if the non-accreting phases are short enough. If accretion is $\lesssim 0.02\,M_\odot/\mathrm{yr}$ for longer periods, SMSs evolve as lower-mass blue hypergiants \citep{Herrington:2022dbu}. An HII region can then form and halt further accretion \citep{kiyuna:2024}. But SMSs may also grow from smaller PopIII stars or slowly accreting SMSs via cold cosmological flows \citep{Latif:2022Natur.607...48L}, even in halos with significant stellar feedback \citep{kiyuna:2024}. \\

If accretion persists, SMSs eventually grow so large that they collapse due to the general relativistic instability (GRI) \citep{Chandrasekhar:1964zz} (also see references in \citealt{Shibata:2024xsl}). The size when this happens depends on the accretion rate, chemical composition and rotation rate, as investigated by \citealt{Shibata:2024xsl,Saio:2024dvx,Nagele:2022lnz,Kehrer:2024dga,Woods:2017ApJ...842L...6W,Haemmerle:2021tcj,Shibata:2016vxy}. The GRI collapse is monolithic and is triggered in the compact SMS core. This is different from core collapses after pressure depletion triggered by photo-disintegration of heavy nuclei, electron capture, or production of electron-positron pairs (see references in \citealt{Uchida:2018ago}). \\

There are multiple evolution channels for collapsing SMSs. In one, a rotating BH will form after the GRI collapse starts. Most of the heavier elements in the stellar core will fall into it \citep{Fujibayashi:2024vnb}. For rapidly rotating SMS cores, a torus will also form around the BH \citep{2002ApJ...572L..39S, Liu:2007cf, Uchida:2017qwn, Fujibayashi:2024vnb}. The outer layers of the SMS core will then fall onto it and bounce off the torus. This leads to highly energetic and massive ejecta moving at $\sim 20\,\%$ of the speed of light on average \citep{Uchida:2017qwn, Fujibayashi:2024vnb}. These ejecta will subsequently sweep through the SMS atmosphere and become visible once they break out of the surface.

Another proposed channel for the explosion and ejecta launch is nuclear burning by the Carbon-Nitrogen-Oxygen (CNO) process that happens in hot dense regions during the collapse \citep{Fricke:1973ApJ...183..941F,Fuller:1986ApJ...307..675F,Montero:2011ps}. This could be relevant for high-metallicity SMSs, in which case no BH remnant is left behind. However for low-metallicity SMSs, the CNO process was found to not be strong enough to trigger an explosion by itself in simulations by \citealt{Fujibayashi:2024vnb}. But He-burning might trigger an explosion \citep{Nagele:2022lnz}. The CNO process might contribute sufficient energy to trigger radial pulsations which eventually disrupt the star, as found by \citealt{Nagele:2024aev}. Massive ejecta are expected to be launched regardless of the explosion scenario. 

The subject of the paper is to study the electromagnetic emissions of the ejecta that are produced after the SMS collapses via the GRI. The main source of electromagnetic signatures will be the direct emission from the hot ejecta itself. From the size of the system and from simulations, we expect explosion energies on the order of $10^{55\mathrm{-}56}\,\mathrm{erg}$ \citep{Fujibayashi:2024vnb,Montero:2011ps} and ejecta masses of several $1000\,M_\odot$ \citep{Matsumoto:2015bjg,Fujibayashi:2024vnb,Nagele:2024aev}. If the surrounding gas is sufficiently dilute, the resulting emissions will be dominated by thermal emission from the gradually cooling expanding ejecta. This scenario has been previously studied by \citealt{Matsumoto:2015bjg,Uchida:2017qwn}. It is physically similar to a hydrogen-rich Type IIp supernova (SN) and features a plateau in the emission (see models by e.g. \citealt{Matsumoto:2015bjg,Matsumoto:2024kqg}). However the involved ejecta masses and energies are much larger for SMSs. This picture might be changed if we take into account the dense circumstellar medium (CSM) of the ACH around the SMS when it collapses. Earlier work \citep{Surace:2018aph} has shown that the presence of a dense cloud might help observability of SMSs during their stellar evolution phase. However a dense CSM could also have a significant impact on the post-explosion light curve \citep{Whalen:2012ib}. The interaction of the CSM and the ejecta through shocks can significantly increase the luminosity \citep{Matsumoto:2022ivk,Suzuki:2018lvv}. This is similar to regular shock interaction-powered Type IIn SNe (see \citealt{Hiramatsu:2024mjz}). We therefore take into account the dense CSM when the SMS explodes and compute the evolution of the ejecta and the electromagnetic radiation emanating from it. The observability of SMS explosions also depends on the involved timescales and on the used telescopes. We here use JWST, EUCLID \citep{Euclid:2022vkk} and the Roman space telescope (RST) \citep{Spergel:2013arXiv1305.5422S}. 

We find that SMS explosions produce light curves that reach a brightness up to $\sim10^{45-47}\,\mathrm{erg}\,$s$^{-1}$ and last $10$--$200$ years in the source frame. The emissions are powered by interactions with an optically thick shock. The strong radiation flux could be sufficient to ionise the CSM using Balmer-continuum photons. Because SMSs explode at redshifts above $\sim 7$--$10$ the light curve could be visible in JWST for over $1000$ yrs in the observer frame. They will then appear as quasi-persistent sources and they could be confused with LRDs or AGN. We further find that sky surveys by EUCLID and RST will be able to detect SMS explosions for redshifts up to $11$--$14$. For example, the EUCLID deep field will be able to constrain SMS explosion rates up to $10^{-10}$\,Mpc$^{-3}$\,yr$^{-1}$. Based on cosmological simulations and observationally inferred star formation rates, we expect it to observe a few hundred SMS explosions. The EUCLID wide field could observe several thousand. 

This paper is structured as follows. In Sec.~\ref{sec:light-curve-model} we explain our light curve model for SMS explosions. We use a semi-analytic model that takes into account the interaction between the ejected material with a surrounding dense CSM. We also describe how to compute the observed brightness in different telescope filters. In Sec.~\ref{sec:Results}, we apply our model to SMS explosions. We first discuss our selection of models in Sec.~\ref{subsec:results:model-selection}. We use initial conditions that are informed by numerical simulations of collapsing SMSs. We then compute the bolometric properties of the SMS explosion light curve in Sec.~\ref{subsec:bolometric-light-curves-of-supermassive-stars}. We discuss the photometric properties of the light curve in different telescope filters of JWST, EUCLID and RST in Sec.~\ref{subsec:results:photometric-observational-properties}. We discuss in detail the (partial) ionisation of the CSM via radiation from the light curve emission and possible Balmer-attenuation in Sec. \ref{subsec:discussion:ionization-of-CSM}. In Sec.~\ref{subsec:results:distinguishing-SMS-Explosions-from-alternative-sources}, we discuss photometric and spectral similarities and differences of SMS explosions and other high-redshift objects such as AGN and LRDs. In Sec.~\ref{subsec:results:Rate-Estimation-of-High-z-Transients} we estimate the expected detectability and detection rates in different upcoming telescope surveys. We close the analysis by discussing future investigation topics and possible improvements in Sec.~\ref{subsec:discussion:further-investigation-and-future-model-improvements}. We conclude this paper in Sec.~\ref{sec:conclusions}. 

Throughout this paper, we use cgs units in our discussion unless stated otherwise. To indicate brightness, we use the AB magnitude system. $G$, $c$, $k_B$ and $\sigma$ denote the gravitational constant, the speed of light, the Boltzmann constant and the Stefan-Boltzmann constant, respectively.

\section{Light Curve Model} \label{sec:light-curve-model}

\begin{figure*}
    \centering
    \includegraphics[width=0.95\textwidth]{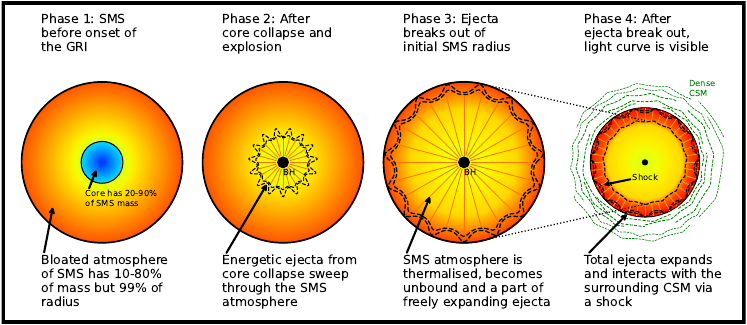}
    \caption{Summary of the main phases of an exploding supermassive star (SMS). \textbf{Phase 1:} Just before the onset of the general relativistic instability (GRI), the SMS (with mass $M_\mathrm{SMS}$) has an approximate two-zone structure consisting of a roughly chemically homogeneous and dense core region with mass $M_\mathrm{c}$ and of a low-density bloated atmosphere with mass $M_\mathrm{atm}$. The mass fraction of the core and atmosphere depend on the accretion rate and the SMS rotation. The core mass fraction is largest for large rotation and small accretion rates. It is smallest for slow rotation and large accretion rates. \textbf{Phase 2:} A black hole (BH) is formed and part of the core mass, $M_\mathrm{eje,c}$, is ejected at mildly relativistic velocities. This can happen either because infalling matter bounces off an BH accretion torus (see \protect\citealt{Fujibayashi:2024vnb}) or from the violent release of fusion energy (see \protect\citealt{Nagele:2024aev}). This material then sweeps through the bloated atmosphere. \textbf{Phase 3:} The ejected material shock wave reaches the initial SMS surface, $R_0$, and has picked up the whole atmosphere material, which is now unbound and homologously expanding. \textbf{Phase 4:} The total ejecta mass, now consisting of the combined core ejecta mass and atmosphere mass ($M_\mathrm{eje}=M_\mathrm{eje,c}+M_\mathrm{atm}$), starts to interact with the surrounding circumstellar medium (CSM) via a shock. The shocked material is hot and optically thick end emits thermal radiation. It cools down and picks up additional mass over time. Later, the material becomes transparent and non-thermal radiation will be released at the shock position.}
    \label{fig:light-curve-model:SMS-explosion-sketch}
\end{figure*}

We here present our model to study the electromagnetic emissions produced after the SMS explodes and the ejected material starts to expand into the surrounding CSM. We adapt a formulation originally developed by \citealt{Suzuki:2016bmq,Suzuki:2018lvv}, which semi-analytically models mildly relativistic supernova ejecta that interact with a dense CSM. Their model has also been verified using hydrodynamics simulations. Although the scales are different, the explosion of SMSs is qualitatively very similar to that of supernova explosions of ordinary massive stars. 

The physical picture studied in this paper is described as follows: The SMS explodes and creates a roughly spherically expanding ejecta. This is valid since, in our previous work \citep{Fujibayashi:2024vnb}, we found that the ejecta will be essentially spherically symmetric, even in the case of rapidly rotating SMS cores. The expanding spherical ejecta will then collide with the surrounding CSM and form shocks. As a consequence, the gas swept up by the forward and reverse shock fronts then forms a geometrically thin shell. The rough physical picture is summarised in the pictures in \figref{fig:light-curve-model:SMS-explosion-sketch}. 

At early times, the shell with shocked material is optically thick. The radiation produced by the shock interaction will be trapped inside the shock shell and will then gradually diffuse out. Over time, the optical depth decreases while the shell loses internal energy mainly due to radiation. Adiabatic cooling will only be relevant at early times, but has only weak effect on the overall light curve evolution, so we neglect it. When the shell eventually becomes optically thin, thermal emission emanating from the un-shocked ejecta will become visible. Additionally, there will also be emissions from the shock front. However we found that the luminosity will be much smaller than that from the optically-thick shock phase for the parameter ranges explored in this work (see a discussion in Sec.~\ref{subsec:energy-evolution-and-emission-from-the-shock-region}). This is why we do not model the second phase in detail. 

Different models were used in the past to model very massive star/SMS explosion light curves. Notably, \citealt{Uchida:2018ago} and \citealt{Matsumoto:2015bjg} used a model for hydrogen-rich Type IIp supernovae developed by \citealt{Nakauchi:2013fja} and \citealt{Dexter:2013ApJ...772...30D} to model the case where the surrounding CSM density is negligible. Numerical approaches have also been employed, e.g., by~\citealt{Whalen:2012ib} and \citealt{Moriya:2021kzc} for exploding SMSs, or \citealt{Nagele:2022odb} for pulsating SMSs.

\subsection{Ejecta and Circumstellar Medium Properties}

After the explosion, the SMS ejecta becomes unbound and is freely expanding. Radiation and gas pressures inside the ejecta are not strong enough to significantly affect its subsequent movement after a few dynamical timescales, which are much shorter than the light curve evolution timescale we are interested in. The velocity distribution of the ejecta thus corresponds to a homologous expansion profile. At a time $t$, the normalised velocity at a radius $r$ within the ejecta is then given by
\begin{align}
    \beta_\mathrm{eje} = \frac{r}{t c} \: .
\end{align}
Then the total kinetic energy and mass of the ejecta can be written as
\begin{align}
    E_\mathrm{kin,eje} &= 4 \pi c^5 t^3 \int_0^{\beta_\mathrm{max}} \rho_\mathrm{eje}(r,t) \Gamma (\Gamma -1) \beta^2 d\beta \: , \label{eq:light-curve-model:ejecta-kinetic-energy} \\
    M_\mathrm{eje} &= 4\pi c^3 t^3 \int_0^{\beta_\mathrm{max}} \rho_\mathrm{eje}(r,t) \Gamma \beta^2 d\beta \: . \label{eq:light-curve-model:ejecta-mass}
\end{align}
Here, $\rho_\mathrm{eje}(r,t)$ is the ejecta density distribution, $\beta_\mathrm{max}$ is the maximum velocity within the ejecta, $\Gamma = 1/\sqrt{1-\beta^2}$ is the Lorentz factor, $\beta$ is the integration variable and is equivalent to $\beta_\mathrm{eje}$. We assume $\rho_\mathrm{eje}(r,t)$ to be a power-law distribution with respect to the four-velocity, given by
\begin{align}
    \rho_\mathrm{eje}(r,t) = \rho_0 \left( \frac{t}{t_0}\right)^{-3} \left( \Gamma \beta_\mathrm{eje} \right)^{- n_\mathrm{eje}} \: , \label{eq:light-curve-model:ejecta-density-profile}
\end{align}
where $\rho_0$ is the scale density, $n_\mathrm{eje} $ is a positive integer and $t_0$ is the time when the interaction between ejecta and CSM starts. In our case, $t_0$ is equivalent to the expansion timescale of the outermost and highest velocity component. The case $n_\mathrm{eje} = 0$ corresponds to a uniform density distribution. Since the density distribution of the ejecta is not well known, we will mostly consider uniform density profiles, for simplicity. This was also done by \citealt{Matsumoto:2015bjg, Uchida:2017qwn} for SMS explosions. In Appendix \ref{sec:appendix:light-curve-model-validation} we give an analysis with different values $n_\mathrm{eje}=1$ and 2, and show that the solution only weakly depends on it. At the time $t_0$, the ejecta fill a region from $r=0$ to $r=c t_0 \beta_\mathrm{max}$ and the outermost layer is adjacent to the CSM at $R_0=c t_0 \beta_\mathrm{max}$, where $R_0$ is the radius of the SMS at the time of explosion.

SMSs are expected to form in ACHs of $M\sim 10^{7-8}\,M_\odot$ and $R\sim1000\,$pc, which collapse nearly isothermally at virial temperatures of around $T_\mathrm{CSM} \approx 8000\,$K (see e.g. \citealt{Inayoshi:2019fun}). ACHs are expected to be transparent because the gas is not in chemical equilibrium and the electron fraction is too low to produce a significant optical depth in the densities of interest ($n<10^{12}\,\mathrm{cm}^{-3}$, see Fig. 4 in \citealt{Omukai:2000ic}). But photons emitted from the shocked region could partially ionise the CSM. We discuss this in detail in Section \ref{subsec:discussion:ionization-of-CSM}. We here assume a primordial composition for the CSM, which is reasonable since fragmentation induced by dust cooling would prohibit the monolithic collapse of ACHs at metallicities $Z\gtrsim 10^{-5}\,Z_\odot$ \citep{Omukai:2008ApJ...686..801O,Hosokawa:2013mba} (we discuss possible signatures of trace metal pollution in Section \ref{subsec:results:distinguishing-SMS-Explosions-from-alternative-sources}).\\
The density and velocity profile of an isothermally collapsing gas cloud was derived by \citealt{Larson:1969mx}. They found that it is possible to obtain the CSM density and velocity purely in terms of $T_\mathrm{CSM}$ (for more details we refer to Appendix C of \citealt{Larson:1969mx}):
\begin{align}
    \rho_\mathrm{CSM} = \frac{\eta_0}{4\pi G} \frac{\mathcal{R}T_\mathrm{CSM}}{r^2} \hspace{0.2cm} , \hspace{0.2cm} v_\mathrm{CSM} = - \sqrt{\mathcal{R} T_\mathrm{CSM}} \; \xi_0 \: , \label{eq:light-curve-model:Larson-CSM-Model}
\end{align}
where $\mathcal{R}=4.733\times 10^7\,\mathrm{erg\,g^{-1}\,K^{-1}}$ is the specific gas constant for primordial gas ($75\,\%$ H and $25\,\%$ He). $\eta_0=8.86$ and $\xi_0=3.28$ were numerically obtained by Larson using asymptotic scaling relations for density and velocity (see Eq.~(C8) in \citealt{Larson:1969mx}). The accretion rate $\Dot{M}_\mathrm{acc} = - \rho v 4\pi r^2 (>0)$ is also easily computed for this model:
\begin{align}
    \Dot{M}_\mathrm{acc} = \frac{\eta_0 \xi_0}{G} (\mathcal{R}T_\mathrm{CSM})^{3/2} \: .
\end{align}
In the case of ACHs ($T_\mathrm{CSM}  \approx 8000\,$K), the accretion rate is approximately $1.6\,M_\odot\mathrm{yr}^{-1}$. This is in the order of the expected accretion rates of $\mathcal{O}(M_\odot \mathrm{yr}^{-1})$ that are necessary to form SMSs \citep{Hosokawa:2013mba}, and is also consistent with results from halo collapse simulations of ACHs \citep{Patrick2023:10.1093/mnras/stad1179}.

\subsection{Evolution of the Shock Shell}

Once the ejecta starts to expand into the CSM, the hydrodynamical interaction between the two media creates two shocks (forward and reverse), which propagate into the CSM and the ejecta, respectively. In this study, we assume that the pre-shocked pressures are sufficiently small compared to post-shock pressures. We also assume that the rest-mass energy dominates over the internal energy ($P/\rho \ll c^2$). In this case, the region between the forward and reverse shocks will be geometrically thin compared to the radius. We can then model the system using the thin-shell approximation, as was done in \citealt{Suzuki:2016bmq,Suzuki:2018lvv}. 

The dynamical evolution of the shell is determined by the following competing effects: Deceleration by the increase in mass of the shock region, by the supply of momentum through ram pressure at the shocks, by the deceleration due to the difference in the post-shock pressures $P_\mathrm{fs}$ and $P_\mathrm{rs}$ at the forward and reverse shocks, respectively. In the following, we refer to the shell of post-shocked medium as the ``shell'' or ``shocked matter/region''. $S_\mathrm{r}$ is the momentum of the shell in the radial direction, $M_\mathrm{s}$ is the shell mass, $R_\mathrm{s}$ is the shell position, $c\beta_\mathrm{s}$ is the shell velocity, and $\Gamma_\mathrm{s}$ is the shell Lorentz factor, respectively. The positions and velocities of the forward and reverse shocks are denoted by $R_\mathrm{fs}$, $c\beta_\mathrm{fs}$ and $R_\mathrm{rs}$, $c\beta_\mathrm{rs}$, respectively. 

Then the evolution of the momentum of the shell is given by the following equation~\citep{Suzuki:2016bmq}\footnote{There is a typo in Eq. (15) of \citealt{Suzuki:2016bmq}, which was corrected here.}:
\begin{align}
    \frac{dS_\mathrm{r}}{dt} = 4\pi R_\mathrm{rs}^2 F_\mathrm{rs} - 4\pi R_\mathrm{fs}^2 F_\mathrm{fs} + 4\pi R_\mathrm{rs}^2 P_\mathrm{rs} - 4\pi R_\mathrm{fs}^2 P_\mathrm{fs} \: . \label{eq:light-curve-model:shock-momentum-ODE}
\end{align}
The first and second terms on the right-hand side represent the momentum fluxes (ram pressure) of gas flowing into the shocked region through the forward and reverse shocks. The third and fourth terms represent the force exerted by the post-shock pressure at the shock fronts. Because the CSM velocity is negligible compared to the shell velocity, we neglect the momentum change due to the CSM and set $F_\mathrm{fs}=0$. The shell momentum is predominately supplied by the ejecta crashing into the reverse shock front. The corresponding flux is given by
\begin{align}
    F_\mathrm{rs} = c^2\rho_\mathrm{eje,rs} \Gamma_\mathrm{eje,rs}^2 \beta_\mathrm{eje,rs} \left( \beta_\mathrm{eje,rs} - \beta_\mathrm{rs} \right)\: ,
\end{align}
where $\rho_\mathrm{eje,rs}$, $\Gamma_\mathrm{eje,rs}$ and $\beta_\mathrm{eje,rs}$ are the density, Lorentz factor and normalised velocity ($\beta_x=v_x/c$ with $x$ an arbitrary index) of the pre-shocked ejecta at the reverse shock position $R_\mathrm{rs}$. 

The mass within the shell will continually increase due to the forward and reverse shock sweeping up mass from the CSM and the ejecta, respectively. The evolution equation of the shell mass is given by
\begin{align}
    \frac{dM_\mathrm{s}}{dt} = 4\pi R_\mathrm{rs}^2 G_\mathrm{rs} - 4\pi R_\mathrm{fs}^2 G_\mathrm{fs} \: , \label{eq:light-curve-model:shock-mass-ODE}
\end{align}
where the mass fluxes flowing through the forward shock, $G_\mathrm{fs}$, and reverse shock, $G_\mathrm{rs}$, are given by
    \begin{align}
    G_\mathrm{fs} &= c\rho_\mathrm{CSM,fs} \Gamma_\mathrm{CSM,fs} \left( \beta_\mathrm{CSM,fs} - \beta_\mathrm{fs} \right) = -c\rho_\mathrm{CSM,fs} \beta_\mathrm{fs} \: , \nonumber \\
    G_\mathrm{rs} &= c\rho_\mathrm{eje,rs} \Gamma_\mathrm{eje,rs} \left( \beta_\mathrm{eje,rs} - \beta_\mathrm{rs} \right)\: .
    \end{align}
Here, $\rho_\mathrm{CSM,fs}$, $\Gamma_\mathrm{CSM,fs}$, and $\beta_\mathrm{CSM,fs}$ are the density, Lorentz factor, and normalised velocity of the pre-shocked CSM at the position of the forward shock $R_\mathrm{fs}$, respectively. We take $\beta_\mathrm{CSM,fs} \approx 0$ because the CSM velocity is much smaller than the velocity of the shell. 

From the shell momentum $S_\mathrm{r}$ and mass $M_\mathrm{s}$, one can compute the shell velocity $c\beta_\mathrm{s}$ and Lorentz factor $\Gamma_\mathrm{s}= 1/\sqrt{1-\beta_\mathrm{s}^2}$ using
\begin{align}
    S_\mathrm{r} = M_\mathrm{s} \Gamma_\mathrm{s}\,c\beta_\mathrm{s} = \sqrt{ S_\mathrm{r}^2 + M_\mathrm{s}^2 } \; c\beta_\mathrm{s} \:\: \Rightarrow\:\: c\beta_\mathrm{s} = S_\mathrm{r} / \sqrt{ S_\mathrm{r}^2 + M_\mathrm{s}^2 } \: . \label{eq:light-curve-model:momentum-shock}
\end{align}
The radius $R_\mathrm{s}$ of the shocked matter shell can then be computed using
\begin{align}
    \frac{dR_\mathrm{s}}{dt} = c\beta_\mathrm{s} \: . \label{eq:light-curve-model:shock-position-ODE}
\end{align}
The time evolution of the forward and reverse shock radii are given by
\begin{align}
    \frac{dR_\mathrm{fs}}{dt} = c\beta_\mathrm{fs} \:\: , \:\: \frac{dR_\mathrm{rs}}{dt} = c\beta_\mathrm{rs} \: . \label{eq:light-curve-model:forward-reverse-shock-position-ODE}
\end{align}
The normalised shock/shell velocities $\beta_\mathrm{s}$, $\beta_\mathrm{fs}$, $\beta_\mathrm{rs}$ and the hydrodynamical variables of the post-shock gas $\rho_\mathrm{fs}$, $P_\mathrm{fs}$, $\rho_\mathrm{rs}$ and $P_\mathrm{rs}$ can be obtained using the relativistic shock jump conditions (see \citealt{Suzuki:2016bmq} for more details). Using these, the shock velocity can be expressed as a function of the upstream ($\beta_\mathrm{u}$, $\Gamma_\mathrm{u}$) and downstream ($\beta_\mathrm{d}$, $\Gamma_\mathrm{d}$) gas velocity and Lorentz factors:
\begin{align}
    \beta_\mathrm{sh}(\beta_\mathrm{u}, \beta_\mathrm{d}) = \frac{ \gamma \Gamma_\mathrm{u} \Gamma_\mathrm{d}^2 \left(\beta_\mathrm{u} - \beta_\mathrm{d} \right) \beta_\mathrm{d} - \left(\gamma-1\right) \left(\Gamma_\mathrm{u} - \Gamma_\mathrm{d}\right) }{ \gamma \Gamma_\mathrm{u} \Gamma_\mathrm{d}^2 \left(\beta_\mathrm{u} - \beta_\mathrm{d} \right) - \left(\gamma-1\right) \left(\Gamma_\mathrm{u}\beta_\mathrm{u} - \Gamma_\mathrm{d}\beta_\mathrm{d}\right) } \: .
\end{align}
Here, $\gamma$ is the adiabatic index of the shocked gas. The shocked material is radiation dominated, and hence $\gamma=4/3$.

\begin{figure}
    \centering
    \includegraphics[width=0.42\textwidth]{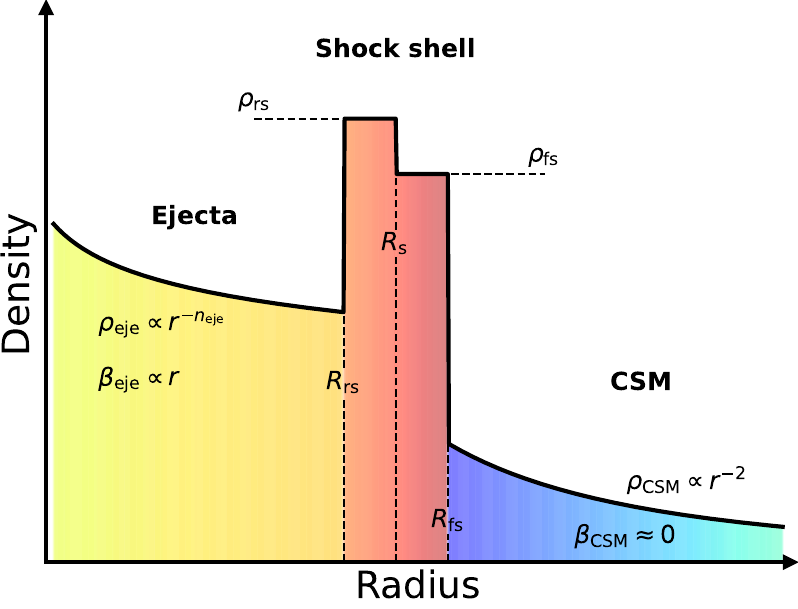}
    \caption{Summary of the radial density distribution of the ejecta (left), of the shocked shell (middle), and of the circumstellar medium (CSM) (right). The densities and velocities of the shock shell at the forward and reverse shock positions ($R_\mathrm{fs}$ and $R_\mathrm{rs}$) are obtained using the shock jump conditions (\eqref{eq:light-curve-model:jump-conditions-forward-shock} and \eqref{eq:light-curve-model:jump-conditions-reverse-shock}).}
    \label{fig:light-curve-model:density-distribution-ejecta-shock-CSM}
\end{figure}

Similar shock jump conditions exist for the density $\rho$ and pressure $P$ (see \citealt{Suzuki:2016bmq} for details). The final jump conditions for the forward shock are
\begin{subequations}
\label{eq:light-curve-model:jump-conditions-forward-shock}
\begin{align}
    \rho_\mathrm{CSM,fs} &= \rho_\mathrm{CSM}(R_\mathrm{fs}) \: , \\
    \beta_\mathrm{CSM,fs} & \approx 0 \: , \\
    \beta_\mathrm{fs} &= \beta_\mathrm{sh}(0 , \, \beta_\mathrm{s}) \: , \\
    \rho_\mathrm{fs} &= \rho_\mathrm{CSM,fs} \frac{\beta_\mathrm{fs}}{\Gamma_\mathrm{s} (\beta_\mathrm{fs} - \beta_\mathrm{s})} \: , \\
    P_\mathrm{fs} &= \rho_\mathrm{CSM,fs} \frac{\beta_\mathrm{s}\beta_\mathrm{fs}\,c^2}{1- \beta_\mathrm{s}\beta_\mathrm{fs}} \: .
\end{align}
\end{subequations}
For the reverse shock, the jump conditions are
\begin{subequations}
\label{eq:light-curve-model:jump-conditions-reverse-shock}
\begin{align}
    \rho_\mathrm{eje,rs} &= \rho_\mathrm{eje}(R_\mathrm{rs},\,t) \: , \\
    \beta_\mathrm{eje,rs} &= R_\mathrm{rs}/ct \: , \\
    \beta_\mathrm{rs} &= \beta_\mathrm{sh}(\beta_\mathrm{eje,rs} , \, \beta_\mathrm{s}) \: , \\
    \rho_\mathrm{rs} &= \rho_\mathrm{eje,rs} \frac{\Gamma_\mathrm{eje,rs} (\beta_\mathrm{eje,rs}- \beta_\mathrm{rs})}{\Gamma_\mathrm{s} (\beta_\mathrm{s} - \beta_\mathrm{rs})} \: , \\
    P_\mathrm{rs} &= c^2\rho_\mathrm{eje,rs} \Gamma_\mathrm{eje,rs}^2 \frac{(\beta_\mathrm{eje,rs}- \beta_\mathrm{rs}) (\beta_\mathrm{eje,rs}- \beta_\mathrm{s})}{1- \beta_\mathrm{s}\beta_\mathrm{rs}} \: .
\end{align}
\end{subequations}
In the above equations, the Lorentz factors are given by $\Gamma_x = 1/ \sqrt{1-\beta_x^2}$. We also provide an overview of the total density structure of the ejecta, shock and CSM in \figref{fig:light-curve-model:density-distribution-ejecta-shock-CSM}.

To summarise, the equations to be integrated are the ordinary differential equations for: the shock momentum $S_\mathrm{r}$ (\eqref{eq:light-curve-model:shock-momentum-ODE}), shock region mass $M_\mathrm{s}$ (\eqref{eq:light-curve-model:shock-mass-ODE}) and the positions of the bulk shock $R_\mathrm{s}$ (\eqref{eq:light-curve-model:shock-position-ODE}), forward shock $R_\mathrm{fs}$ and reverse shock $R_\mathrm{rs}$ (\eqref{eq:light-curve-model:forward-reverse-shock-position-ODE}). \\
The evolution of these quantities are determined from the initial values of $E_\mathrm{kin,eje}$, $M_\mathrm{eje}$ and $R_0$. We discuss the initial conditions in detail in Sec. \ref{subsec:initial-conditions}.

\subsection{Energy Evolution and Emission From the Shock Region}
\label{subsec:energy-evolution-and-emission-from-the-shock-region}

In the early stages of the evolution, the shocked region is optically thick. Due to the shock interactions, kinetic energy from the ejecta and the shell are converted into internal energy of the shocked region. This internal energy then gradually diffuses out of the shocked region and is emitted as electromagnetic radiation. The aim of our model is to compute the bolometric light curve and spectra of the shocked region in the optically thick phase. We do not model the radiation which is expected to become relevant once the region becomes transparent. 

As with the shock dynamics, our model is based on the model by \citealt{Suzuki:2016bmq,Suzuki:2018lvv}. Suzuki's model does not manifestly conserve the total energy. This is because they make the thin-shell approximation, but the shell has actually finite thickness. We thus modify their model to be manifestly energy-conserving. 

We write the evolution equation of the total energy as
\begin{eqnarray}
    \frac{d}{dt} E_\mathrm{tot} &=& \frac{d}{dt} \left( E_\mathrm{kin,sh} + E_\mathrm{kin,eje} + E_\mathrm{int,sh} + E_\mathrm{int,eje} \right) \nonumber \\&=& - L_\mathrm{diff} \: ,
\end{eqnarray}
where $E_\mathrm{kin,sh}$, $E_\mathrm{kin,eje}$, $E_\mathrm{int,sh}$ and $E_\mathrm{int,eje}$ are the kinetic and internal energies of the shocked region and of the pre-shocked ejecta. $L_\mathrm{diff}$ is the energy loss due to photon diffusion/radiation. We hereafter neglect the contributions from the internal energy of the ejecta $E_\mathrm{int,eje}$. This is allowed since it quickly becomes much smaller than $E_\mathrm{kin,eje}$ after only a few dynamical times $t_0$ (see the discussion after \eqref{eq:light-curve-model:optical-depth-shock-total}).

The evolution equation for the internal energy of the shock region therefore becomes:
\begin{align}
    \frac{d}{dt} E_\mathrm{int,sh} = - \Dot{E}_\mathrm{kin,sh} - \Dot{E}_\mathrm{kin,eje} - L_\mathrm{diff} \: . \label{eq:light-curve-model:shock-internal-energy-ODE}
\end{align}
Here the kinetic energy of the shell is given by $E_\mathrm{kin,sh} = \left( \Gamma_\mathrm{s} - 1 \right) M_\mathrm{s} c^2$, from which the time derivative can be easily computed (using \eqref{eq:light-curve-model:momentum-shock}):
\begin{align}
    \Dot{E}_\mathrm{kin,sh} = \left( \Gamma_\mathrm{s} - \Gamma_\mathrm{s} \beta_\mathrm{s}^2 - 1 \right) \Dot{M}_\mathrm{s}\,c^2 + c\beta_\mathrm{s} \Dot{S}_\mathrm{r} \: .
\end{align}
$\Dot{M}_\mathrm{s}$ and $\Dot{S}_\mathrm{r}$ are the time derivatives of the shock mass and momentum given by \eqref{eq:light-curve-model:shock-mass-ODE} and \eqref{eq:light-curve-model:shock-momentum-ODE}, respectively. The derivative of the pre-shocked ejecta kinetic energy, $E_\mathrm{kin,eje}$ (\eqref{eq:light-curve-model:ejecta-kinetic-energy}), with the power-law density profile (\eqref{eq:light-curve-model:ejecta-density-profile}), is
\begin{align}
    \Dot{E}_\mathrm{kin,eje} = 4\pi c^5 \rho_0 t_0^3 \left( \Gamma_\mathrm{max} -1 \right) \Gamma_\mathrm{max} \beta_\mathrm{max}^2 \left(\Gamma_\mathrm{max} \beta_\mathrm{max}\right)^{-n_\mathrm{eje}} \Dot{\beta}_\mathrm{max}  \: ,
\end{align}
where $\beta_\mathrm{max} = R_\mathrm{rs}/ct$ (with $\Dot{\beta}_\mathrm{max} = \Dot{R}_\mathrm{rs}/ct - R_\mathrm{rs}/(c\,t^2)$) and $\Gamma_\mathrm{max} = 1/\sqrt{1-\beta_\mathrm{max}^2}$ stand for the velocity and Lorentz factor of the un-shocked ejecta at the position of the reverse shock. Here we assume that everything beyond the reverse shock has already flown into the shock region.

The energy loss due to radiation is given by the diffusion radiation, $L_{\rm diff}$. It depends on the energetics of the shell and is determined in a more complicated way, as we discuss as follows. During the evolution, the shocked shell will gradually cool down and the hydrogen in the shell could partially recombine. This modifies the evolution of the luminosity and optical depth. Additionally, we found that in some high-luminosity cases, the ions and photons in the shell might not be fully thermalised. We address these points in the following and add corrections to our model. We first discuss thermalisation because it also affects recombination. 

We base the following discussion on \citealt{NakarSari:2010ApJ...725..904N}. Thermalisation depends on the production rate of photons. If it is smaller than what is needed to maintain a blackbody spectrum, the radiating region will be out of equilibrium. The radiation energy will be distributed between fewer photons, which raises the average photon energy $\sim 3 k_B T_\mathrm{ph}$, and thus the photon temperature $T_\mathrm{ph}$. $T_\mathrm{ph}$ is also called ``colour temperature''. It can be related to the ion temperature if some timescales are met. 

We therefore briefly discuss relevant timescales based on Sec. 4.2.4 in \citealt{Suzuki:2016bmq}. For our cases, the Coulomb timescale is shorter than the dynamical timescale. Electrons and ions therefore have the same temperature, $T_\mathrm{e}$. This is important because this allows for quick distribution of the energy released by the shock-interaction throughout the shocked matter. The Compton cooling timescale is also much shorter than the evolution time of the system. Electrons thus can exchange energy efficiently with photons and we have $T_\mathrm{ph}=T_\mathrm{e}$. The number of photons is also much larger than the number of electrons. These conditions are essential for the validity of taking $u_{\rm s,rad}\approx E_{\rm int,sh}/V_{\rm s}$ (with $V_\mathrm{s} = \frac{4}{3}\pi (R_\mathrm{fs}^3-R_\mathrm{rs}^3)$) for the radiation energy density. 

According to \citealt{NakarSari:2010ApJ...725..904N}, the thermal coupling coefficient, $\eta$, determines whether the radiation is in equilibrium ($\eta<1$) or out of equilibrium ($\eta>1$). It is given by:
\begin{align}
    \eta \approx \frac{4\e{5}\,\mathrm{s}}{\min\{t, t_\mathrm{d}\}} \left( \frac{\rho_\mathrm{av}}{10^{-10}\,\mathrm{g\,cm}^{-3}} \right)^{-2} \left( \frac{T_\mathrm{BB,bulk}}{10^6\,\mathrm{K}} \right)^{7/2} \: . \label{eq:light-curve-model:eta-parameter-definition}
\end{align}
Here $\rho_\mathrm{av}= M_\mathrm{s}/(\Gamma_\mathrm{s}V_\mathrm{s})$ is the average shell density, $T_\mathrm{BB,bulk}=\sqrt[4]{u_\mathrm{s,rad}/a_\mathrm{r}}$ (with $a_\mathrm{r}=4\sigma/c$) is the ``blackbody temperature'' of the shock bulk, $t$ is the time and $t_\mathrm{d} = (R_\mathrm{fs}-R_\mathrm{rs})/v_\mathrm{diff}$ is the diffusion timescale of the thin shell (with $v_\mathrm{diff}$ as in \eqref{eq:light-curve-model:diffusion-velocity-shock}). The blackbody temperature is the temperature that a fully thermalised radiation field would have.

The physical meaning of $\eta$ is the ratio of the number of photons that are necessary to maintain thermalisation and the number of photons that are produced within one diffusion time (see \citealt{NakarSari:2010ApJ...725..904N}, Eq. 9). Another interpretation is the photon consumption rate divided by the photon production rate within the radiating region. The photon consumption depends on the total luminosity, and the production rate depends on the production channel (in our case free-free emission). If photon production is too low to maintain a blackbody spectrum at a given luminosity, then the radiation is out of equilibrium and thus $\eta>1$. The total energy will then be distributed over fewer photons and the average energy per photon will increase, thus leading to a bluer spectrum.

Following \citealt{NakarSari:2010ApJ...725..904N}, we can use $\eta$ to compute the bulk electron temperature, $T_\mathrm{e,bulk}$, which is relevant for recombination. It is given by
\begin{align}
    T_\mathrm{e,bulk} = T = T_\mathrm{BB,bulk} 
    \begin{cases}
    \eta^2 / \xi(T)^2 \: , &  \eta>1 \: , \\
    1 \: , &  \eta<1 \: . 
  \end{cases}
  \label{eq:light-curve-model:T_ion_colour-calculation}
\end{align}
Here, $\xi(T) = \max\{1,\; \frac{1}{2} \ln[y_\mathrm{max}] \left(1.6 + \ln[y_\mathrm{max}] \right) \}$ and
\begin{align}
    y_\mathrm{max} \approx 2 \left( \frac{\rho_\mathrm{av}}{10^{-9}\,\mathrm{g\,cm}^{-3}} \right)^{-1/2} \left( \frac{T}{10^6\,\mathrm{K}} \right)^{9/4} \: .
\end{align}
$T_\mathrm{e,bulk}$ is then found via root-finding.

To assess hydrogen recombination in the shell and determine the colour temperature of the diffused photons, we need to estimate not only $T_{\rm e,bulk}$ but also the electron temperature at the shell surface, $T_{\rm e,surf}$. For the case of $\eta<1$, we assume that photons at the shell surface are well-thermalised. $T_{\rm e,surf}$ can then be estimated by employing the Stefan Boltzmann law, $T_{\rm e,surf}\approx T_{\rm BB,surf}\equiv \sqrt[4]{L_\mathrm{diff}/4\pi R_\mathrm{s}^2 \sigma}$. For $\eta>1$, photons at the shell surface are expected to be out of thermal equilibrium. Since it is not trivial how $T_{\rm e,surf}$ can be given under the one-zone approximation of the shell, for simplicity, we employ a crude prescription that $T_{\rm e,surf}$ can be estimated by $T_{\rm e,surf}\approx(\eta(T_{\rm e,bulk})/\xi(T_{\rm e,bulk}))^2 T_{\rm BB,surf}$, using $\eta$ and $\xi$ obtained in the bulk (\eqref{eq:light-curve-model:eta-parameter-definition}--\eqref{eq:light-curve-model:T_ion_colour-calculation}). This prescription of course introduces systematic errors, since generally $\eta$ and $\xi$ at the shell surface do not necessarily coincide with that in the bulk. Hence, our spectra model for emitted photons in non-thermalised cases will suffer in accuracy. Thus, for $\eta>1$, we focus only on general trends and order of magnitude of the solutions, while being aware of that the non-thermalised cases have some (possibly large) errors.

The energy loss due to radiation is given by the diffusion luminosity emanating from the thin shell:
\begin{align}
    L_\mathrm{diff} = 4\pi R_\mathrm{s}^2 u_\mathrm{s,rad} v_\mathrm{diff} \: , \label{eq:light-curve-model:diffusion-luminosity-shock}
\end{align}
where $u_\mathrm{s,rad} = E_\mathrm{int,sh} / V_\mathrm{s}$ (with $V_\mathrm{s} = \frac{4}{3}\pi (R_\mathrm{fs}^3-R_\mathrm{rs}^3)$) is the radiation energy density of the shell. $v_\mathrm{diff}$ is the diffusion velocity, given by (see \citealt{Suzuki:2018lvv})
\begin{align}
    v_\mathrm{diff} = \frac{c (1- \beta_\mathrm{s}^2)}{ (3 + \beta_\mathrm{s}^2) \tau_\mathrm{s} + 2 \beta_\mathrm{s} } \: . \label{eq:light-curve-model:diffusion-velocity-shock}
\end{align}
Here, $\tau_\mathrm{s}$ is the optical depth of the shock. If the shell is completely ionised, $\tau_\mathrm{s}$ is given by
\begin{align}
    \tau_\mathrm{s} = \tau^0_\mathrm{s} = \frac{\kappa M_\mathrm{s}}{4\pi R_\mathrm{s}^2} \: . \label{eq:light-curve-model:optical-depth-shock}
\end{align}
Here, as $\kappa$, we take the Thompson opacity of primordial gas ($X=75\,\%$ H and $Y=25\,\%$ He) and therefore $\kappa = 0.2\, (1+X)\,\mathrm{cm^2\,g^{-1}} \approx 0.35\,\mathrm{cm^2 \,g^{-1}}$. If $v_\mathrm{diff}$ computed using the above equation exceeds
\begin{align}
    v_\mathrm{diff,max} = c (1 - \beta_\mathrm{s}) \: ,
\end{align}
this would mean that radiation would escape from the shell superluminally. We therefore set $v_\mathrm{diff}  = v_\mathrm{diff,max}$ if it exceeds this value.

If $T_\mathrm{e,surf}$ is higher than the ionisation temperature of hydrogen, $T_\mathrm{ion}$, the optical depth is given by \eqref{eq:light-curve-model:optical-depth-shock}. We take $T_\mathrm{ion} = 6000\,$K as in \citealt{Matsumoto:2015bjg}, which is applicable for the typical shock region densities of $\sim 10^{-11}$--$10^{-13}\,$g cm$^{-3}$ when recombination becomes relevant \citep{Goldfriend:2014arXiv1404.6313G}. 

However, if $T_\mathrm{e,surf}$ drops below $T_\mathrm{ion}$, before the shock becomes transparent ($\tau_\mathrm{s}^0<1$), a part of the shock region would recombine (starting from the outside to the inside) and become transparent to radiation. This happens even though the average electron temperature of the bulk shock region $T_\mathrm{e,bulk} > T_\mathrm{ion}$. 

Recombination in the shell happens once the condition of $T_{\rm e,surf}\leq T_{\rm ion}$ is satisfied. Then, the ionisation front -- the surface where the gas remains ionised in the shell -- moves inwards in the shell surface and the optical depth of the shell $\tau_{\rm s}$ decreases until the electron temperature there ($T_\mathrm{e,if}$) becomes sufficiently high. The condition for $\tau_{\rm s}$ can be determined as follows: The ionisation front is characterised by $T_\mathrm{e,if}=T_\mathrm{ion}$ and coincides with the layer where radiation is released. While the expression of \eqref{eq:light-curve-model:diffusion-luminosity-shock} still holds, the diffusion luminosity can also be given using the effective blackbody temperature at this layer, $T_\mathrm{BB,if}$, by
\begin{align}
    L_\mathrm{diff} = 4\pi R_\mathrm{s}^2 \sigma T_\mathrm{BB,if}^4 \: . \label{eq:light-curve-model:diffusion-luminosity-shock-modified}
\end{align}
Here, under the thin shell approximation, we approximate that the position of the ionisation front is the same as $R_\mathrm{s}$. Employing the same prescription as for the surface temperature, $T_{\rm e,if} =T_{\rm ion}\approx (\eta(T_{\rm e,bulk})/\xi(T_{\rm e,bulk}))^2 T_{\rm BB,if}$, $\tau_\mathrm{s}$
in this phase can be obtained by setting $\sigma T_\mathrm{BB,if}^4 \stackrel{!}{=} u_\mathrm{s,rad} v_\mathrm{diff}$ and we solve this expression for the optical depth of the ionised component of the shock region:
\begin{align}
    \tau_\mathrm{s} = \left[ \left( 1 - \beta_\mathrm{s}^2 \right) \frac{u_\mathrm{s,rad}\,c}{\sigma T_\mathrm{BB,if}^4}  - 2 \beta_\mathrm{s} \right]  \left( 3 + \beta_\mathrm{s}^2 \right)^{-1} \: . \label{eq:light-curve-model:optical-depth-shock-modified}
\end{align}
This takes into account the partial ionisation of the shock region. Finally, we take into account the temperature dependence of the opacity $\kappa$ by setting it zero if $T_\mathrm{e,bulk}<T_\mathrm{ion}$. This is a good approximation because $\kappa$ steeply decreases for $T<T_\mathrm{ion}$ as $\propto T^{-11}$, see \cite{Goldfriend:2014arXiv1404.6313G}. 

We note that, since $T_{\rm e,surf}$ is dependent on $L_{\rm diff}$ through $T_{\rm BB,surf}$, the modification in $\tau_{\rm s}$ and hence $L_{\rm diff}$ also changes the condition for the recombination ($T_{\rm e,surf}\leq T_{\rm ion} $). For simplicity, when computing $T_\mathrm{e,surf}$, we always use $\tau^0_{\rm s}$ (see \eqref{eq:light-curve-model:optical-depth-shock}) for $\tau_\mathrm{s}$ and neglect the back-reaction of the recombination to $T_\mathrm{e,surf}$ to avoid complex numerical root finding.

The observed colour temperature of the diffused photons, $T_{\rm col}$, will be given by the electron temperature at the shell surface or the ionisation front, that is
\begin{align}
    T_\mathrm{col} = \begin{cases}
     T_\mathrm{e,surf} \: , &  T_\mathrm{e,surf} > T_\mathrm{ion} \: , \\
    T_\mathrm{ion} \: , &  T_\mathrm{e,surf} \leq T_\mathrm{ion} \: , \label{eq:light-curve-model:color-temperature-shock-surface}
  \end{cases}
\end{align}\noindent
with $T_{\rm e,surf}={\rm max}[\eta(T_{\rm e,bulk})/\xi(T_{\rm e,bulk}),1]^2 T_{\rm BB,surf}$. The spectrum in cases of $\eta>1$ will however be different from a blackbody spectrum and we discuss this in Sec. \ref{subsec:computation-of-radiation-flux-and-apparent-magnitude}. 

To summarise the previous discussion, the optical depth evolves as:
\begin{align}
    \tau_\mathrm{s} = \begin{cases}
     \text{\eqref{eq:light-curve-model:optical-depth-shock}} \: , &  T_\mathrm{e,surf} > T_\mathrm{ion} \: , \\
    \text{\eqref{eq:light-curve-model:optical-depth-shock-modified}} \: , &  T_\mathrm{e,surf} \leq T_\mathrm{ion} \: . \label{eq:light-curve-model:optical-depth-shock-total}
  \end{cases}
\end{align}
$L_\mathrm{diff}$ is always given by \eqref{eq:light-curve-model:diffusion-luminosity-shock}. But note that $v_\mathrm{diff}$ is a function of $\tau_\mathrm{s}$, which is given by \eqref{eq:light-curve-model:optical-depth-shock-total} and in the case of $T_\mathrm{e,surf}<T_\mathrm{ion}$, $L_\mathrm{diff}$ reduces to \eqref{eq:light-curve-model:diffusion-luminosity-shock-modified} with $T_{\rm BB,if}={\rm max}[\eta(T_{\rm e,bulk})/\xi(T_{\rm e,bulk}),1]^{-2} T_{\rm ion}$.
 
After $\tau_\mathrm{s}$ drops below unity, the light curve enters an optically thin shock phase. But the interactions at the forward and reverse shocks may keep converting kinetic energy into non-thermal radiation. The bolometric luminosity of this radiation can be estimated using $L_\mathrm{bol,nt} = - \varepsilon \Dot{E}_\mathrm{kin,eje}$, where $\varepsilon \approx 0.1$ is a common assumption for the energy conversion efficiency to photons and is also consistent with more complete treatments (see e.g. \cite{Tsuna:2019ApJ...884...87T} and references therein). We do not model this radiation here in detail since it is expected to be much smaller than the emission in the previous optically-thick-shock phase for the parameters that are relevant for SMS explosions. It can become dominant only if the ejecta rest-mass energy is larger than the ejecta kinetic energy, see Appendix \ref{sec:appendix:light-curve-model-validation}. We therefore need to keep in mind the range of validity of our model. This will be indicated when relevant. 

A similar reasoning can be applied to why we do not model possible non-thermal emissions in the phase where the shock is only partially ionised. We would expect that non-thermal radiation is released at the forward shock front, which is located in an optically thin region outside the ionisation front. However, again, the non-thermal emissions should be negligible compared to the thermal emissions because the lower photon conversion efficiency $\varepsilon \sim 0.1$ and because most of the kinetic energy/momentum is provided to the shock through the reverse shock.

Additionally, diffusing radiation from the un-shocked ejecta at $r<R_\mathrm{rs}$ becomes visible after $\tau_\mathrm{s}<1$. This diffusion radiation part will be powered by the release of internal energy from the un-shocked region. We also neglect this radiation contribution in this work. This is because we expect the internal energy to be much smaller than the kinetic energy of the ejecta at this time in the evolution. The internal energy can be estimated using the assumption of homologous expansion and adiabatic energy loss. The internal energy density evolves as $u_\mathrm{int,eje} \propto (t/t_0)^{-4}$ for a radiation dominated gas (see \cite{Suzuki:2018lvv}), where $t_0 = R_0/c\beta_\mathrm{max}$ is the dynamical timescale. For our models $t_0\approx3-30$ days (see Sec. \ref{subsec:results:model-selection}). In comparison, the kinetic energy density is expected to decrease like $u_\mathrm{kin,eje} \propto (t/t_0)^{-3}$. At the time when the shock becomes transparent at $t \sim \mathcal{O}(10\,{\mathrm{yr}})$, $t/t_0 \gtrsim 100$--$1000$ and the internal energy of the un-shocked ejecta will have decreased so much that any released thermal radiation will be negligible compared to possible non-thermal radiation created at the shock fronts. 

\subsection{Initial Conditions}
\label{subsec:initial-conditions}

At the initial stage of the evolution, the system can be characterised by the following quantities: The kinetic energy of the ejecta, $E_\mathrm{kin,eje}$, the density distribution and total mass of the ejecta, $M_\mathrm{eje}$, and the radius of the SMS, $R_0$, at the time of explosion. All initial conditions for the ordinary differential equations of the shock momentum $S_\mathrm{r}$, mass $M_\mathrm{s}$ and positions $R_\mathrm{s}$, $R_\mathrm{fs}$, $R_\mathrm{rs}$ can be deduced from these quantities. 

At the initial time, the SMS radius is related to the initial interaction time $t_0$ and the initial maximal velocity inside the ejecta, $\beta_\mathrm{max,0}$, through $R_0=c t_0 \beta_\mathrm{max,0}$. For a given density distribution (\eqref{eq:light-curve-model:ejecta-density-profile}), the initial kinetic energy (\eqref{eq:light-curve-model:ejecta-kinetic-energy}) and mass of the ejecta (\eqref{eq:light-curve-model:ejecta-mass}) can be written as
\begin{align}
    E_\mathrm{kin,eje} &= 4 \pi c^5 t_0^3 \rho_0\, I_{n_\mathrm{eje}}(\beta_\mathrm{max,0}) \: , \label{eq:light-curve-model:initial-conditions-ejecta-kinetic-energy} \\
    M_\mathrm{eje} &= 4\pi c^3 t_0^3 \rho_0\, J_{n_\mathrm{eje}}(\beta_\mathrm{max,0}) \: , \label{eq:light-curve-model:initial-conditions-ejecta-mass}
\end{align}
where the integrals are defined as
\begin{align}
    I_n(\Tilde{\beta}) &= \int_0^{\Tilde{\beta}} \left( \Gamma \beta \right)^{- n} \Gamma (\Gamma -1) \beta^2 d\beta \: , \\
    J_n(\Tilde{\beta}) &= \int_0^{\Tilde{\beta}} \left( \Gamma \beta \right)^{- n} \Gamma \beta^2 d\beta \: .
\end{align}
These integrals $I_n$ and $J_n$ can be solved analytically for some integer $n$. We report the analytic expressions in Appendix \ref{sec:appendix:analytic-form-of-the-ejecta-energy-and-mass}. 

We use the above relations and the known initial values of $E_\mathrm{kin,eje}$, $M_\mathrm{eje}$ and $R_0$ to compute $t_0$, $\rho_0$ and $\beta_\mathrm{max,0}$ as follows: We take the ratio of the initial kinetic energy and the ejecta mass and get
\begin{align}
    \frac{E_\mathrm{kin,eje}}{M_\mathrm{eje}} = c^2 \frac{I_{n_\mathrm{eje}}(\beta_\mathrm{max,0})}{J_{n_\mathrm{eje}}(\beta_\mathrm{max,0})} \: .
\end{align}
We can then perform root-finding to obtain $\beta_\mathrm{max,0}$. From this, we obtain the initial interaction time through $t_0= R_0 / ( c \beta_\mathrm{max,0} )$ and the scale density $\rho_0$ by simply re-arranging either \eqref{eq:light-curve-model:initial-conditions-ejecta-kinetic-energy} or \eqref{eq:light-curve-model:initial-conditions-ejecta-mass}. 

For the case with very massive ejecta, it can become non-relativistic with $\beta_\mathrm{max,0} < 10^{-3}$. To avoid numerical problems in the integrals $I_n$ and $J_n$, we then instead use the Newtonian expressions to solve the initial conditions for $\beta_\mathrm{max,0}$, $\rho_0$ and $t_0$: 
\begin{align}
    E_\mathrm{kin,eje} &= \frac{3-n_\mathrm{eje}}{2(5-n_\mathrm{eje})} M_\mathrm{eje} (c \beta_\mathrm{max,0} )^2 \: , \label{eq:light-curve-model:initial-conditions-non-rel-ejecta-kinetic-energy} \\
    M_\mathrm{eje} &= \frac{4\pi \rho_0}{3-n_\mathrm{eje}} \frac{R_0^3}{(\beta_\mathrm{max,0})^{-n_\mathrm{eje}}} \: . \label{eq:light-curve-model:initial-conditions-non-rel-ejecta-mass}
\end{align}

For the initial conditions of the shock quantities, we use the same prescription as described in~\cite{Suzuki:2016bmq}. At the initial time, the radii $R_\mathrm{s}$, $R_\mathrm{fs}$ and $R_\mathrm{rs}$ coincide with the SMS radius: $R_0 = R_\mathrm{s} = R_\mathrm{fs} = R_\mathrm{rs}$. For numerical stability, we found that it is better to initialise the forward and reverse shock positions as $R_\mathrm{fs} = R_\mathrm{s}(1+\epsilon_\mathrm{s})$ and $ R_\mathrm{rs} = R_\mathrm{s}(1-\epsilon_\mathrm{s})$, where $\epsilon_\mathrm{s}$ is a small number (we used $\sim 10^{-10}$). 

The initial shock momentum $S_\mathrm{r}$ and mass $M_\mathrm{s}$ are obtained using the fact that initially, the pressure at the forward ($P_\mathrm{fs}$, \eqref{eq:light-curve-model:jump-conditions-forward-shock}) and reverse shocks ($P_\mathrm{rs}$, \eqref{eq:light-curve-model:jump-conditions-reverse-shock}) are balanced. This is because the shock interfaces coincide at $t_0$. Thus, we solve
\begin{align}
    P_\mathrm{fs} \stackrel{!}{=} P_\mathrm{rs} \: .
\end{align}
This equation can be solved analytically by evaluating the pre-shocked density and velocities of the ejecta and the CSM at $r=R_0$ and taking $c\beta_\mathrm{s} = c\beta_\mathrm{fs} = c\beta_\mathrm{rs}$ (because these velocities are the same initially). We obtain a quadratic equation for the shell velocity $\beta_\mathrm{s}$:
\begin{align}
     \left[1 - \frac{\rho_\mathrm{CSM,fs}}{\rho_\mathrm{eje,rs}} \left(1-\beta_\mathrm{eje,rs}^2 \right) \right] \beta_\mathrm{s}^2 - 2 \beta_\mathrm{eje,rs}\; \beta_\mathrm{s} + \beta_\mathrm{eje,rs}^2 = 0 \: ,
\end{align}
where $\rho_\mathrm{CSM,fs}$, $\rho_\mathrm{eje,rs}$ and $\beta_\mathrm{eje,rs}$ are defined as in \eqref{eq:light-curve-model:jump-conditions-forward-shock} and \eqref{eq:light-curve-model:jump-conditions-reverse-shock}. We take the negative root of $\beta_\mathrm{s}$ in the quadratic formula because it could be superluminal otherwise. 

Setting the initial shock mass $M_\mathrm{s}$ to zero might cause numerical problems. We therefore take one small time step $\Delta t$ of \eqref{eq:light-curve-model:shock-mass-ODE} to compute the initial mass as $M_\mathrm{s} = \Dot{M}_\mathrm{s} \Delta t$. For this we also set $\beta_\mathrm{s} = \beta_\mathrm{fs} = \beta_\mathrm{rs}$. 

With the initial values of $M_\mathrm{s}$ and $\beta_\mathrm{s}$ now known, we compute the initial momentum using $S_\mathrm{r} = M_\mathrm{s} \Gamma_\mathrm{s} c\beta_\mathrm{s}$ (\eqref{eq:light-curve-model:momentum-shock}). The initial internal energy of the shock region is zero: $E_\mathrm{int,sh} = 0$. We initialise it with some very small non-zero number $>0$ for the sake of numerical stability. 

With the initial conditions in place, we can integrate the system \eqref{eq:light-curve-model:shock-momentum-ODE}, \eqref{eq:light-curve-model:shock-mass-ODE}, \eqref{eq:light-curve-model:shock-position-ODE}, \eqref{eq:light-curve-model:forward-reverse-shock-position-ODE} and \eqref{eq:light-curve-model:shock-internal-energy-ODE}.

\subsection{Computation of Radiation Flux and Apparent Magnitude} 
\label{subsec:computation-of-radiation-flux-and-apparent-magnitude}

\begin{table}
\centering
\begin{tabular}{c|c} 
 filter name & Filter bandwidth [$\mu$m] \\ [0.5ex] 
 \hline
     JWST F090W & $0.795$--$1.005$ \\ 
     JWST F115W & $1.013$--$1.282$ \\
     JWST F150W & $1.331$--$1.668$ \\ 
     JWST F200W & $1.755$--$2.228$ \\
     JWST F277W & $2.422$--$3.131$ \\
     JWST F356W & $3.136$--$3.981$ \\
     JWST F444W & $3.881$--$4.982$ \\
     \hline
     EUCLID J-band & $1.1676$--$1.5670$ \\
     EUCLID H-band & $1.5215$--$2.0214$ \\
     \hline
     RST J-band & $1.131$--$1.454$ \\
     RST H-band & $1.380$--$1.774$ \\
\end{tabular}
\caption{Telescope filter transmission functions employed in this work. The specifications for each filter can be found on the web site (\protect\url{https://jwst-docs.stsci.edu/jwst-near-infrared-camera/nircam-instrumentation/nircam-filters}[accessed 2024–11–22]) for JWST, in \protect\cite{Euclid:2022vkk} for EUCLID, and in \protect\cite{Spergel:2013arXiv1305.5422S} for the Roman Space Telescope (RST) Wide Field Instrument (WFI).}
\label{tab:light-curve-model:telescope-filter-functions}
\end{table}

Explosions of SMSs are expected to occur in the early universe at redshift $z>13$--$20$ (see, e.g., \citealt{Patrick2023:10.1093/mnras/stad1179}). The source spectrum will therefore be shifted due to cosmological redshift when observed in the present day. Technical constraints posed by telescope sensors must also be considered to assess whether a given source is observable using the available hardware. 

In the previous sections, we laid out how to compute the expected luminosity of an SMS explosion. In the first phase of the light curve, the shock region is optically thick and the radiation is caused by diffusion out of the photosphere. The size of the photosphere $R_\mathrm{ph}(t)$ is given approximately by the shock position $R_\mathrm{s}(t)$ and the observed surface temperature is the colour temperature $T_\mathrm{col}$ (\eqref{eq:light-curve-model:color-temperature-shock-surface}). When the shock region is thermalised ($\eta<1$), it radiates like a black body emitter. But otherwise the spectrum will change to a modified blackbody spectrum. For simplicity, we here approximate this by a shifted and re-scaled blackbody spectrum (we comment on the effect of this approximation below). We define the spectrum as follows:
\begin{align}
    B_\mathrm{mod}(\nu,T_\mathrm{col}) = \frac{2h}{c^2} \frac{\nu^3}{\exp\left( h\nu / k_B T_\mathrm{col} \right) - 1} \times \min \left\{1\,,\: \left( \frac{T_\mathrm{BB,surf}}{T_\mathrm{col}}\right)^4 \right\} \: , \label{eq:light-curve-model:Planck-spectral-radiance}
\end{align}
where $h$ is Planck's constant. $B_\mathrm{mod}$ is defined in a way that for $\eta<1$ it is equal to a blackbody spectrum. Otherwise, the normalization by $(T_\mathrm{BB,surf}/T_\mathrm{col})^4 = (L_\mathrm{diff}/4\pi\sigma R_\mathrm{s}^2 T_\mathrm{col}^4)$ makes sure that the total luminosity is always equal to $L_\mathrm{diff}$. \eqref{eq:light-curve-model:Planck-spectral-radiance} is in general different from a modified blackbody spectrum and will under-estimate the flux in the red part of the spectrum. But this is not too relevant because (i) the spectrum is even more sensitive to small changes in $\eta$ (recall $T_\mathrm{col}\propto\eta^2$), (ii) our model is anyway not very accurate when $\eta>1$ due to the assumption that $\eta$ is constant throughout the shock shell and (iii) with $\eta>1$ we are only interested in the over-all qualitative features and trends accurate to within one or two magnitudes.

Using $B_\mathrm{mod}$, we can compute the expected radiation flux that can be captured by current telescopes like the JWST and others. The flux as a function of frequency in the observer frame is given by (see, e.g., \citealt{Hogg:2002yh,Chowdhury:2024wvm})
\begin{align}
    F_{\nu_\mathrm{obs}} = \pi (1+z) \left( \frac{R_\mathrm{ph}}{d_L} \right)^2 B_\mathrm{mod}\left( (1+z)\nu_\mathrm{obs}, T_\mathrm{col} \right) \: , \label{eq:light-curve-model:flux-observer-frame}
\end{align}
where $R_\mathrm{ph}$ is the size of the photosphere of the source object and $d_L$ is the luminosity distance. It is a function of the redshift and depends on the cosmological model chosen (see \citealt{Hogg:1999ad}):
\begin{align}
    d_L(z) = (1+z) \frac{c}{H_0} \int_0^z \frac{dz'}{\sqrt{\Omega_m(1+z')^3 + \Omega_\Lambda}} \: . \label{eq:light-curve-model:def-luminosity-distance}
\end{align} 
We assume the standard model of cosmology and adopt the values from the Planck collaboration for the cosmological parameters $H_0=67.4\, \mathrm{km\,s^{-1}\,Mpc^{-1}}$, $\Omega_m=0.315$, $\Omega_\Lambda=0.685$ and $\Omega_k=0$ \citep{Planck:2018vyg}. We also set the universe radiation density $\Omega_r=0$ because its contribution is negligible at $z \sim 10$--$20$. 

In this work, we use the AB magnitude system to measure brightness. The apparent magnitude $m_\mathrm{AB}$ of a source at redshift $z$, seen by an observer is given by \citep{Hogg:2002yh}:
\begin{align}
    m_\mathrm{AB} := -2.5 \log_{10} \left( \frac{ \int_0^\infty F_\nu\, a(\nu) \, t(\nu) \,\nu^{-1} d\nu}{ \int_0^\infty 3631\,\mathrm{Jy}\, t(\nu) \,\nu^{-1} d\nu } \right) \: . \label{eq:light-curve-model:def-AB-magnitude}
\end{align}
The integral is performed over the observer frequency $\nu=\nu_\mathrm{obs}$. Here, $F_\nu$ is the flux in the observer frame (\eqref{eq:light-curve-model:flux-observer-frame}), $a(\nu)$ is the absorption function (due to, e.g., intergalactic absorption), and $t(\nu)$ is the transmission function of a given telescope filter. 

Modern telescopes operate different transmission filters, which only let through specific parts of the spectrum. In this work, we compute $m_\mathrm{AB}$ as seen through different filters of JWST, EUCLID and the Roman Space Telescope. We approximate the filters as step functions, which are zero outside the filter width and equal to the transmittance inside this range. We here assume a constant transmittance, so it will cancel out when computing $m_\mathrm{AB}$. The specifications of the filters used are summarised in \tabref{tab:light-curve-model:telescope-filter-functions}. 

When observing objects in the early universe, the absorption of light by the intergalactic medium (IGM) needs to be taken into account. We do this using the absorption function $a(\nu)$. The present universe is transparent to electromagnetic radiation over essentially the whole spectrum. This is because densities are small and the IGM is fully ionised and thus light can pass through. In the early universe before redshifts of $z\sim 5$--$6$, however, the IGM was not yet fully ionised.
Photons above the excitation (and ionisation) energy of hydrogen will then be absorbed. This leads to an absorption trough (Gunn-Peterson trough), where essentially all flux at frequencies $\nu_\mathrm{obs} > \nu_\mathrm{Ly\alpha}/(1+z)$ is absorbed ($\nu_\mathrm{Ly\alpha}$ is the frequency of the Lyman-$\alpha$ line of hydrogen). Thus, we take $a(\nu)=0$  if $\nu > \nu_\mathrm{Ly\alpha}/(1+z)$, and  $a(\nu)=1$ otherwise. 

Furthermore, light could be absorbed or attenuated by the CSM itself. Although it is transparent at first (recall the discussion above \eqref{eq:light-curve-model:Larson-CSM-Model}), the large amount of photons released from the shocked region could partially ionise the CSM. There could also be attenuation of Balmer-continuum photons under certain conditions. We discuss these points in detail in Section \ref{subsec:discussion:ionization-of-CSM}.

\section{Results} \label{sec:Results}
\subsection{Model Selection} \label{subsec:results:model-selection}

\begin{table}
\centering
\begin{tabular}{c|c|c|c} 
 model name & $R_0\,[R_\odot]$ & $E_\mathrm{kin,eje}\,[\mathrm{erg}]$ & $M_\mathrm{eje}\,[M_\odot]$ \\ [0.5ex] 
 \hline
     FujH1 & $10^4$ & $9.5\e{54}$ & $23613$ \\
     FujH2 & $10^4$ & $5.1\e{55}$ & $37255$ \\
     FujH4 & $10^4$ & $1.9\e{56}$ & $82467$ \\
     FujDif1 & $10^4$ & $2.8\e{56}$ & $110322$ \\
     FujHe1 & $10^4$ & $1.7\e{54}$ & $5611$ \\
     FujHe2 & $10^4$ & $8.6\e{54}$ & $8199$ \\
     FujHe4 & $10^4$ & $4.2\e{55}$ & $19378$ \\
     \hline
     NagCol1 & $10^4$ & $8.49\e{54}$ & $29737$ \\
     NagCol2 & $10^4$ & $1.22\e{55}$ & $96731$ \\
     NagPul1 & $10^4$ & $1.31\e{53}$ & $3051$ \\
     NagPul2 & $10^4$ & $1.73\e{53}$ & $4312$ \\
     NagExp & $10^4$ & $1.03\e{55}$ & $298600$ \\
\end{tabular}
\caption{Initial SMS radius $R_0$, ejecta kinetic energy $E_\mathrm{kin,eje}$ and total ejecta mass $M_\mathrm{eje}$. The parameters are for collapsing/pulsating rapidly accreting SMSs. The values are informed by stellar evolution calculations and numerical simulations; see main text.}
\label{tab:results:initial-parameters-accreting-SMS}
\end{table}

Our light curve model is computed for inputs of the SMS radius $R_0$, the kinetic energy of the ejecta from the explosion $E_\mathrm{kin,eje}$ and the ejecta mass $M_\mathrm{eje}$. Due to this simple parametrisation, our model is quite versatile and we therefore can apply it also to scenarios for PopIII stars that are surrounded by a dense envelope, e.g., because they are located inside an ACH where typical densities reach $\sim 10^{10-11}\,$cm$^{-3}$ near the centre $r<10^{-3}\,$pc (see \eqref{eq:light-curve-model:Larson-CSM-Model}). One example is the explosion of a ``very massive stars'' of several $100$--$1000\,M_\odot$ as studied, e.g., in \citealt{Uchida:2017qwn,Volpato:2023ApJ...944...40V}. 

Explosion and mass ejection scenarios for rapidly accreting SMSs ($\Dot{M}>0.02\,M_\odot\mathrm{yr^{-1}}$) have some variety. First, there is a class of explosion scenarios where the GRI is triggered and the SMS explodes. The first option is that, as density and temperature rise during the collapse, rapid efficient nuclear fusion of hydrogen or helium is activated. This will release enough energy to completely disrupt the star without leaving behind a BH. Numerical simulations of this scenario have been performed by \citealt{Nagele:2024aev}. The other collapse scenario is that the collapsing SMS forms a BH with a torus around it. Infalling matter will then bounce off the torus and produce highly energetic ejecta, as simulated in our previous work \citep{Fujibayashi:2024vnb}. Secondly, SMSs can also run into a pulsational instability due to the GRI \citep{Nagele:2022odb,Nagele:2024aev}. These pulsations will eject a significant amount of matter from the SMS but without destroying it \citep{Nagele:2024aev}. Pulsations might therefore be repeating. This scenario can be studied using our model, which is why we also investigate this case here. 

For these scenarios, we select models in the following way: Rapidly accreting SMSs all feature a bloated atmosphere with large radius. It depends on the accretion rate but, at collapse onset, the radius will be in the range of $1$--$2\e{4}\,R_\odot$ (see \citealt{Herrington:2022dbu,Hosokawa:2013mba}). Since our light curve model only depends weakly on the initial radius (see Appendix \ref{sec:appendix:light-curve-model-validation}), we here opt for $R_0=1\e{4}\,R_\odot$ for our fiducial accreting SMS radius. 

For $E_\mathrm{kin,eje}$ and $M_\mathrm{eje}$ we make choices that are informed from numerical simulations. In a previous work \citep{Fujibayashi:2024vnb}, we simulated the collapse of rapidly rotating SMS cores. We take the value of $E_\mathrm{kin,eje}$ as reported in their Table 2. For $M_\mathrm{eje}$ we take the ejecta mass from the core (also from their Table 2) and add to that the mass of the bloated atmosphere. The atmosphere is neglected in their work. But since the models of \citealt{Fujibayashi:2024vnb} are rapidly rotating, the core likely makes up a large fraction ($>90\,\%$) of the total SMS mass (see \citealt{Haemmerle:2021tcj}). We here take a core mass fraction of $90\,\%$. The atmosphere thus makes up the remaining $10\,\%$ of the SMS mass. 

We also use the simulation results from \citealt{Nagele:2024aev}. We here take the explosion energy and ejecta mass from their models with $0.1$--$1\,M_\odot\mathrm{yr^{-1}}$. For their pulsating and exploding models, we take the values as in their Table 1. For the collapsing models, we assume that the relation between the SMS core mass, ejecta from the core and explosion energy found in \citealt{Fujibayashi:2024vnb} holds: we take as the ejecta mass $1\%$ of the core mass ($M_\mathrm{eje,c}\sim 0.01 M_\mathrm{core}$) and add to that the mass outside of the core. For the explosion energy, we use that $E_\mathrm{kin,eje} \sim 0.5 M_\mathrm{eje,c} (0.2c)^2 \sim 0.5 \times 0.01 M_\mathrm{core} \times (0.2c)^2$ \citep{Fujibayashi:2024vnb}. We collect the model parameters that we use in our work in \tabref{tab:results:initial-parameters-accreting-SMS}. 

The chemical composition of the ejecta should be roughly primordial, even in the cases that the GRI is encountered while the SMS has a helium core. This is because during the collapse most of the core region will fall into the newly formed BH (see \citealt{Fujibayashi:2024vnb}). The total ejecta is composed of the ejected material from the core plus the material from the bloated atmosphere, which has essentially primordial composition. The bloated envelope will thus dominate the chemical composition and we assume that all ejecta for rapidly accreting SMS are hydrogen rich with roughly primordial composition ($75\,\%$ H, $25\,\%$ He).

For slower accreting SMSs ($\Dot{M}<0.02\,M_\odot\mathrm{yr^{-1}}$), the stars grow to several thousand $M_\odot$ and eventually collapse due to the pair-instability (see \citealt{Shibata:2024xsl,Uchida:2018ago}). Their stellar evolution is also different. Slowly accreting SMSs evolve as blue hypergiants \citep{Herrington:2022dbu} and the strong UV radiation might ionise or blow away the surrounding CSM \citep{kiyuna:2024}. This could have significant impact on the dynamics of the ejecta and the observability to distant observers. In this work we focus on the cases where the CSM is dense and mostly optically thin. We therefore focus on the more rapidly accreting SMSs, where a dense un-ionised CSM is more likely to exist at the time of collapse. During the light curve evolution, the CSM might get self-ionised by the radiation, which changes the optical depth. Balmer absorption could also be relevant. We discuss these points in detail in section \ref{subsec:discussion:further-investigation-and-future-model-improvements}.

Whether an explosion happens depends strongly on the rotation rate. In the above mentioned simulations, it was assumed that the SMSs are rapidly rotating. But it should be additionally noted that SMSs may not rotate at significant rates. This is due to the $\Omega\Gamma$-limit \citep{Lee:2016ApJ...820..135L,Haemmerle:2018}. It prevents accretion of angular momentum due to a barrier of the centrifugal force and radiation pressure. However there is some evidence that SMSs might be differentially rotating with a slowly rotating outer layer and an inner core region which can rotate at significant fractions of the Keplerian rotation rate \citep{Haemmerle:2018}. Additionally, it was shown by \citealt{Kimura:2023ApJ...950..184K} in a simulation of an accreting SMS protostar of $<10\,M_\odot$ that a high rotation rate can be maintained even while rapidly accreting. But longer simulations are needed to test whether the rotation rate can be maintained up until the collapse.

\subsection{Bolometric Light Curve Properties of Supermassive Star Explosions} \label{subsec:bolometric-light-curves-of-supermassive-stars}

\begin{figure*}
    \centering
    \includegraphics[width=0.49\textwidth]{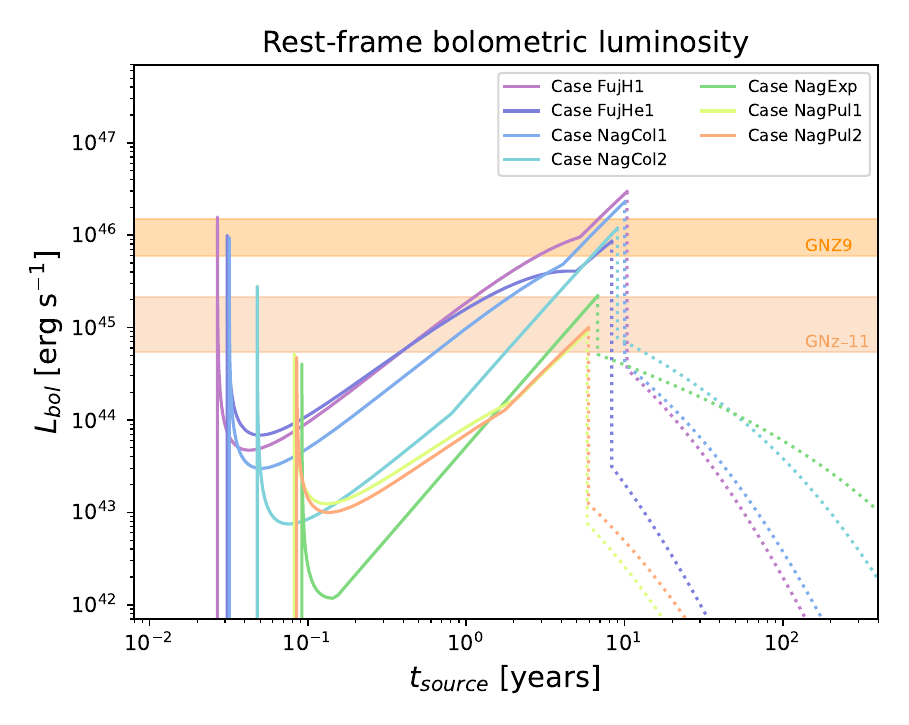}
    \includegraphics[width=0.49\textwidth]{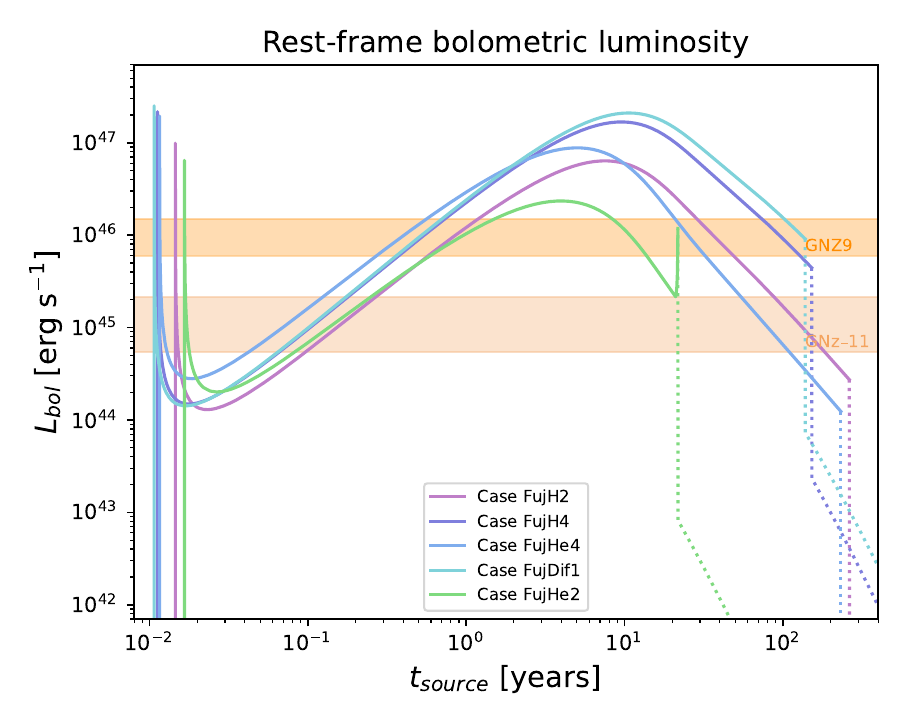}
    \caption{Bolometric luminosity as a function of time of the SMS models from \tabref{tab:results:initial-parameters-accreting-SMS} in the source frame. \textbf{Left panel:} Models where the radiation is always fully thermalised (i.e. $\eta<1$). The phase where the shock is optically thick is marked using a continuous line. Dotted lines denote the phase where the shock is optically thin and we observe non-thermal radiation. The shaded regions mark observed luminosities of two high-redshift AGN; we put them here as a comparison.
    \textbf{Right panel:} Same as the left panel, but for models where the radiation is not fully thermalised at some point during the evolution (i.e. $\eta>1$).}
    \label{fig:results:Lbol-all-SMS-models}
\end{figure*}

We here discuss the expected light curves of SMS explosions from our light curve models. We compute the bolometric light curve as a function of time for the models in \tabref{tab:results:initial-parameters-accreting-SMS}. The model parameters were informed by previous numerical simulations and stellar evolution calculations of rapidly accreting SMSs. 

In \figref{fig:results:Lbol-all-SMS-models}, we show the bolometric luminosity in the source frame as a function of time for our selected models. We first focus on the left panel, where we show all models for which the radiation is always thermalised, i.e. $\eta<1$ at all times. It is found that the light curves follow the same general trend. At the beginning at $t=t_0$, there is a sharp luminosity peak with a duration of around one dynamical time $t_0$. An initial radiation burst due to a shock breakout may occur in reality \citep[see e.g.,][]{Moriya:2021kzc}. However, the early emissions cannot be reliably modelled using our assumptions (e.g., diffusion approximation and simplified ejecta matter profile). Hence, we neglect this feature in the following discussion.

After the initial peak, the luminosity increases gradually. In this stage, the shock gains internal energy by conversion of kinetic energy at the forward and reverse shock fronts. This energy then gradually diffuses out of the shock region due to radiation loss. Since the optical depth in this phase is quite high, the energy loss is smaller than the energy gained. The internal shock energy therefore increases and the luminosity increases as well. At the same time, the shock radius expands and the shock over all cools down. This continues until the effective surface temperature reaches the ionisation temperature of hydrogen ($T_\mathrm{ion} \approx 6000\,$K). The hydrogen in the shock region then starts to recombine. At this point, the shock effective surface temperature stays constant at $T_\mathrm{ion}$ and there is a kink in the light curve. The luminosity then increases $\propto R_\mathrm{s}^2 \approx (c \beta_\mathrm{s} t)^2$ until all the internal energy of the shock is released and the shock becomes transparent.

After that, the bolometric luminosity is produced by non-blackbody radiation from the optically thin shock interacting with the CSM. For our models, these emissions are much smaller than the previous emissions from the optically thick phase. Hence, we do not model them in detail here. 

For some cases, the overall peak in the light curve can come before the recombination phase. We found that this peak happens when the diffusion timescale ($t_\mathrm{d} \approx 3\tau_\mathrm{s}(R_\mathrm{fs}-R_\mathrm{rs})/c$) of photons out of the shocked shell becomes comparable to the dynamical timescale, $t_\mathrm{dyn} \approx R_\mathrm{s} / (c\beta_\mathrm{s})$. Most of the radiation will then be released within one diffusion timescale. Additionally, we find that the deceleration timescale is often similar to the diffusion timescale. It is given by the time when the shock has accreted a similar amount of CSM mass compared to the initial ejecta mass, $M_\mathrm{eje}$. The light curve maximum can then have increased luminosity because the conversion of kinetic energy to internal energy is most efficient when there is a large amount of kinetic energy loss. 

Generally, the length of the optically thick shock phase is on the order of $10$ yrs and the luminosity can reach $10^{45}\,\mathrm{erg\,s^{-1}}$ to a few times $10^{46}\,\mathrm{erg\,s^{-1}}$. The length is maximised if the diffusion peak coincides with the end of the optically thick phase. The peak luminosity depends on the initial ejecta kinetic energy and on the ratio of kinetic energy to ejecta mass. See also Appendix \ref{sec:appendix:light-curve-model-validation} for more information on systematic and scaling behaviour of the model. 

In the right panel of \figref{fig:results:Lbol-all-SMS-models}, we show the cases which have $\eta>1$ at some point in the evolution. The general trends are the same as in the $\eta<1$ cases. The early--time short luminosity peak and the luminosity rise are the same as those found for the other cases. But for the cases here, the radiation becomes out of equilibrium and $\eta$ goes above unity during the evolution. This happens around the time of the luminosity peak for our cases. This leads to an increased electron temperature in the radiative layer, which prevents recombination and thus tends to prevent the plateau phase with constant surface temperature. Instead, the luminosity steadily falls after the peak and eventually the shock region becomes transparent because it becomes too dilute and $\tau_\mathrm{s}$ drops. The only outlier is the case FujHe2, where the recombination phase is reached and the internal energy is subsequently rapidly released, leading to a late--time luminosity peak. However, we note that this could also be an artifact of our modelling and should not be taken at face value.

The optically thick phase length and peak luminosity are larger for the non--thermalised cases ($\eta>1$) and reach up to $\sim 200\,$yrs and $10^{47}\,\mathrm{erg\,s^{-1}}$, respectively. This is not surprising because the higher luminosity cases need more photons to remain thermalised. In these cases, photon production cannot keep up, which makes full thermalisation less likely. Hence, we would expect very energetic cases to not thermalise. The lack of thermalisation then pushes up the average photon energy and electron temperature, which then prevents recombination of the shell, thus lengthening the optically thick shock phase.

The pulsating models (NagPul1, NagPul2) and the exploding model (NagExp) have peak bolometric luminosities of $\sim 10^{45}\,\mathrm{erg\,s^{-1}}$. This is comparable to the luminosity of some AGN observed by JWST in the early universe at redshift $z\sim 10$ (e.g., GNz–11, \citealt{Maiolino:2023zdu}). The lower-mass variants of the exploding hydrogen and helium core models (FujH1, FujHe1, NagCol1, NagCol2) reach a peak luminosity on the order of a few $10^{46}\,\mathrm{erg\,s^{-1}}$. The models with higher mass and explosion energy (FujH2, FujH4, FujHe2, FujHe4, FujDif1) have luminosities over $10^{47}\,\mathrm{erg\,s^{-1}}$. This is larger than some of the brightest high--redshift AGN. Hence, those models should be bright enough to be clearly observable even at very high redshift $z>10$.

A point to consider for observational prospects is the possible self-ionisation of the CSM by ionising photons from the shock region. We discuss this point, along with possible attenuation of Balmer-continuum photons, in detail in Section \ref{subsec:discussion:ionization-of-CSM}. In the following, we therefore focus on discussing the intrinsic light curves of SMS explosions, while keeping in mind that the ultimate visibility could be subject to change based on the ionisation of the CSM.

\begin{figure}
    \centering
    \includegraphics[width=0.498\textwidth]{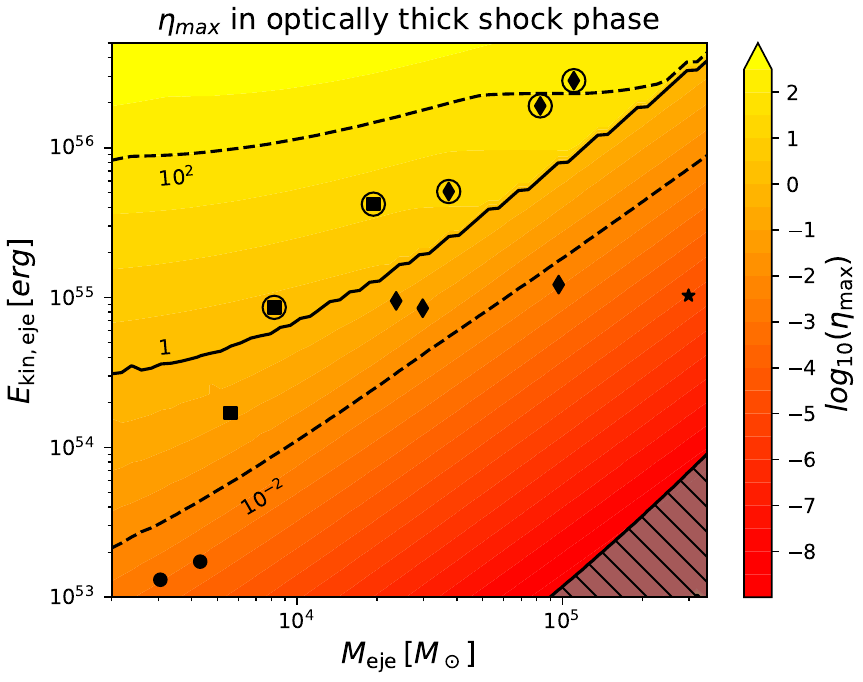}
    \caption{Maximum value of $\eta$ that can be achieved during the optically thick shock phase (excluding the initial sharp peak) as a function of the total ejecta mass $M_\mathrm{eje}$ and the kinetic energy of the ejecta $E_\mathrm{kin,eje}$. The symbols mark the SMS models from \tabref{tab:results:initial-parameters-accreting-SMS}: diamonds denote collapsing H models, squares denote collapsing He models, the star denotes the full SMS fusion powered explosion, and circles denote pulsating models. The circled symbols mark models where $\eta>1$ during their evolution. Black curves indicate contours of equal values of $\eta$. The solid curve marks $\eta=1$. The gray striped/shaded region marks configurations where the shock luminosity in the optically-thin phase is larger than the diffusion luminosity from the optically thick phase. It thus marks the range of validity of our model.}
    \label{fig:results:eta-parameter-grid-Ekin-Meje}
\end{figure}

Next, we scan the parameter space using our light curve model to get a better understanding of the photometric properties of typical SMS explosions. We first focus on the thermal coupling coefficient $\eta$, because it indirectly delineates the parameter space where our model is more reliable ($\eta<1$) or less reliable ($\eta>1$).

In Figure \ref{fig:results:eta-parameter-grid-Ekin-Meje}, we show the maximum value of $\eta$ that can be achieved during the optically thick shock phase of the light curve. The symbols mark the positions of our models from \tabref{tab:results:initial-parameters-accreting-SMS} in the parameter space. The symbols with circles around them highlight models where $\eta>1$ during their evolution (see \figref{fig:results:Lbol-all-SMS-models}). The grey striped/shaded area marks the region where the bolometric luminosity in the optically-thin phase dominates that in the optically-thick phase. We therefore exclude this part since our model could incur significant errors in this parameter range. We note that all of our chosen SMS configurations lie well within the region where our model is applicable. 

As discussed before, models with $\eta>1$ have phases where the radiation is out of equilibrium and where the spectrum is bluer compared to a blackbody. In the region where $\eta>1$, our model is capable of showing trends and rough order of magnitude results, but it should not be relied upon for precise quantitative results. We see that the models FujHe2, FujHe4, FujH2, FujH4 and FujDif1 fall in the region where $\eta>1$. Due to the large flux of ionising photons, the surrounding CSM could be fully ionised, as discussed in section \ref{subsec:discussion:ionization-of-CSM}.

\begin{figure}
    \centering
    \includegraphics[width=0.498\textwidth]{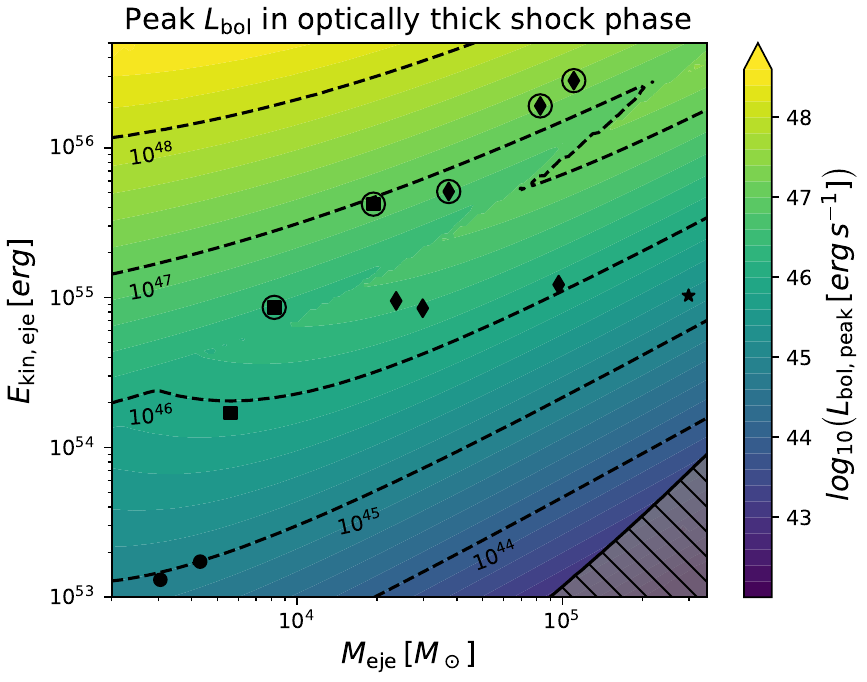}
    \caption{Peak bolometric luminosity $L_\mathrm{bol,peak}$ in the optically thick shock phase (excluding the initial sharp peak) as a function of the total ejecta mass $M_\mathrm{eje}$ and the kinetic energy of the ejecta $E_\mathrm{kin,eje}$. The symbols mark the SMS models from \tabref{tab:results:initial-parameters-accreting-SMS}. They have the same meaning as in \figref{fig:results:eta-parameter-grid-Ekin-Meje}. Dashed curves indicate contours of equal peak bolometric luminosity. The gray striped/shaded region marks configurations outside the range of validity of our model (same as \figref{fig:results:eta-parameter-grid-Ekin-Meje}).}
    \label{fig:results:Lbol-peak-grid-Ekin-Meje}
\end{figure}

Figure \ref{fig:results:Lbol-peak-grid-Ekin-Meje} shows the peak luminosity of SMS explosion light curves with different initial ejecta masses and kinetic energies. We excluded the early time peak, since it is likely unphysical and would not be visible due to the short timescale, regardless. The symbols with circles around them highlight models where the emitting region is out of equilibrium ($\eta>1$) at some point during the light curve evolution. If this is the case at the luminosity peak, the increased ionising photon flux could form an ionised bubble, as discussed in detail in section \ref{subsec:discussion:ionization-of-CSM}. This could modify the value of the peak luminosity to be somewhat different from the predicted value.

There is zig-zag-shaped a bump in the peak luminosity on the line from $(2\e{3}\,M_\odot,2\e{54}\,\mathrm{erg})$-- $(4\e{5}\,M_\odot,5\e{56}\,\mathrm{erg})$. This pattern comes from two effects that happen close in the parameter space. The first effect is that the position of the luminosity peak changes from the diffusion peak (above this line) to the late-time recombination phase (below this line). The second effect is that for some models, $\eta$ rises above unity during the recombination phase. This raises the electron temperature above $T_\mathrm{ion}$. This ultimately results in a short-lived luminosity bump, which may be larger than the diffusion peak. 

The general trends in the figure are as follows. The peak luminosity $L_\mathrm{bol,peak}$ generally decreases when increasing the ejecta mass but it increases when increasing the kinetic energy, except for the aforementioned zig-zag line. But the proportionality is different for the lower region of the plot, where $\eta<1$ and for the upper region where $\eta>1$ (cf. Fig. \ref{fig:results:eta-parameter-grid-Ekin-Meje}). When varying the kinetic energy and the ejecta mass in the region $\eta<1$, the luminosity roughly scales like $L_\mathrm{bol,peak}\propto E_\mathrm{kin,eje}^{1.4}$ and $L_\mathrm{bol,peak}\propto M_\mathrm{eje}^{-1.2}$, respectively. For the region $\eta>1$, we find a scaling behaviour of $L_\mathrm{bol,peak}\propto E_\mathrm{kin,eje}^{1.2}$ and $L_\mathrm{bol,peak}\propto M_\mathrm{eje}^{-0.6}$. Because the scaling relations are not exact, the exponents can change by $\pm 0.2$ depending on where we read them off in the figure. The peak in the $\eta>1$ region can be understood if it coincides with the diffusion timescale $t_\mathrm{diff} \propto M_\mathrm{eje}^{0.75} E_\mathrm{kin,eje}^{-0.25}$ \citep[][Eq. (24)]{Uchida:2018ago} where most radiation has been released, thus $L_\mathrm{bol,peak} \sim E_\mathrm{kin,eje}/t_\mathrm{diff} \propto M_\mathrm{eje}^{-0.75} E_\mathrm{kin,eje}^{1.25}$. This analytic estimate is roughly consistent with our results.


\begin{figure}
    \centering
    \includegraphics[width=0.498\textwidth]{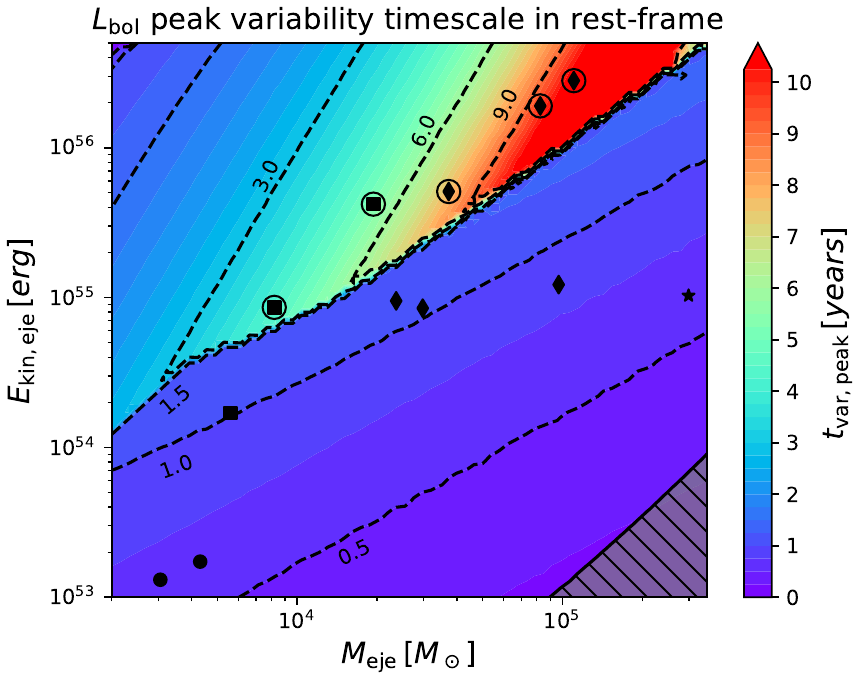}
    \caption{Luminosity variability timescale $t_\mathrm{var,peak}$ around the light curve peak (excluding the initial sharp peak), as defined in the main text. The symbols mark the SMS models from \tabref{tab:results:initial-parameters-accreting-SMS}. They have the same meaning as in \figref{fig:results:eta-parameter-grid-Ekin-Meje}. Dashed curves indicate contours of equal luminosity variability timescale. The gray striped/shaded region marks configurations outside the range of validity of our model (same as \figref{fig:results:eta-parameter-grid-Ekin-Meje}).
    The sharp transition line marks the point where the light curve peak changes from the diffusion peak to the recombination phase. The typical variability time of the sources lies in the range $0.5$--$10\,$yrs.}
    \label{fig:results:time_variability-grid-Ekin-Meje}
\end{figure}

Because of their formation environment in collapsing ACHs, SMSs are expected to exist only at redshifts larger than $z\sim 10$--$20$ (see, e.g., \citealt{Patrick2023:10.1093/mnras/stad1179}). In non-standard scenarios, their formation might be somewhat delayed, see \citealt{kiyuna:2024}. But the redshift should always be larger than $z> 6$--$7$. Cosmological time dilation is significant at these redshifts and needs to be taken into account when assessing observational prospects of SMS explosions. 

The time dilation will increase the timescale of the observed SMS explosion light curves by a factor $(1+z)$. The timescale of the optically thick shock phase -- for $\eta<1$ around $\sim10\,$yrs and for $\eta>1$ around $\sim100\,$yrs --  will then lead to transient timescales on the order of $100$ ($\eta<1$) or $1000$ ($\eta>1$) yrs. The optically thin phase could last even longer. This has implications on the observed variability of these transient sources in high-redshift telescope data. If the variability of the transient is small enough, it might be confused photometrically with a persistent high-redshift source. This is aggravated by the fact that telescopes like JWST or EUCLID started their observations quite recently (since July 2020 and February 2023, respectively), so that long-term transients might be classified as persistent sources due to the lack of elapsed observation time.

Prominent high-redshift sources, such as little red dots (LRDs), have been observed to show low variability. More specifically, the surveys by, e.g., \citealt{Tee:2024lackrestframeuvvariability} and \citealt{Zhang:2024arXiv241102729Z} show a source variability of LRDs of at most $\Delta m_\mathrm{AB}\sim 0.2$ over $1$--$2$\,yrs. \citealt{OBrian:2024ApJS..272...19O} did a similar study that searched for transients in combined data from the Hubble space telescope and JWST, extending the time range up to 4\,yrs for some objects. 

In terms of luminosity, a magnitude variability of $\Delta m = 0.2$ corresponds to $L_\mathrm{bol}$ is $\Delta L/L = 10^{\pm \Delta m / 2.5}$, i.e., roughly $20\,\%$. Based on this, we define the luminosity variability timescale. This is the timescale, in the source's rest-frame, where the luminosity changes by a factor of $\Delta L/L$. Within this time, the transient source will show at most a similar variability to LRDs and will therefore appear to be persistent.

In Figure~\ref{fig:results:time_variability-grid-Ekin-Meje}, we show the luminosity variability timescale around the light curve peak, $t_\mathrm{var,peak}$, as a function of ejecta mass and ejecta kinetic energy. As before, we excluded the initial sharp light peak at the beginning to not skew our results, and because $t_\mathrm{var,peak}$ around this peak would be very short regardless. As in the previous Figure \ref{fig:results:Lbol-peak-grid-Ekin-Meje}, the markers show the cases from \tabref{tab:results:initial-parameters-accreting-SMS}. The stripped/shaded region marks configurations which lie outside the range of validity of our light curve model. The sharp transition line marks the point where the light curve peak changes from the diffusion peak to the recombination phase. The typical variability time of the sources lie in the range $0.5$--$10$\, yrs in the source-frame. The observed timescales would be longer by a factor $(1+z)$. When the luminosity peak lies in the recombination phase, the variability timescale is around $0.5$--$1.5$\,yrs. It is significantly longer ($>2\,$yrs), if the luminosity peak is located at the diffusion peak. 

Even though the figure shows a sharp transition between the variability timescale length (because the peak position changes), it does not mean that the observed changes would be that drastic. Around the transition line, the diffusion peak and the late-time peak have a similar brightness. Both peaks could therefore be observed separately with different luminosities. Both peak phases would then be separated by a dimmer period. Therefore, a smoking gun detection for SMS explosions would be a high-$z$ quasi-persistent source that shines for several years, then vanishes (or dims) for a few years and then reappears a few decades later with a different luminosity.

\begin{figure*}
    \centering
    \includegraphics[width=0.32\textwidth]{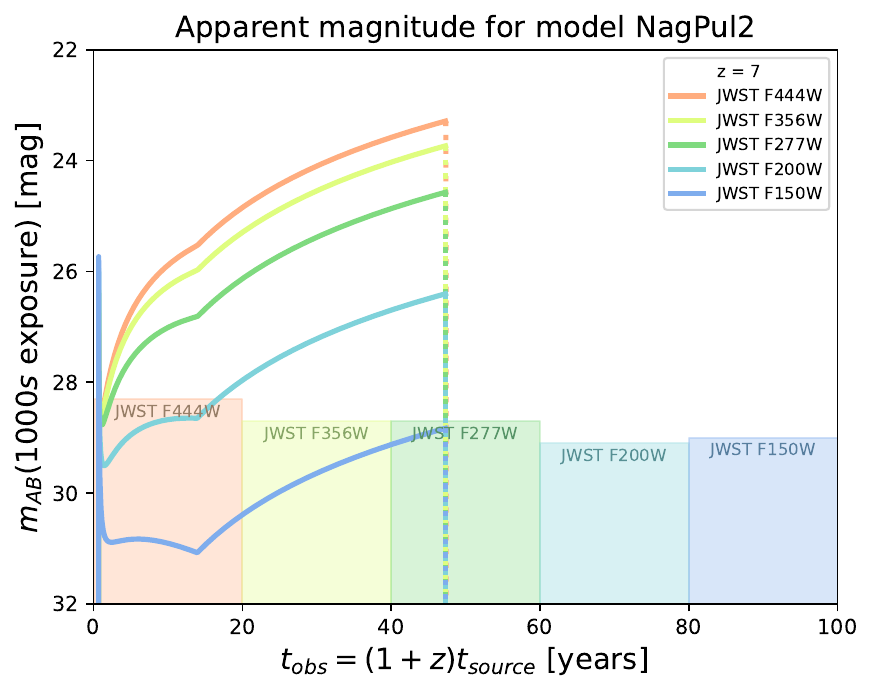}
    \includegraphics[width=0.32\textwidth]{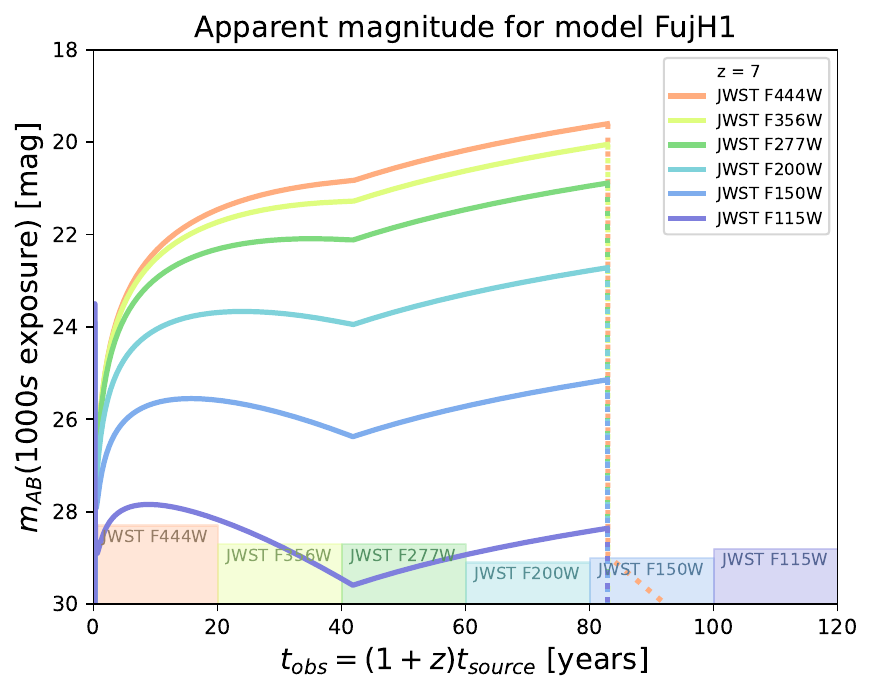}
    \includegraphics[width=0.32\textwidth]{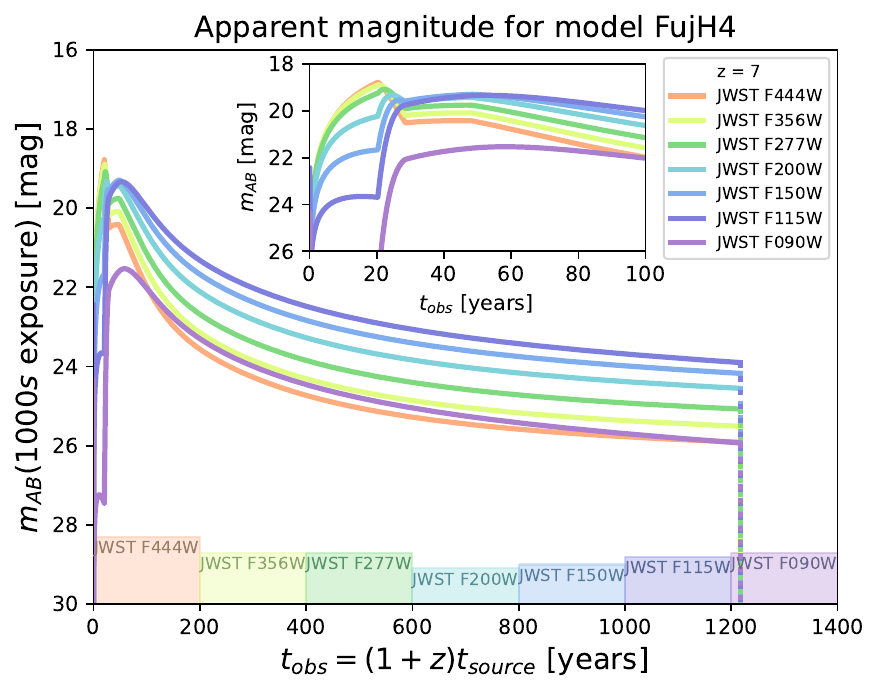}

    \includegraphics[width=0.32\textwidth]{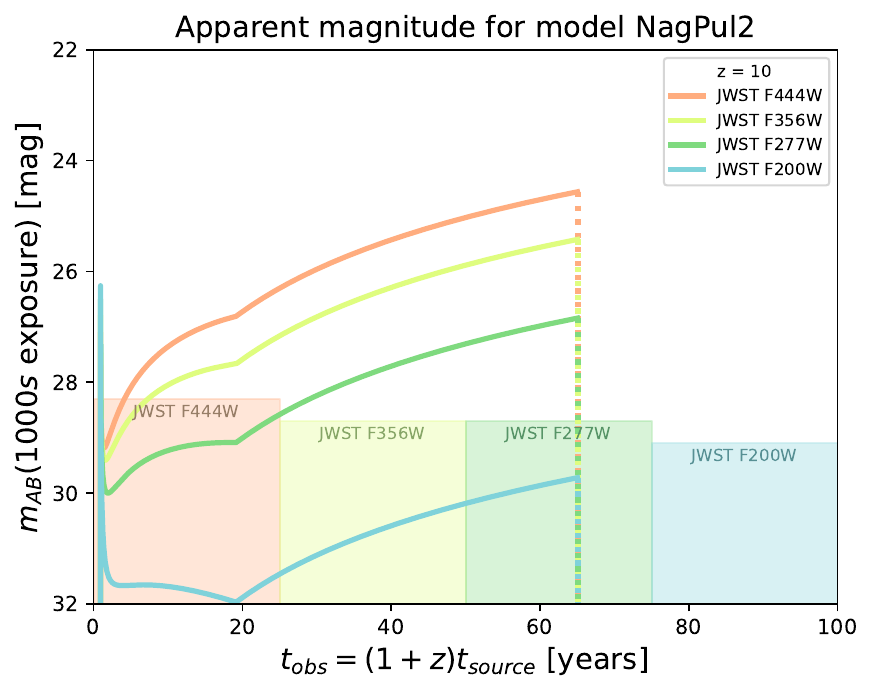}
    \includegraphics[width=0.32\textwidth]{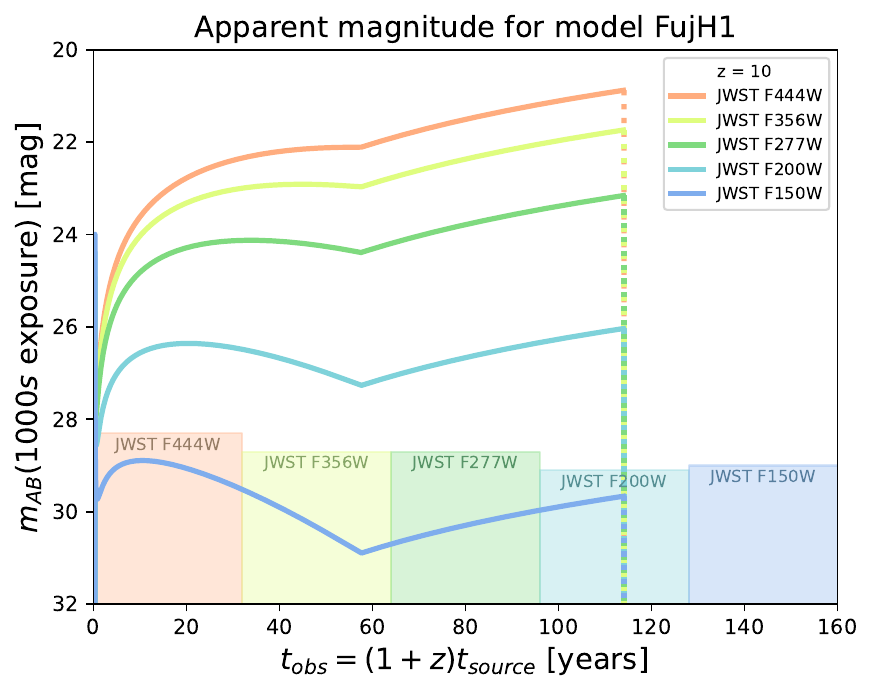}
    \includegraphics[width=0.32\textwidth]{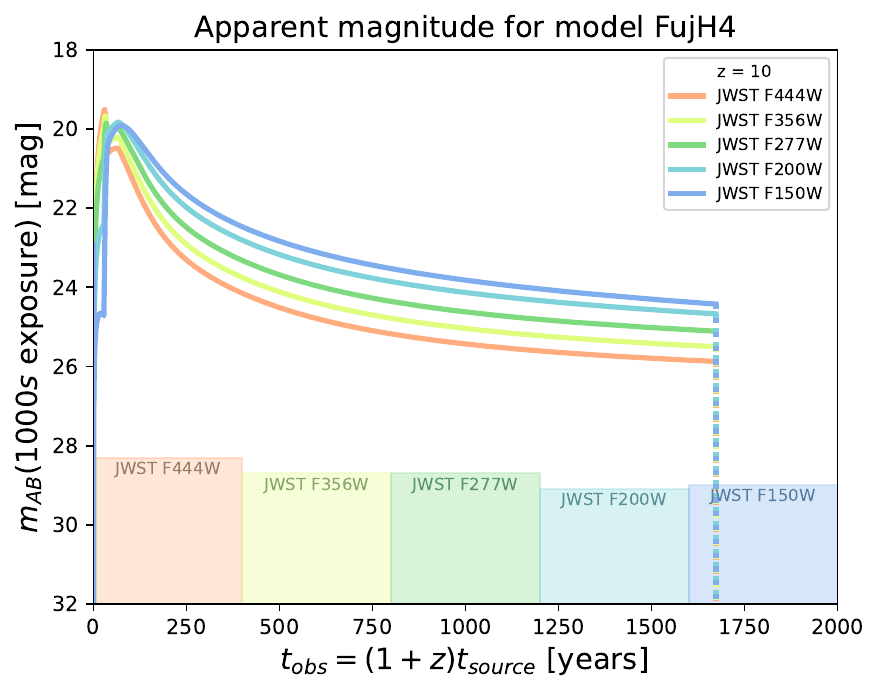}

    \includegraphics[width=0.32\textwidth]{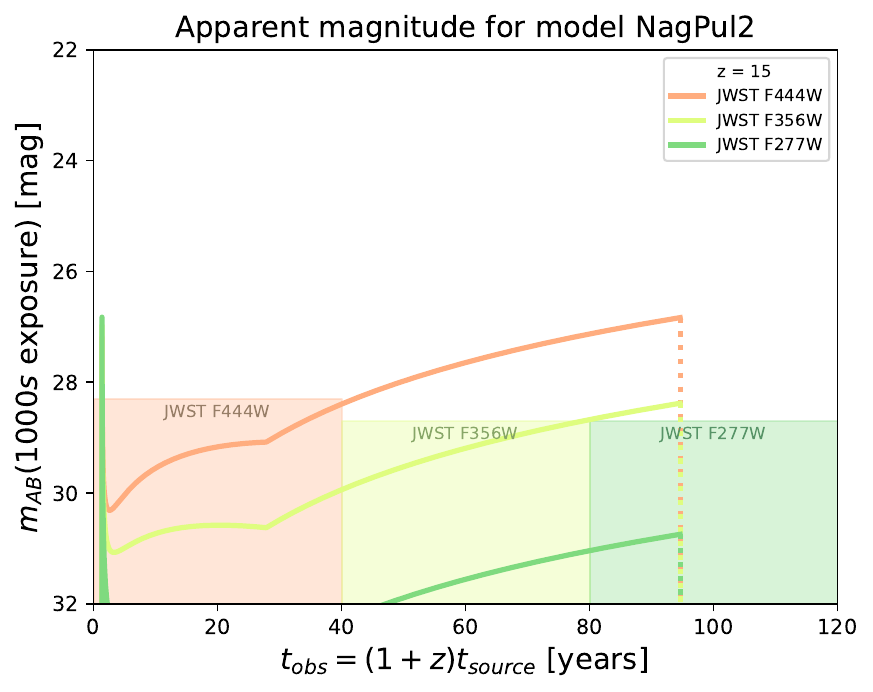}
    \includegraphics[width=0.32\textwidth]{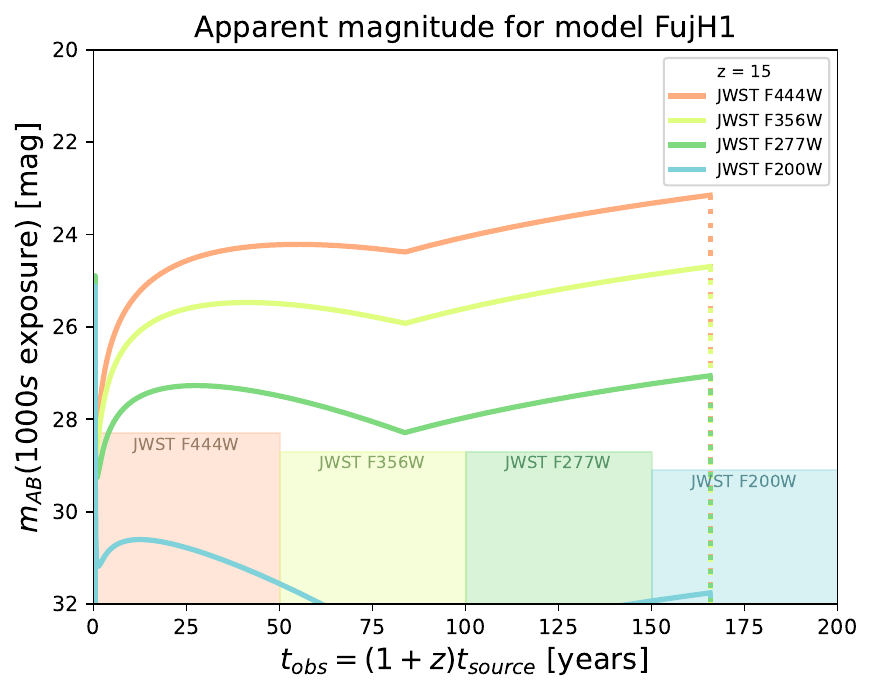}
    \includegraphics[width=0.32\textwidth]{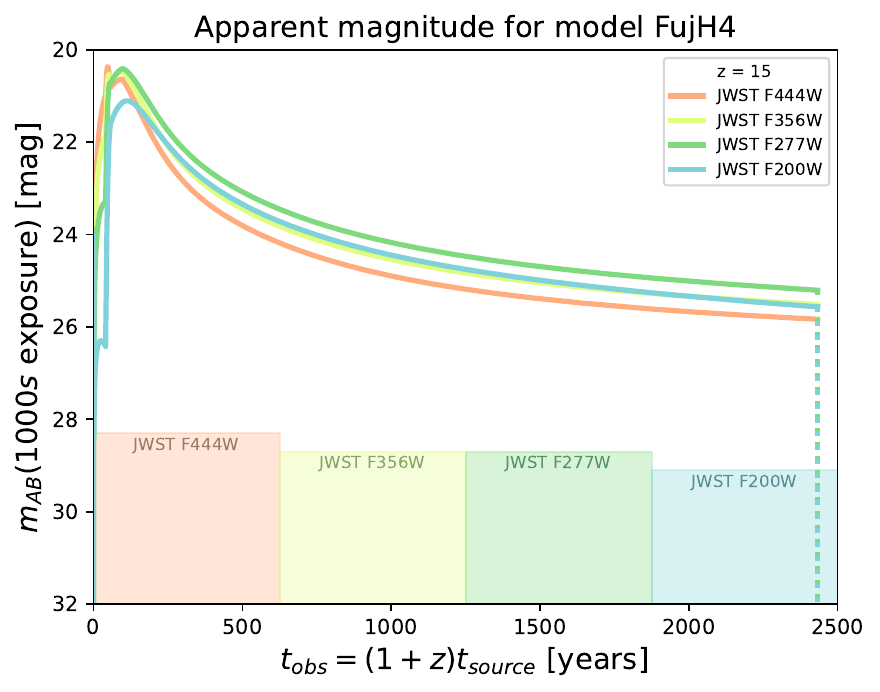}

    \includegraphics[width=0.32\textwidth]{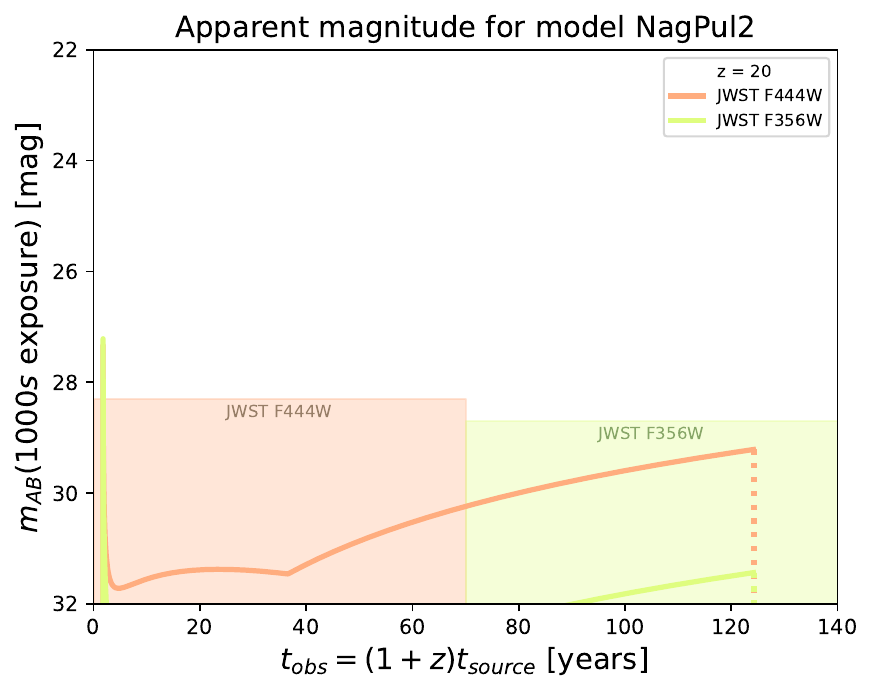}
    \includegraphics[width=0.32\textwidth]{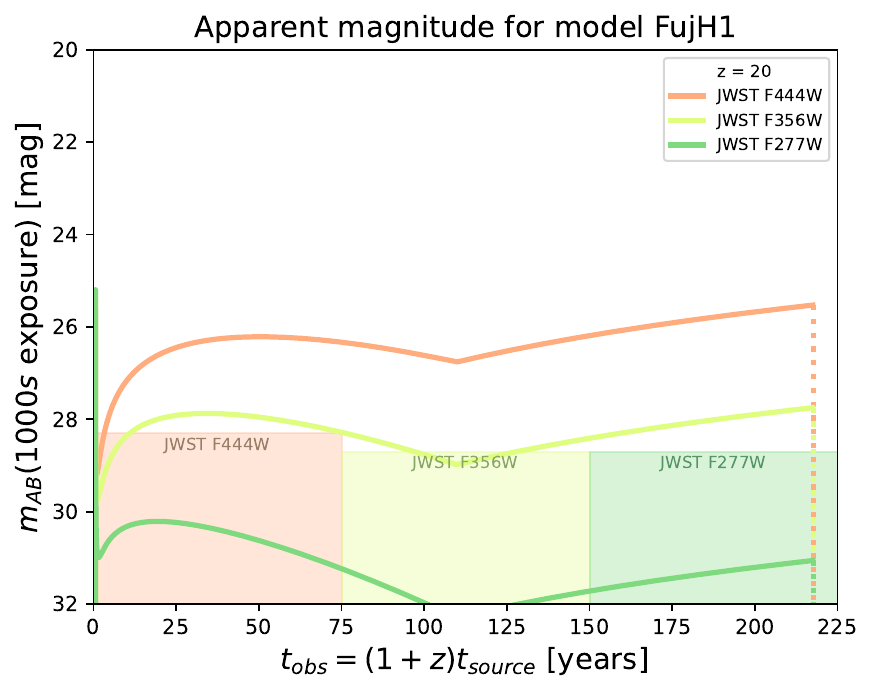}
    \includegraphics[width=0.32\textwidth]{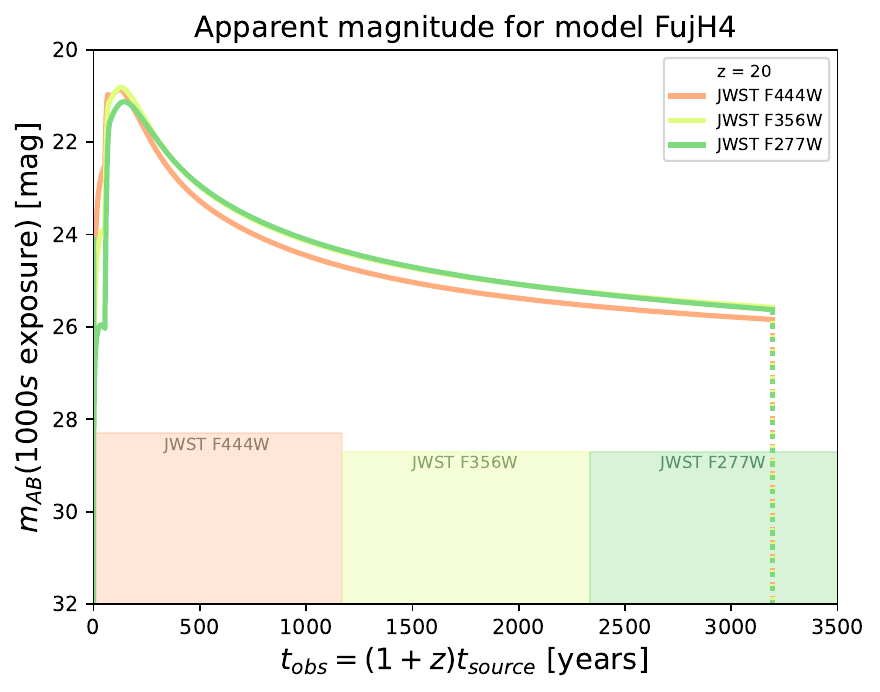}
    
    \caption{AB magnitude as a function of time in the observer frame for three representative models, as observed in different filters of the JWST near infrared camera (NIRCam) at different redshifts $z$. The models NagPul2, FujH1 (both $\eta<1$) and FujH4 ($\eta>1$) represent low, medium and high mass and explosion energy configurations, respectively. The coloured bars represent the detection bound for a given filter after $1000\,\mathrm{s}$ of exposure time. We only show filters where the source would be visible at some point. Solid curves denote the optically thick phase of the light curve. After that, the AB magnitude drops considerably because the bolometric luminosity drops at this point as well. Dotted curves denote the optically-thin emission phase. The observability is best in long wavelength filters for the $\eta<1$ cases. The $\eta>1$ case is brighter in short wavelengths due to the blue spectrum. Some signals could be bright enough also for shorter wavelengths filters, but light with $\lambda<(1+z)\times\lambda_\mathrm{Ly\alpha}\sim 0.121(1+z)\,\mu$m is absorbed by the neutral intergalactic hydrogen for $z>5-6$.}
    \label{fig:results:AB-magnitude-three-models}
\end{figure*}

\subsection{Photometric Observational Properties} \label{subsec:results:photometric-observational-properties}

In the previous section, we discussed the intrinsic bolometric luminosity and light curve properties of SMS explosions in the source frame. When telescopes observe the sky, they only pick up a fraction of the available light and spectrum. This depends on the telescope camera hardware and the available spectral filters. We now show SMS explosion light curves as seen by JWST, EUCLID and the Roman Space Telescope (RST) through different filters. We use the AB magnitude system to denote brightness.

\begin{figure*}
    \centering
    \includegraphics[width=0.49\textwidth]{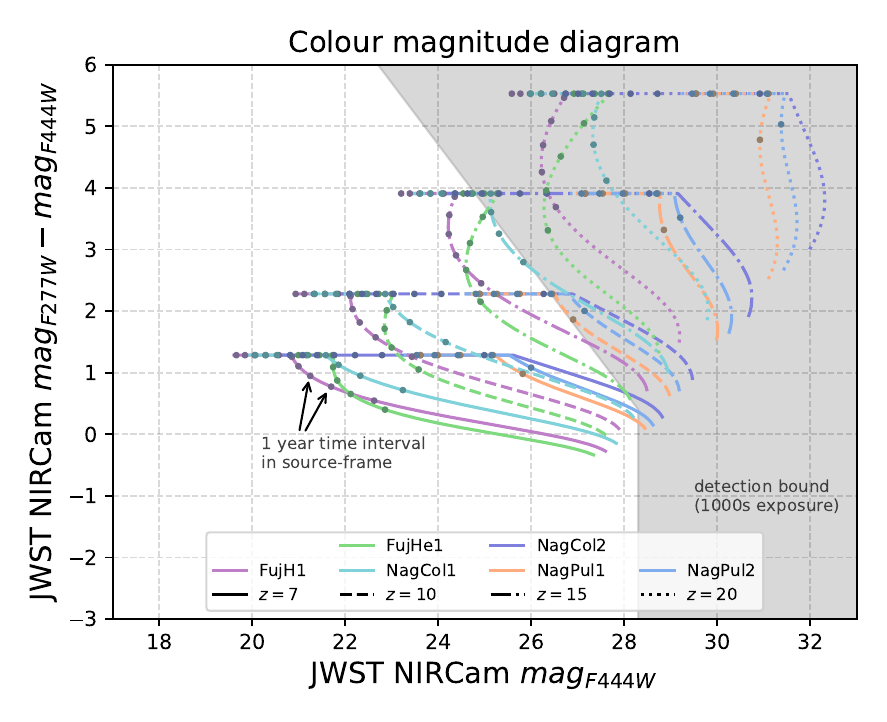}
    \includegraphics[width=0.49\textwidth]{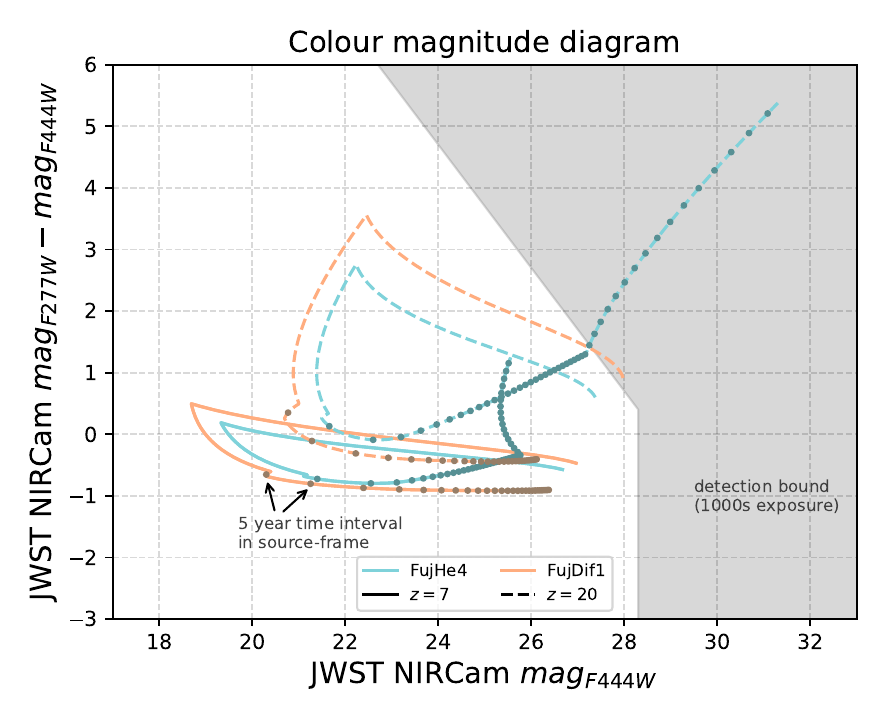}
    \caption{\textbf{Left panel:} Tracks of different SMS configurations with $\eta<1$ in a colour magnitude diagram for the JWST NIRCam filters F277W and F444W (cf. Fig. 4 in \protect\citealt{Price:2024arXiv240803920P}). As the SMS light curve evolves, it moves upwards in the diagram. In the recombination phase, the source moves horizontally to the left in this diagram because the surface temperature is $T_\mathrm{ion}$. Due to the large redshift, the time evolution will be very slow in the observer frame. The circles mark time intervals of one year in the source frame. The source will thus move from one point to the next in $(1+z)$ yrs. The early luminosity peak has been cut out in this figure; see the discussion of \figref{fig:results:Lbol-all-SMS-models}.
    \textbf{Right panel:} Same as left panel, but for selected models with $\eta>1$. The light curve starts on the right and moves to the left and upwards. At the kink, $\eta=1$, and after that the signal becomes bluer and subsequently moves downwards and to the right, until the shock region becomes transparent. The circles mark time intervals of five years in the source frame. The gray region marks the region where a source is too faint to be detected with $1000\,$s exposure time. }
    \label{fig:results:colour-magnitude-diagrams}
\end{figure*}

In Figure \ref{fig:results:AB-magnitude-three-models}, we show the AB magnitude as a function of time in the observer frame as observed in different filters of the JWST near infrared camera (NIRCam) at different redshifts. We show the light curve for three representative configurations (see \figref{fig:results:Lbol-peak-grid-Ekin-Meje}) at four different redshifts of $z=\{7,10,15,20\}$: NagPul2 and FujH1 represent cases with low and medium masses and explosion energies, that also have $\eta<1$; FujH4 represents a high mass and explosion energy case, that also has $\eta>1$.

The coloured bars represent the detection bound for a given filter after $1000\,$s of exposure time. We only show filters that have a non--zero flux for a given redshift and source. Solid lines denote the optically thick phase of the light curve. The dotted lines denote the optically thin emission phase of the light curve. The dotted line appears vertical because the signal in the optically thin phase is below the range of the plot. The subsequent evolution of $m_\mathrm{AB}$ can be seen slightly for model FujH1 at $z=7$. Since we do not model the emissions in the optically thin phase in this work, we crudely estimated the spectrum of these emissions as a black body with radius of the shock position and a luminosity of the bolometric shock luminosity (see discussion after \eqref{eq:light-curve-model:optical-depth-shock-total}). For all cases, the optically thin shock phase emissions are considerably fainter than the emissions in the optically thick phase. This explains the sharp drop in AB magnitude. 

Generally, we see that SMS explosions would be visible up to high redshifts through multiple filters of the JWST NIRCam. The time frame of visibility extends from a few hundred (cases $\eta<1$) to a few thousand (cases $\eta>1$) years if the explosion occurs at redshifts $z\sim15$--$20$, which is typical for SMS formation. But even at lower redshifts of $z\sim 7$, which are common for, e.g., LRDs, the transient would be observable for several decades. The amount of bands in which the sources are visible decreases with increasing redshift. This is due to Lyman-alpha absorption by neutral hydrogen in the intergalactic medium. Also see the discussion in section \ref{subsec:discussion:ionization-of-CSM} about the CSM ionisation properties and possible absorption of Balmer-continuum photons by the CSM. Nevertheless, many SMS transients will be observable in multiple filters. Pulsating SMSs could be visible until redshifts of $\sim15$, according to their intrinsic brightness. Exploding SMSs in their hydrogen burning phases could easily be visible up to $z\sim20$. Helium burning SMSs would be visible to similarly high redshifts as well. 

For models with $\eta<1$, the best visibility will be in the long-wavelength filters because the spectral peak is located at longer wavelengths of $\lambda_\mathrm{peak}\sim 0.48(6000\,\mathrm{K}/T)(1+z)\,\mu$m. This is due to the high redshift but also due to the cool surface temperature $\sim T_\mathrm{ion}=6000\,$K in the recombination phase. For models with $\eta>1$, the spectrum is much bluer and thus the signal will be brighter in bluer filters. The observability in blue filters is here constrained by intergalactic Lyman-alpha absorption (and possible partial absorption by the CSM, as discussed in sec. \ref{subsec:discussion:ionization-of-CSM}), even though the intrinsic brightness of these objects would be high enough to be detected in all JWST bands. For example, at $z\sim7$, FujH4 can be observed in the F090W filter, but at $z\sim15$, the source will be undetected in filters with shorter wavelength than filter F150W.

The AB magnitude in different filters reveals information about the colour of the observed objects. This can best be presented in a colour-magnitude diagram. This is also done with present bulk observation data from surveys, such as UNCOVER \citep{Price:2024arXiv240803920P}, and other analyses of photometric data, e.g., by \citealt{Zhang:2024arXiv241102729Z}. 

In the left panel of Figure \ref{fig:results:colour-magnitude-diagrams}, we show a colour-magnitude diagram using the JSWT filters F277W and F444W. We plot the observed magnitudes of the different sources with $\eta<1$ at different redshifts over time as tracks. These cases also correspond to the cases with moderate explosion energies ($<10^{55}\,$erg). We here neglect the early time peak, as done before (see Appendix \ref{sec:appendix:colour-magnitude-plot-with-early-time-peak} for a version including the early-time peak). The dots on each line show time intervals of one year in the source-frame. A given SMS explosion light curve will move in the diagram from one point to the next in $(1+z)$ yrs, depending on the redshift of the transient. The gray region marks the region where a source is too faint to be detected with $1000\,$s exposure time in both filters. 

During the light curve evolution, a given source will move through the diagram over time. The evolution starts in the lower part of the diagram and then moves left and upwards. This corresponds to a brightening and reddening of the source. The track becomes horizontal in the end and the evolution then is sidewards to the left. The horizontal movement corresponds to the recombination phase of the light curve where the effective surface temperature is constant at $T_\mathrm{ion}=6000\,$K and the luminosity increases as $L_\mathrm{bol}\propto (c\beta_\mathrm{s} \times t)^2$. We find that these SMS transients would be visible in JWST until $z\sim17$ in both filters and until $z\sim24$ for the longer-wavelength F444W filter.

In the right panel of \figref{fig:results:colour-magnitude-diagrams}, we show selected models with $\eta>1$, which correspond to cases with high explosion energies ($>10^{55}\,$erg). We only show two models because the other models have very similar evolution tracks in this diagram and the plot would look too busy otherwise. The dots on each line show time intervals of five years in the source-frame. The light curve starts on the right and moves to the left and upwards. At the kink, $\eta=1$, and after that the signal becomes bluer and subsequently moves downwards and to the right, until the shock region becomes optically thin. The evolution in this second phase is very slow ($<0.2/(1+z)\,\mathrm{mag}\,\mathrm{yr}^{-1}$) and significant variability of the sources could be difficult to detect. Over all, the models with $\eta>1$ are bright enough to be seen even to higher redshift. The constraint here is the spectral range of the telescope filters. Hence, these SMSs could be easily visible for $z<24$ in both filters, and for $z<35$ in the F444W filter alone.

\begin{figure*}
    \centering
    \includegraphics[width=0.49\textwidth]{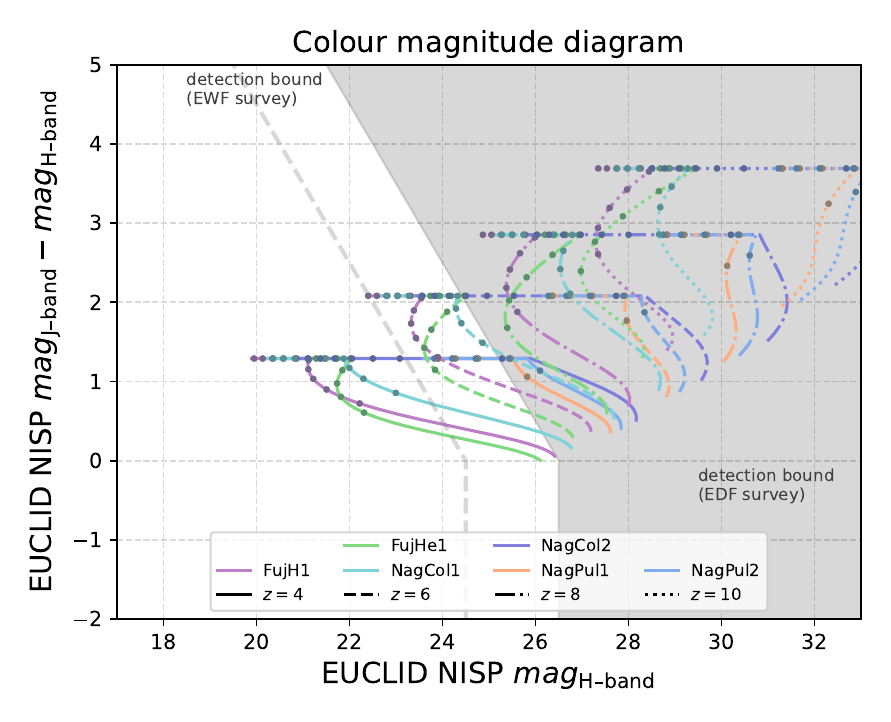}
    \includegraphics[width=0.49\textwidth]{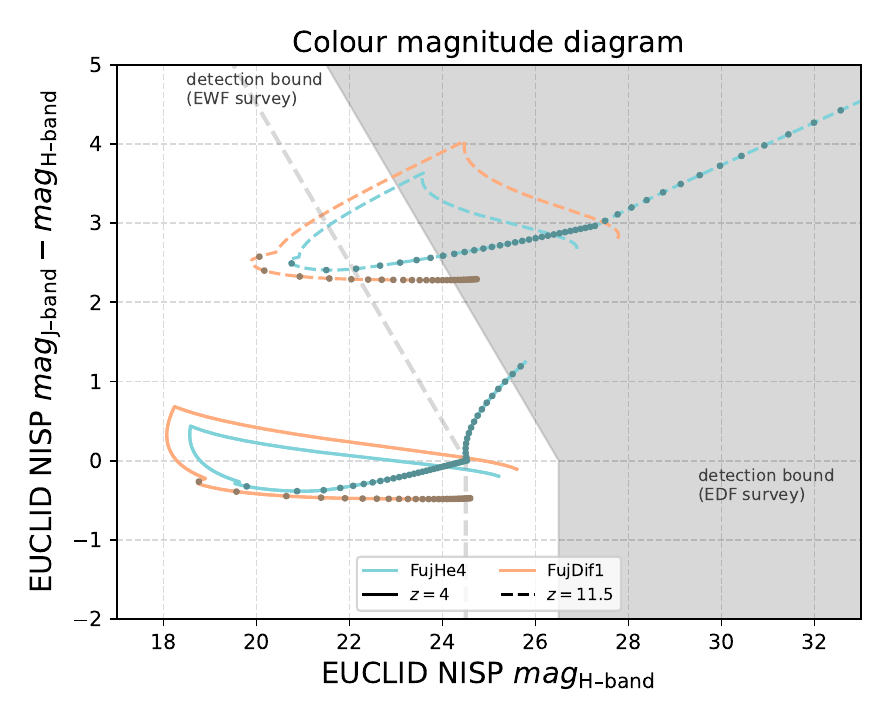}
    \includegraphics[width=0.49\textwidth]{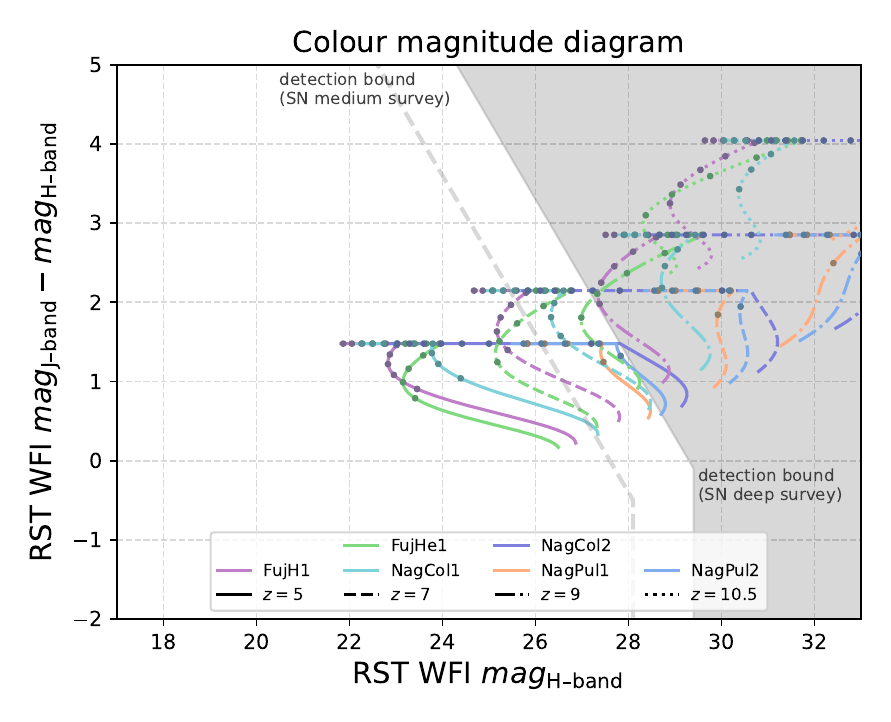}
    \includegraphics[width=0.49\textwidth]{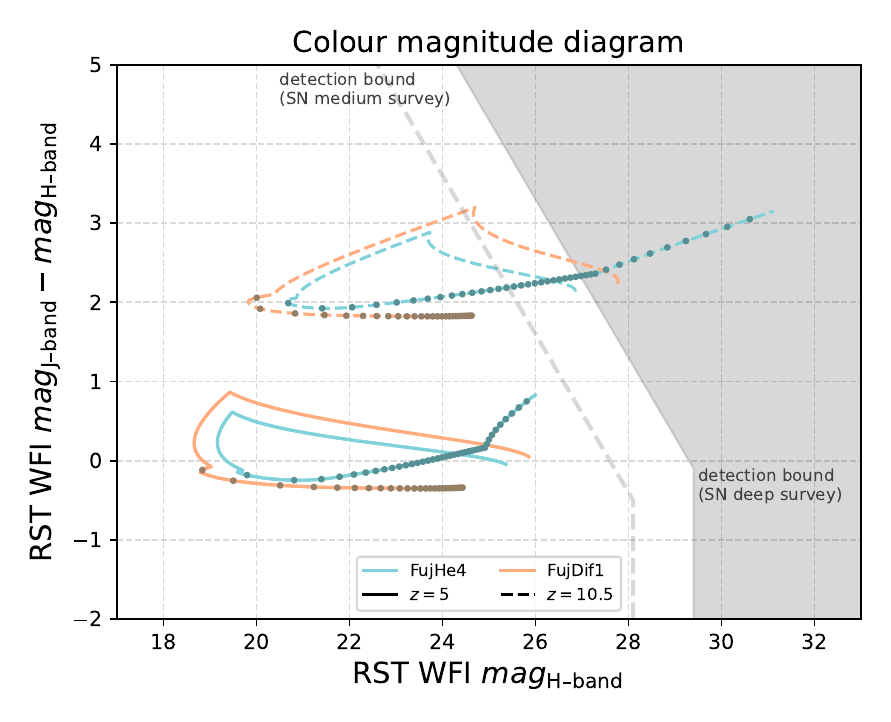}
    \caption{\textbf{Upper panels:} Same as Figure \ref{fig:results:colour-magnitude-diagrams} but for filters of the EUCLID telescope's Near-Infrared Spectrometer and Photometer (NISP) instrument. For the detection bounds, we used the design target magnitude for the EUCLID wide field (EFW) and EUCLID deep field (EDF). The dots in the left panels again indicate source-frame intervals of $1$ yr and points in the right panels of $5$ yrs. \textbf{Bottom panels:} Same as above but for filters of the Roman Space Telescope (RST) Wide field instrument (WFI). For the detection bounds, we used the observational bounds of the supernova (SN) medium and deep surveys.}
    \label{fig:results:colour-magnitude-diagrams-EUCLID-ROMAN}
\end{figure*}

We now investigate SMS observability in EUCLID and the Roman space telescope (RST). Even though these telescopes are less sensitive than JWST and host filters with shorter wavelengths, they might still be suitable tools to observe high-redshift SMS explosion transients. This is mainly because of the difference in the mission profile. JWST focuses on single, deep observations with a narrow field of view. EUCLID and RST on the other hand are designed for large sky surveys, conducted over several years, which will cover significant areas on the night sky. The increased sky coverage could thus offset the smaller depth and lead to meaningful observational opportunities.

In Figure \ref{fig:results:colour-magnitude-diagrams-EUCLID-ROMAN}, we show colour magnitude diagrams for EUCLID (upper panels) and for RST (bottom panels), same as in \figref{fig:results:colour-magnitude-diagrams}. The coloured lines show the tracks that the light curves will move on during their evolution. The ticks again show time intervals of 1 year (left) and five years (right) in the source frame. The grey coloured surface and dashed line show the observational detection limits based on the planned surveys for EUCLID and RST.

For EUCLID, we first focus on the upper left panel, which shows the cases where $\eta<1$. We find that those SMS explosion transients will be visible until $z\sim 5.5$ in the EUCLID wide field (EWF) survey and up to $z\sim 7$ for the EUCLID deep field (EDF). A single-band detection in the H-band is possible up to $z\sim9$ in the deep field. Pulsating SMSs would be visible in EUCLID until $z\sim 4$ in the deep field survey (but this could be changed due to absorption of Balmer-continuum photons, see discussion in sec. \ref{subsec:discussion:ionization-of-CSM}). 
For the more energetic cases with $\eta>1$ (see upper right panel), we find that they would be very bright and clearly visible in the EWF and EDF surveys -- even considering our model uncertainties. Their visibility is constrained by the redshift range of the filters. Above a redshift of $z=11.8$, SMS explosions would only be visible in the H-band due to Lyman-alpha absorption. Slowly varying sources that are observed only in this band would be interesting signatures for high-redshift transients. However, only the brightest cases would be bright enough to be good candidates at these redshifts.

For RST, the situation is more favourable due to the higher depth of the planned surveys of up to $29.4\,\mathrm{mag}$. Here, lower-energy exploding SMSs (with $\eta<1$, see lower left panel) will be visible in the SN medium survey in both filters up to $z\sim 7$--$8$ and in the SN deep survey up to $z\sim 9$. Some events might also be marginally detectable in the H-band until $z\sim 10.5$. The brighter $\eta>1$ cases (see bottom right panel) will be clearly visible in both surveys for $z<10.5$. Due to their brightness, their visibility is constrained by the spectral range of the filters because of intergalactic Lyman-alpha absorption. At larger redshifts of $z>10.5$--$10.8$, these transients will only be visible in the longer-wavelength H-band filter until around $z=13$.

\section{Discussion}
\label{sec:discussion}

\subsection{Ionisation of the CSM} \label{subsec:discussion:ionization-of-CSM}

In the previous section, we found that SMS explosions have high intrinsic luminosities of $10^{45-47}\,$erg s$^{-1}$. We therefore expect a significant flux of ionising radiation, especially in the cases $\eta>1$. But it is unclear whether their number is enough to ionise the dense circumstellar medium (CSM) around the exploding SMS. The possible ionisation of the CSM has strong implications on the light curve visibility via optical depth from electron scattering and also Balmer-attenuation. 

We therefore calculate the amount of ionising photons. We find that the CSM can be partially ionised by $>13.6\,$eV photons only for cases with $\eta>1$. For all other cases, the number of $>13.6\,$eV photos is too low to significantly ionise the CSM. But due to the special environment of ACHs at $T_\mathrm{CSM}=8000\,$K, we find that ionisation via Balmer-continuum photons $>3.4\,$eV can be viable. We discuss this in the following. 

Hydrogen can be ionised from the ground state ($n=1$) by Lyman-continuum photons. Hydrogen at the $n=2$ level can also be ionised by Balmer-continuum photons. The CSM will reach the ionisation-recombination equilibrium if a sufficient amount of atoms are in the $n=2$ excited state and $n=2$ atoms are replenished quickly enough, i.e., the collisional excitation timescale is similar or smaller than the dynamical timescale. This is the case here, at least in the central region of the halo (cf. Fig. 12 in \citealt{Soon:1992ApJ...394..717S} and \citealt{Scholz:1990MNRAS.242..692S}), and we assume for simplicity that this is given in the whole CSM (but see discussion at the end of this section). Note that, although hydrogen atoms in states with $n\geq 3$ may contribute comparably to those in the $n=2$ level, we neglect these in the following discussion for simplicity, as their inclusion does not essentially change the results.

In equilibrium, the ionisation rate is balanced with the recombination rate to all levels. But any recombination to levels $n=1$ and $n=2$ will produce a photon that can ionise another atom in the CSM because the CSM is (marginally) optically thick for both Lyman and Balmer continuum photons, as we see later. Thus, recombination to these levels will have no impact on the total degree of ionisation of the CSM. We then assume local balance between recombination to the $n=1$ and $n=2$ levels and the corresponding locally emitted ionizing photons, which is called the on-the-spot approximation (see \citealt{Osterbrock:2006agna.book.....O}, Chapter 2). According to this approximation, the total rate density of photoionisation by the photons emitted from the shocked shell will be balanced with the recombination rate density to states $n\geq 3$:
\begin{align}
     n_{\gamma,\mathrm{Ly}} n_1 \sigma_\mathrm{bf,1} c  +  n_{\gamma,\mathrm{Ba}} n_2 \sigma_\mathrm{bf,2} c = n_e n_p \alpha_C \: . \label{eq:discussion:ionised-CSM:total-ionisation-rate}
\end{align}
$n_{\gamma,\mathrm{Ly}}$ and $n_{\gamma,\mathrm{Ba}}$ are the number densities of photons emitted from the shocked shell with $>13.6\,$eV and $>3.4\,$eV, respectively. $n_1$ and $n_2$ are the number densities of neutral hydrogen in the $n=1$ and $n=2$ states, respectively. $\sigma_\mathrm{bf,1}$ and $\sigma_\mathrm{bf,2}$ are the bound-free cross-sections for hydrogen in the $n=1$ and $n=2$ states, respectively. $n_e$ and $n_p$ are the electron and proton number densities and $\alpha_C = 2.425\e{-13}\,$cm$^3$s$^{-2}$ (at $T_\mathrm{CSM}$, interpolated from \citealt{Martin:1988ApJS...66..125M}, Table 1) is the recombination parameter excluding direct recombination to the ground state and $n=2$ level.

Because the amount of Lyman-continuum photons is quite low, we find that, $n_{\gamma,\mathrm{Ly}}n_1$ is much smaller than $n_{\gamma,\mathrm{Ba}}n_2$ in our case, even though $n_1\gg n_2$. We can therefore neglect the ionisation by Lyman-continuum photons, $n_{\gamma,\mathrm{Ly}} n_1 \sigma_\mathrm{bf,1} c$, in the left hand side of \eqref{eq:discussion:ionised-CSM:total-ionisation-rate}.
We employ $\sigma_\mathrm{bf,2}=1.58\e{-17}\,$cm$^2$ (see \citealt{Rybicki-Lightman:1979rpa..book.....R}, chapter 10.5) as the bound-free cross-section for atoms in the $n=2$ state at the Balmer limit ($\lambda=0.364\,\mu$m) in the following.

For partially ionised hydrogen, $n_e=n_p= f_\mathrm{ion} n_\mathrm{tot}$, where $f_\mathrm{ion}$ is the ionisation fraction and $n_\mathrm{tot}=A_\mathrm{L}/m_\mathrm{H} r^2$ ($A_\mathrm{L}=\eta_0\mathcal{R}T_\mathrm{CSM}/4\pi G = 4\e{18}\,$g cm$^{-1}$, see \eqref{eq:light-curve-model:Larson-CSM-Model}) is the total number density. We assume a pure hydrogen CSM, instead of a primordial CSM, for simplicity -- this will under-estimate the ionisation fraction, and over-estimate the optical depth of Balmer photons by around $25\,\%$. From \eqref{eq:discussion:ionised-CSM:total-ionisation-rate} we then get:
\begin{align}
    f_\gamma f_2 \frac{\sigma_\mathrm{bf,2} c}{\alpha_C} = f_\mathrm{ion}^2 \: . \label{eq:discussion:ionised-CSM:f2-fraction-balance-equation}
\end{align}
We have defined $f_\gamma = n_{\gamma,\mathrm{Ba}}/n_\mathrm{tot}$ and $f_2 = n_2/n_\mathrm{tot}$ as the $>3.4\,$eV-photon and $n=2$ hydrogen fractions relative to the total number density. Since $f_\mathrm{ion} \leq1$ and $\displaystyle\frac{\sigma_\mathrm{bf,2} c}{\alpha_C} \approx 1.95\e{6}$ we already see that $f_2$ is bounded by the photon fraction $f_\gamma$ and is $f_2<10^{-6}$ if $f_\gamma>1$. Given the number rate of ionising photons, $Q_\gamma$, the number density is $n_{\gamma,\mathrm{Ba}}= Q_\gamma/4\pi c r^2$ (neglecting attenuation, which is justified for an order-of-magnitude estimate, as we see later) and $f_\gamma = Q_\gamma m_\mathrm{H} / 4\pi c A_\mathrm{L}$. Also, we rewrite $f_2$ as
\begin{align}
    f_2 = \frac{n_2}{n_\mathrm{tot}} = \frac{n_2}{n_1}\frac{n_1}{n_\mathrm{tot}} = \frac{n_2}{n_1} (1 - f_\mathrm{ion}) \: ,
\end{align}
where we approximate that $n_\mathrm{tot} \approx n_1 + n_p$. As a first approximation, the ratio $n_2/n_1$ can be broadly estimated using the expression from local thermal equilibrium as 
\begin{equation}
\frac{n_2}{n_1} = \frac{g_2}{g_1} \exp\left( -10.2\,\mathrm{eV}/k_\mathrm{B} T_\mathrm{CSM} \right) = 1.5\e{-6}\label{eq:discussion:n2n1-ratio}
\end{equation}
($g_2=8$, $g_1=2$ are the statistical weights). Employing this value, we finally find
\begin{align}
    f_\mathrm{ion}^2 = \frac{\sigma_\mathrm{bf,2} c}{\alpha_C} \frac{n_2}{n_1} f_\gamma  (1 - f_\mathrm{ion}) = 2.93 \times f_\gamma (1 - f_\mathrm{ion})  \: .
\end{align}
For a photon fraction $f_\gamma = \{1, 10, 100\}$ we find ionisation fractions of $f_\mathrm{ion} \approx \{0.79, 0.97, 0.997\}$. For our models, we find that $\eta>1$ models (FujH2, FujH4, FujHe2, FujHe4, FujExp) have $f_\gamma \sim 1500-4000$ around the light curve peak. Exploding $\eta<1$ models (FujH1, FujHe1, NagCol1, NagCol2, NagExp) have $f_\gamma \sim 100-500$ and the pulsating models (NagPul1, NagPul2) have $f_\gamma \sim 10-20$ around the light curve peak. We therefore find that the CSM can be significantly ionised using Balmer continuum photons. 

We note that, however, that the ratio $n_2/n_1$, may not be given by the thermal equilibrium value (\eqref{eq:discussion:n2n1-ratio}) and might even be higher due to irradiation of Lyman-$\alpha$ photons (see \citealt{Omukai:2000ic}, footnote 3). We also point out that, while a temperature of $T_\mathrm{CSM}=8000\,$K is used in \eqref{eq:discussion:n2n1-ratio}, the actual temperature of the CSM may deviate from this value after the interaction with photons. If, for example, the CSM temperature is determined by a cooler colour temperature of the radiation, such as $6000$ K, the ratio $n_2/n_1$ could be reduced by a factor of $100$. Nevertheless, for either case, the ionisation fraction is still likely to remain close to unity for the models with peak luminosities exceeding $10^{46}\,{\rm erg~s^{-1}}$.

The size of the ionised region depends on the incoming photon rate $Q_\gamma$ and can be estimated by calculating the Strömgen sphere radius $R_\mathrm{str}$ (see \citealt{Osterbrock:2006agna.book.....O}, chapter 2). It can be derived assuming that $Q_\gamma$ is equal to the recombination rate:
\begin{align}
    Q_\mathrm{rec} = \int_{R_\mathrm{fs}}^{R_\mathrm{str}} n_e n_p \alpha_C dV = \frac{4\pi \alpha_C A_\mathrm{L}^2}{m_\mathrm{H}^2} f_\mathrm{ion}^2\left( \frac{1}{R_\mathrm{fs}} - \frac{1}{R_\mathrm{str}} \right) \: , \label{eq:discussion:recombination-rate}
\end{align}
with $n_e=n_p=f_\mathrm{ion}n_\mathrm{tot}$ (with $f_\mathrm{ion}\approx 1$) and $\alpha_C$ as above. $R_\mathrm{str}$ is defined in the steady state, which is applicable if the recombination timescale $\tau_\mathrm{rec}\sim 1/n_e \alpha_C$ ($\ll 100$ days for our case) is shorter than the system evolution time $\sim$yrs. \eqref{eq:discussion:recombination-rate} can be solved for $R_\mathrm{str}$:
\begin{align}
    R_\mathrm{str} = R_\mathrm{fs} \left( 1 - \frac{Q_\gamma}{Q_\mathrm{thr}} \right)^{-1} \: . \label{eq:discussion:stroemgen-sphere-radius}
\end{align}
We find a threshold photon rate $Q_\mathrm{thr} = f_\mathrm{ion}^24\pi \alpha_C A_\mathrm{L}^2/m_\mathrm{H}^2 R_\mathrm{fs} \approx 10^{56}\mathrm{s}^{-1} f_\mathrm{ion}^2 (0.05\,\mathrm{pc}/R_\mathrm{fs})$, above which $R_\mathrm{str}$ diverges and the CSM will be fully ionised. Excess photons will leave the ACH and could contribute to reionisation of the intergalactic medium.

\begin{figure}
    \centering
    \includegraphics[width=0.498\textwidth]{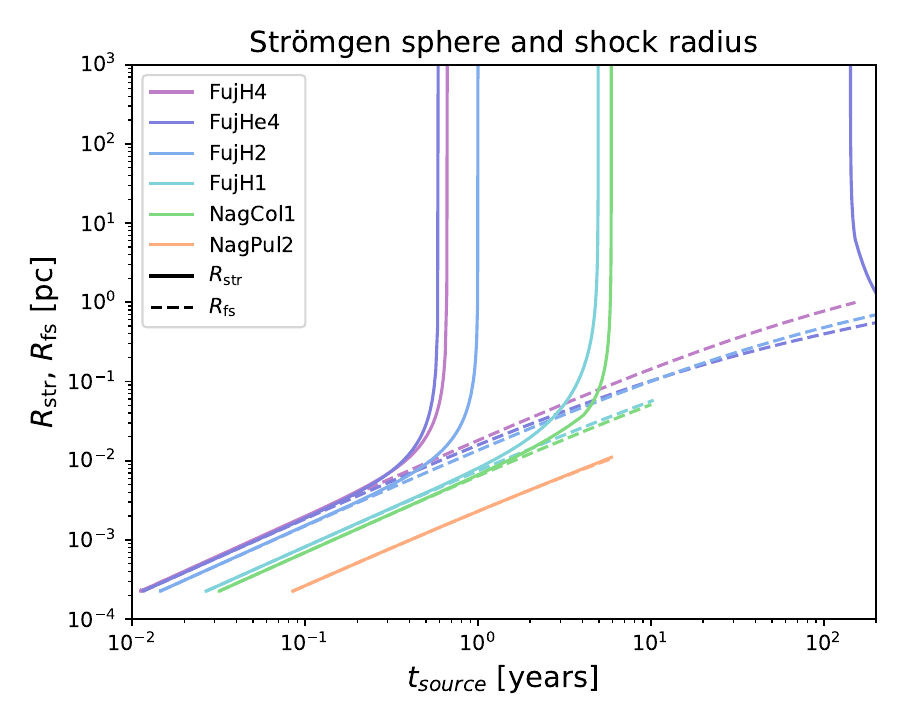}
    \caption{Strömgen sphere radius $R_\mathrm{str}$ (solid lines) and forward shock radius  $R_\mathrm{fs}$ (dashed lines) for several representative models. We show the evolution only for the optically thick shock phase. The Strömgen sphere radius diverges after some time, indicating full ionisation of the halo. The time when this happens is $\simeq5$--$6\,\mathrm{yrs}$ for thermalised ($\eta<1$) models. For the non-thermalised ($\eta>1$) models FujH4, FujHe4, FujHe2, the ionised bubble forms after $<0.8\,\mathrm{yrs}$. This is before the luminosity peak in all cases. The pulsating models do not significantly ionise the CSM.}
    \label{fig:discussion:Stroemgen-sphere-ionised-CSM}
\end{figure}

In Fig. \ref{fig:discussion:Stroemgen-sphere-ionised-CSM}, we show the Strömgen sphere radius $R_\mathrm{str}$ and forward shock radius $R_\mathrm{fs}$ (equivalent to the photosphere radius) as a function of time for some representative models. We find that all models except for the pulsating model (NagPul2) produce enough Balmer-continuum photons to fully ionise the CSM before the light curve peak. Due to the form of \eqref{eq:discussion:stroemgen-sphere-radius}, the Strömgen radius sharply increases once the photon flux $Q_\gamma$ gets close to the threshold for full ionisation. The Strömgen sphere would recede for model FujHe4 according to \eqref{eq:discussion:stroemgen-sphere-radius}. But this is not too relevant here because this would happen at late times when the total luminosity is much lower than the peak. Also, the recombination rate might be too small in the outer layers, such that most of the CSM would stay ionised regardless within the light curve timescale. 

We discuss the attenuation of Balmer continuum photons with $>3.4\,$eV, which can be absorbed by $n=2$ neutral hydrogen. The optical depth depends on the number density $n_2$ (as above) and is given by
\begin{align}
    \tau_\mathrm{Bal} = \int_{R_\mathrm{fs}}^{\infty} n_2 \sigma_\mathrm{bf,2} dr = \frac{\sigma_\mathrm{bf,2}A_\mathrm{L}}{m_\mathrm{H}R_\mathrm{fs}} f_2  \: .
\end{align}
Here, we obtain $f_2$ from \eqref{eq:discussion:ionised-CSM:f2-fraction-balance-equation}. In the case of a non-ionised CSM ($f_\gamma\ll1$, $f_\mathrm{ion}=0$ and $f_2 = n_2/n_\mathrm{tot} \approx n_2/n_1 = 1.5\e{-6}$) and for a typical value of $R_\mathrm{fs}\sim0.05\,$pc around the light curve peak, we find $\tau_\mathrm{Bal}\simeq 367 \times (0.05\,\mathrm{pc}/R_\mathrm{fs})$, which leads essentially to total extinction. But for an ionised CSM with sufficient Balmer photon flux of $f_\gamma \sim 100$ (e.g., the exploding $\eta<1$ models), $f_\mathrm{ion}\gtrsim 0.997$, and the optical depth drops to $\tau_\mathrm{Bal}\sim 1.25\times (0.05\,\mathrm{pc}/R_\mathrm{fs})$; for $f_\gamma\sim500$, $f_\mathrm{ion}\gtrsim 0.9993$ and we get $\tau_\mathrm{Bal}\sim 0.25\times (0.05\,\mathrm{pc}/R_\mathrm{fs})$. This is consistent with our previous assumption that the CSM is moderately optically thick.

Naively taking $\tau_\mathrm{Bal}$ to compute the Balmer attenuation, we would expect an attenuation of $e^{-\tau_\mathrm{Bal}}\approx0.29-0.78$ around the Balmer frequency, which would decrease the signal by $\Delta m_\mathrm{AB}\simeq 0.3-1.3$.
But the actual strength of Balmer attenuation might be significantly lower due to the difference in CSM temperature and incoming photon colour temperature $T_\mathrm{col}$. If $T_\mathrm{col}$ is similar to, or smaller than, $T_\mathrm{CSM}$, attenuation will be negligible because the CSM will intrinsically radiate photons at $T_\mathrm{CSM}$. This will be the case for the $\eta<1$ cases around and after the luminosity peak, especially during the recombination phase. If $T_\mathrm{col}>T_\mathrm{CSM}$, we might expect attenuation by that logic. But the only cases where this is fulfilled around the peak are our cases with $\eta>1$ (with $T_\mathrm{col}\gtrsim 10^5$). But in those cases, the flux of ionising photons will be very high $f_\gamma\gg 1500$, thus $f_2$ is much lower compared to the $\eta<1$ cases, and $\tau_\mathrm{Bal}$ will be negligible.

The above discussed points have strong implications on the visibility of our models. In the early parts of the light curve, before the CSM is ionised, we expect significant attenuation for frequencies blueward of the Balmer line. For the pulsating models, where the CSM cannot be ionised, the Balmer attenuation will likely make these sources unobservable for $z>2$ in the J- and H-bands (RST and EUCLID). But for models where the CSM can be fully ionised, we do not expect magnitudes to be lowered significantly. This will be true around the luminosity peak and also after, while the CSM stays ionised. The general visibility of the light curve may therefore not be significantly affected by the Balmer attenuation. Therefore, we do not consider attenuation in the following discussions of the light curve visibility. But we note that more comprehensive and detailed modelling of the CSM will be necessary in the future, to validate our estimates.

Another point to be discussed regarding ionisation of the CSM is that the fully ionised CSM might be optically thick due to electron scattering. The SMS explosion might then not be directly visible to a distant observer. Instead, it might be fainter because photons will scatter and slowly diffuse out of the ionised bubble over time from a larger photosphere radius (radius of last scattering). If the diffusion timescale out of the ionised CSM $t_\mathrm{diff,CSM}$ is significantly longer than the variability timescale of the radiation emission, the variability of the source is smeared out and the shape of the light curve changes. We therefore estimate the optical depth and diffusion timescale of photons out of the ionised CSM (i.e. the region between the shocked shell and the outer CSM radius at $\sim 100$--$1000\,$pc) due to electron scattering:
\begin{align}
    \tau_\mathrm{CSM,es} = \int_{R_\mathrm{fs}}^{\infty} \kappa \rho_\mathrm{CSM} dr = \frac{\kappa A_\mathrm{L}}{R_\mathrm{fs}} f_\mathrm{ion}  \approx 9.1 \left( \frac{0.05\,\mathrm{pc}}{R_\mathrm{fs}} \right) \: ,
\end{align}
with $\kappa = 0.35\,$cm$^2$g$^{-1}$ and $f_\mathrm{ion}\approx1$ for a fully ionised CSM. The diffusion timescale is approximately given by $t_\mathrm{diff,CSM} \approx 3\tau_\mathrm{CSM,es} (\Delta R)/c$, where $\Delta R$ is the distance between the shock radius and the radius of last scattering $\sim0.45\,$pc. At $t\sim10\,$yrs we find that $\tau_\mathrm{CSM,es}\sim4-9$ and $t_\mathrm{diff,CSM}\sim24$--$40\,$yrs. After around $t\sim20$\,yrs, these values have dropped to $\tau_\mathrm{CSM}\sim 0.5$--$4$ and $t_\mathrm{diff,CSM}\sim 1$--$8\,$yrs, with the trend continuing to shorter $\tau_\mathrm{CSM,es}$ and $t_\mathrm{diff,CSM}$. Note that $\tau_\mathrm{CSM,es}$ depends on the column density of hydrogen between the shock region and the CSM outer radius. It becomes smaller simply because the shock radius increases and the distance to the CSM radius decreases. 

The diffusion time is therefore comparable to or shorter than the dynamical timescale of the system at the time of peak brightness. This suggests that the ionised bubble, with its moderate optical depth, allows for the light emitted from the shock region to leak out relatively quickly, leading only to minor changes in the light curve shape and brightness around the peak time. 

We further note that the growing HII region may expel the gas in the halo. This could reduce the CSM density and decrease the luminosity (see Appendix, Fig. \ref{fig:appendix:varying-CSM-density}). However, we find that the characteristic velocity of ionised gas, broadly estimated from its internal energy acquired by the irradiation, is lower than the shock shell expansion velocity by around a factor of $3$--$5$. Hence, the characteristic timescale for the CSM to be blown away should be longer than the light curve timescale of $10$--$100\,$yrs and the effect on the CSM density should be minor. But we stress that more detailed radiation-hydrodynamics/radiation transport simulations will be necessary to quantify this effect.

In the above discussion, we assumed that the collisional excitation timescale for hydrogen, $\tau_\mathrm{col}$, is always shorter than the dynamical timescale throughout the whole CSM. But since $\tau_\mathrm{col} \approx 1/(\alpha_\mathrm{col}\,n)$, it implicitly depends on the radius. With excitation parameters of around $\alpha_\mathrm{col}\approx 10^{-13}\,$cm$^3$s$^{-1}$ \citep{Soon:1992ApJ...394..717S,Scholz:1990MNRAS.242..692S}, we expect our assumption to not hold for radii $\gtrsim5\,$pc. Hence, at large radii, the rate of hydrogen creation in the $n=2$ state will not be sufficient to fully ionise the CSM within the system evolution time. Lyman-alpha continuum photons might excite hydrogen atoms but it is not clear whether the flux is high enough to then fully ionise the CSM at larger radii. Because the photon fraction $f_\gamma$ is large, the system might then be starved of $n_2$ atoms. Once all $n_2$ atoms have been consumed, insufficient amounts of $n_2$ will be created and thus the ionisation will be limited to the initial $n_2/n_\mathrm{tot}$-fraction of $\approx 1.5\e{-6}$. Attenuation of Balmer photons will therefore be smaller because the atoms in the $n=2$ state will be exhausted. Also, because of less ionisation, the electron scattering opacity will be smaller as well. These points could improve over-all visibility of the source. But the effect may not be significantly different from the estimation above because the main contribution to the optical depth will come from the central part of CSM.

\subsection{Distinguishing SMS Explosions From Alternative Sources} \label{subsec:results:distinguishing-SMS-Explosions-from-alternative-sources}

In Section~\ref{sec:Results}, we discussed the observability of SMS explosions and possible typical observational signatures. One of our findings is that these explosions are slowly varying transients which appear as quasi-persistent sources during much of their evolution time. The question is whether some of the previously observed objects in the early universe might instead be such slowly varying SMS explosions. Especially in the part around the emission peak, the variability of these sources can be so small that they could be confused with persistent sources.

The most prominent high-redshift objects are LRDs and AGN. LRDs appear as very red unresolved point–like sources (JWST $mag_\mathrm{F444W} \sim 28$--$23$ and $mag_\mathrm{F277W}-mag_\mathrm{F444W} \sim 0$--$2$, see Fig. 1 in \citealt{Zhang:2024arXiv241102729Z}). Due to their position in the colour-magnitude diagram, they might therefore be confused photometrically with SMS explosions, if only data from some long-wavelength filters is available. In fact, it was already speculated that some LRDs might be high-redshift supernovae \citep{Yan:2023wwh}. A similar reasoning will apply to SMS explosions as well. In addition, LRDs were found to have relatively small variability in their brightness of $\sim0.2\,\mathrm{mag}$ within a time period of around 2 yrs in the observer frame (see the studies by \citealt{Tee:2024lackrestframeuvvariability,Zhang:2024arXiv241102729Z,OBrian:2024ApJS..272...19O}). We computed the variability of SMS explosion light curves in section \ref{subsec:bolometric-light-curves-of-supermassive-stars} and found that they would have a variability comparable to LRDs for timescales of $0.5$--$9 \times(1+z)\,$yrs. At redshifts of $z\sim7$--$10$, which are typical for LRDs, this implies that we do not expect to detect any significant variability of SMS explosions in the time frame of $5$--$100$\,yrs, if we detect the light curve near its emission peak. Variability can be stronger if the light curve is observed near the beginning of its evolution or when the shock becomes transparent and the luminosity drops.

Only having access to photometric data might thus be too little to identify and/or discriminate a LRD and a SMS explosion transient. But spectrally, they can be quite different, as we discuss below. Having access to a spectrum will therefore be of great importance to distinguish them. But even with a spectrum available, there is the possibility that the red part of the spectrum might be explained using a gas–covered SMS explosion and the bluer part by a surrounding young stellar population. A similar idea was discussed by \citealt{Naidu:2025arXiv250316596N}, where the red part of the spectrum could be an AGN enshrouded in a dense cloud of hydrogen. Such cloud would produce similar observations in both cases.

SMS explosions may be regarded as large--scale Type IIn supernovae with large ejecta mass and explosion energy. The ejecta velocities lie in the range of a few percent of $c$. We therefore expect similar spectral properties to these Type IIn supernovae with some differences: First, we discuss the number of spectral lines. SMS explosions will have very metal poor spectra, lacking heavy element spectral lines and instead showing strong hydrogen and helium lines. This can be easily understood for the cases where a BH is formed: heavy elements would form near the centre during the SMS collapse and promptly fall into the BH (as found in simulations by \citealt{Fujibayashi:2024vnb}). But even for exploding models, where the heavier elements are released, the final spectrum should be metal poor. This is because the photospheric features are determined by the shock region, which mostly consists of the near--primordial CSM ($Z<10^{-5}\,Z_\odot$, \citealt{Omukai:2008ApJ...686..801O}) and metal-poor outer layers of the SMS atmosphere. A primordial composition should therefore be a good approximation regardless of the explosion mechanism, while the shock shell is opaque and the CSM consists of primordial gas.

After the shock shell becomes optically thin, spectral lines from heavier elements in the SMS ejecta, which are synthesized during the SMS hydrostatic evolution and/or during its explosion, may become visible. If the SMS explosion is triggered by runaway nuclear burning, we also expect a higher abundance of heavier elements such as Mg and Si (see \citealt{Nagele:2020xqq}) in the spectrum. This is because then the heavy elements will be ejected since no BH forms in this case.

Regarding the shape of the spectral lines, we expect some Doppler broadening of the lines on the order of $3000\,\mathrm{km\,s^{-1}}\: (\sim0.01c)$ from the elements within the shock front. Because we only observe the part of the envelope which travels towards us, this broadening should be asymmetric and be stronger towards the blue colours. But line broadening by electron scattering due to an ionised CSM (see section~\ref{subsec:discussion:ionization-of-CSM}), could also be present and smear out the asymmetry. The dense expanding envelope could also lead to a P-Cygni spectral line shape. These features have already been observed in LRD candidates by \citealt{Rusakov:2025arXiv250316595R}, and would also be consistent with exploding SMSs, according to our findings.

For SMSs to form, a dense isothermal gas halo is needed, with typical sizes of $r\sim1\,$kpc \citep{Patrick2023:10.1093/mnras/stad1179}. Even if it is mostly transparent and does not absorb the radiation coming from inside, we still expect scattering due to gas \citep{Naidu:2025arXiv250316596N}, and also some dust attenuation to occur. Trace amounts of dust could be present due to outflows from blue stars on the outskirts and outside of the halo. Such stars could be necessary, in some SMS formation scenarios, to provide the Lyman-alpha background needed for the isothermally collapsing halo to be realised (see references in the introduction). For SMSs to form, an environment of low metallicity $Z<10^{-5}\,Z_\odot$ is necessary \citep{Omukai:2008ApJ...686..801O}. At these metallicities, the effect of pollutants on the halo opacity will be minor (see \citealt{Cox:1976ApJS...31..271C}). But some weak metal absorption features might appear in the final observed spectrum, as discussed by \citealt{Kasen:2011ApJ...734..102K}. The result could be a reddened spectrum with some absorption features and broadened spectral lines not too different from previously observed reddened AGN or LRDs.

Other spectral features could appear in the recombination phase of the light curve. There, the shock shell region is only partly ionised and the radiative region lies below a layer of recombined material. Due to this, a Balmer break could appear in the spectrum as well \citep[see Fig.11 in][]{Kasen:2011ApJ...734..102K}, which may change the detectability. However, the strength of the Balmer break strongly depends on the detailed temperature structure and excitation state of the recombined shell. Therefore, a more sophisticated spectral modelling is required for quantitative predictions.

As we discussed in the previous section \ref{subsec:discussion:ionization-of-CSM}, the CSM could be almost fully ionised, which might lead to a moderate optical depth to electron scattering. This could lead to a lowered luminosity because photons trapped in the ionised shell will diffuse out within the diffusion timescale out of the ionised region. The CSM might also be partially opaque to Balmer-continuum photons, which might create a visible Balmer break in the final spectrum of around one magnitude. Without ionisation of the CSM, the Balmer attenuation could be so severe that it leads to essentially complete absorption. In these cases, the strong Balmer break might be photometrically miss-identified as a Lyman break, leading to a wrong redshift estimate for those sources.

SMS explosions could also be confused with high-$z$ AGNs. High-redshift AGNs are often identified photometrically through filter-dropout. Once an AGN is identified, one usually uses single-epoch observations of the luminosity to estimate the AGN mass while assuming Eddington luminosity. Since SMS explosions would be similarly affected by filter-dropout, it is possible to miss-classify them as AGNs. If we then assume an AGN instead of an SMS explosion transient, the mass estimation of the remnant BH will be wrong by several orders of magnitude. This can be understood because the luminosity of an SMS explosion is not limited by the Eddington luminosity of the remnant BH. 

As an example we take an SMS explosion with observed peak luminosity of $10^{45-46}\,\mathrm{erg\,s^{-1}}$. If we assume this to come from an AGN that radiates with Eddington luminosity $L_\mathrm{edd}=1.4\e{38}\,\mathrm{erg\,s^{-1}}\,(M/M_\odot)$, we would infer a central BH mass of $10^{7-8}\,M_\odot$. The corresponding SMS remnant BH however would only have a mass on the order of $10^{5-6}\,M_\odot$. By confusing an SMS explosion with an AGN we therefore could overestimate the masses of some early SMBHs. Taking this possible bias into account could thus alleviate the tension for SMBH formation caused by high-$z$ AGNs with extremely massive SMBHs. 

There is also the possibility of confusion with brown dwarfs. SMS explosions with $\eta<1$  appear as black bodies with $T\sim 6000/(1+z)\,$K in the recombination phase. This is similar to the surface temperature of brown dwarfs. SMS explosions and brown dwarfs could therefore populate similar regions in the colour-magnitude and colour-colour diagrams. But this confusion can be avoided if filter-dropout is considered or if multi--band data is available.

SMS explosions can also leave behind more indirect evidence, which would be visible long after the explosion happens. One signature is bursty star formation within a relatively small area of the host galaxy/halo, for which there is evidence in compact high-$z$ galaxies \citep{Castellano:2024jwst,Zavala:2024detection}. This could happen if an SMS explodes inside a halo and the ejected material then sweeps through it, forming shocks and compressing matter. Another way is strong AGN feedback, that is launched when the halo material accretes on the newly formed BH. These channels would then lead to enhanced star formation. The abundance of young blue stars would then lead to a bluer spectrum with a strong Balmer break. This would be detectable even if the young central BH is obscured by the host proto-galaxy (see \citealt{Silk:2024rsf}). Bursty star formation also drives chemical enrichment of the AGN and could explain the observed rapid chemical evolution of early AGN \citep{Shen:2018ojb}. Taken together, these indirect signatures would likely lead to a combined spectrum of a young SMS-remnant AGN with a young blue stellar population associated with bursty star formation. This morphology corresponds to a number of previously observed LRDs, where the red part could come from an accreting AGN and the blue part from a young stellar population.

We might additionally expect metal-poor and/or super-Eddington accreting AGNs. Since SMSs have a low metallicity and are surrounded by a roughly primordial halo, an accretion torus would form out of this material around the remnant BH. This would be observable as a metal-poor AGN, as observed for example by \citealt{Furtak:2024}. The AGN could also be obscured by the rapidly infalling matter, as suggested by \citealt{Kido:2025arXiv250506965K}.
Since SMSs form in high-density regions, super-Eddington accretion might occur temporarily, which might be observed at lower redshift \citep{Suh:2024jbx}.

Other observational signatures from collapsing SMSs include the emission of gravitational waves (GWs) during BH formation itself \citep{Uchida:2017qwn,Shibata:2016vzw}; the ringdown signal could be observed at redshifts $z<5$ by the laser interferometry space antenna (LISA) \citep{Shibata:2016vzw}. LISA could also see a clump of gas in the accretion disk around the BH \citep{Uchida:2017qwn}. Binary intermediate-mass BHs created from collapsing SMSs can also easily form in the protostellar disk \citep{Patrick2023:10.1093/mnras/stad1179,Latif:2020kto}, which are also observation targets for LISA \citep{LISA:2017pwj}. Accreting pairs could be visible as a binary AGN. Some candidates have already been observed by JWST \citep{Ubler:2023mxt}.

\subsection{Rate Estimation of High-z SMS Transients} \label{subsec:results:Rate-Estimation-of-High-z-Transients}

\begin{table*}
\resizebox{\textwidth}{!}{
\begin{tabular}{c|c|c|c|c|c} 
 survey and method & sky area & transient time & $z$-range & comoving volume [Mpc$^3$] & detectable rate [Mpc$^{-3}$yr$^{-1}$] \\ [0.5ex] 
 \hline
 \hline
     EUCLID EWF survey & $14500\,\mathrm{deg}^2$ & $100\:(10)\,\mathrm{yr}$ & $7$--$11.5\:(5$--$5.5)$ & $2.30(0.73)\e{11}$ & $0.43(13.7)\e{-13}$ \\ [0.5ex] 
     \hline
     EUCLID EWF+single-filter & $14500\,\mathrm{deg}^2$ & $100\:(10)\,\mathrm{yr}$ & $7$--$14\:(5$--$6.5)$ & $6.11(2.06)\e{11}$ & $0.16(4.85)\e{-13}$ \\ [0.5ex]  
     \hline
     EUCLID EDF survey & $53\,\mathrm{deg}^2$ & $100\:(10)\,\mathrm{yr}$ & $7$--$11.5\:(5$--$7)$ & $1.59(0.98)\e{9\:}$ & $0.63(10.2)\e{-11}$ \\ [0.5ex]  
     \hline
     EUCLID EDF+single-filter & $53\,\mathrm{deg}^2$ & $100\:(10)\,\mathrm{yr}$ & $7$--$14\:(7$--$9)$ & $2.23(0.79)\e{9\:}$ & $0.45(12.6)\e{-11}$ \\ [0.5ex] 
     \hline
     RST SN med+deep & $14\,\mathrm{deg}^2$ & $100\,\mathrm{yr}$ & $7$--$10.5$ & $3.41\e{8\:}$ & $2.93\e{-11}$ \\ [0.5ex] 
     \hline
     RST SN med+deep+single-filter & $14\,\mathrm{deg}^2$ & $100\,\mathrm{yr}$ & $7$--$13$ & $5.26\e{8\:}$ & $1.90\e{-11}$ \\ [0.5ex]
     \hline
     \hline
     RST SN med & $9\,\mathrm{deg}^2$ & $(10)\,\mathrm{yr}$ & $(7$--$8)$ & $(7.06\e{7\:})$ & $(1.42\e{-9\:})$ \\ [0.5ex] 
     \hline
     RST SN med+single-filter & $9\,\mathrm{deg}^2$ & $(10)\,\mathrm{yr}$ & $(7$--$9.5)$ & $(1.64\e{8\:})$ & $(6.08\e{-10})$ \\ [0.5ex]
     \hline
     RST SN deep & $5\,\mathrm{deg}^2$ & $(10)\,\mathrm{yr}$ & $(7$--$9)$ & $(7.46\e{7\:})$ & $(1.34\e{-9\:})$ \\ [0.5ex] 
     \hline
     RST SN deep+single-filter & $5\,\mathrm{deg}^2$ & $(10)\,\mathrm{yr}$ & $(7$--$10.5)$ & $(1.22\e{8\:})$ & $(8.20\e{-10})$ \\ [0.5ex]
     \hline
     \hline
\end{tabular}
}
\caption{For different surveys of EUCLID and RST (see main text), we show the covered sky area, the transient timescale in the source frame, the redshift range in which SMSs can be observed, the comoving volume covered, and the detectable SMS explosion rate. If the intrinsic SMS explosion rate is above the detectable rate, we expect at least one detection in a given survey. The numbers are given based on the brightest SMS explosions in our study (up to luminosities $L_\mathrm{bol}\gtrsim10^{47}\,\mathrm{erg\,s^{-1}}$ and with $\eta>1$); the values in round brackets correspond to moderately bright SMS explosions with luminosities $L_\mathrm{bol}\sim10^{46}\,\mathrm{erg\,s^{-1}}$ and $\eta<1$. The lower redshift-cutoff was chosen as $z=7$ because we don't expect many SMS to form by that time. The redshift range was extended for the less bright cases. For a comparison to theoretical estimates and calculations of the SMS explosion rate, see \figref{fig:results:SMS-explosion-rate-estimation}.}
\label{tab:results:telescope-survey-SMS-detection-rates}
\end{table*}

In this section, we focus on the expected observable rate of SMSs. Surveys of the high-$z$ universe will be paramount to this and allow us to constrain a significant range of the SMS formation and evolution channels through their signatures when they explode. Based on current and planned telescope hardware, JWST, EUCLID and RST will be useful to detect SMS explosions. EUCLID and RST specifically will be able to put strong constraints on the lower-redshift SMS formation channels at $z<7$--$10$. They are designed to make long-duration surveys in the infrared band during their missions. 
We here focus on EUCLID and RST, mainly due to their large sky area coverage and thus larger possibility to observe SMS explosions. Detection bounds for JWST will be taken from other works to serve as a comparison.

EUCLID will perform two main sky surveys (see \citealt{Euclid:2021icp}): The EUCLID wide field (EWF) covers an area of $14500\,\mathrm{deg}^2$ (roughly $1/3$ of the sky) in the Y-, J- and H-bands at a depth of $24.5\,\mathrm{mag}$. The sky map will be produced over the span of $6\,$yrs, but with limited overlap between each observed sky patch. After the design mission has been completed, EUCLID could revisit some of the previously visited patches. This will lead to enhanced opportunities of detecting slowly varying transients such as SMS explosions. But even with single-epoch observations it could be possible to observe SMS transients, as was already attempted by \citealt{Moriya:2023jfq}. The other important survey is the EUCLID deep field (EDF). It will cover an area of $53\,\mathrm{deg}^2$ in the Y-, J- and H-bands with a depth of $26.5\,\mathrm{mag}$. Each frame will be visited for 40--50 times over the span of the six-year mission and thus allow good opportunities to spot variations of high-redshift transients. The higher magnitude depth will also allow to observe fainter objects that might be missed in the EWF. 

The RST has a specifically designed survey to spot high-z supernova (SN) transients. The RST SN survey (see \citealt{WFIRST:2013Roman:arXiv1305.5422S}) will be performed in three variations of wide, medium and deep. They cover sky areas of $\sim (27.4, 9, 5)\,\mathrm{deg}^2$ and observe in the band filters (Y\&J, J\&H, J\&H) and at depths of (Y = 27.1 \& J = 27.5, J = 27.6 \& H = 28.1, J = 29.3 \& H = 29.4)$\,\mathrm{mag}$, respectively. Each patch is visited every 5 days distributed over the span of $2\,\mathrm{yrs}$. 

Early data from some JWST deep fields has also been used to perform a similar transient search \citep{Moriya:2023jfq} spanning $88.7\,\mathrm{arcmin}^2\sim0.025\,\mathrm{deg}^2$ with a depth of $\sim29\,\mathrm{mag}$.

There will also be some time overlap between the EUCLID, RST and JWST missions. There are therefore additional opportunities to cross-correlate images of overlapping sky patches to extend the detection range in magnitude and timescale. 

Ideally, a survey to find SMS transients should cover a large sky area, have a deep magnitude reach and run over a large timescale of at least $\sim 0.5$--$1.5\,(1+z)\,\mathrm{yrs}$ (for the medium bright cases with $\eta<1$) to allow for the intrinsic light curve variability to manifest. Based on the available surveys, we expect the EDF and the RST SN medium/deep surveys to provide the best opportunities of detecting SMSs in the redshift range $z<10$. But even the EWF could be useful if it is possible to reliably identify SMS explosions from single-epoch observations and/or if it is possible to detect them using data of a single spectral filter. This at least sounds plausible due to the high SMS explosion brightness and extremely red characteristic colour and redshift. We therefore take these surveys as a starting point and compute the number rates of SMS explosions that we expect to find within them. 

We use our light curve calculations to determine the range in which SMS explosions will be visible within each survey. We then use this knowledge to constrain SMS explosions at the low-redshift end and subsequently constrain the intrinsic number density of SMSs. In a later study, using this, one could also estimate the amount of expected SMBH seeds and compare this with the actually observed high-$z$ AGN. 

We estimate the visibility range of SMS explosions using the colour–magnitude diagrams in \figref{fig:results:colour-magnitude-diagrams-EUCLID-ROMAN} for EUCLID and RST as follows. We look at the individual light curves and read off whether they are inside of the detectable range of the survey. The signals become less bright in the filters at higher redshift. We then mark the maximum redshift $z_\mathrm{max}$ when they become invisible and this marks the detection range ($z<z_\mathrm{max}$) for SMS explosions of a given brightness. The maximal redshift ranges should be understood with an error of $\Delta z \sim0.5-1$. We here neglect the possible attenuation by the CSM, as discussed in section \ref{subsec:discussion:ionization-of-CSM}, because the effect would be minor $\Delta\,\mathrm{mag}_\mathrm{AB}\ll 1\,$mag and a change of this order lies well within our model uncertainty.

As a first step, we consider the most optimistic case where we can see any of the SMS explosion light curves. The brightest models (e.g. FujH4, FujDif1, FujHe4\footnote{Note the caveats regarding ionisation of the CSM discussed in sec. \ref{subsec:discussion:ionization-of-CSM}}) with luminosities $L_\mathrm{bol}\gtrsim10^{47}\,\mathrm{erg\,s^{-1}}$ (and also $\eta>1$) then become the benchmark for observability. We find that the observability is limited by the spectral range of the filters due to the Lyman absorption and not by the intrinsic brightness of the sources. But if most of the filter range is absorbed, the observed AB magnitude will at some point become too faint to be detected in the respective filters. We then find that SMS explosion transients are visible in the surveys as follows: in EWF for $z<11.5$, in EDF for $z<11.5$, in EWF with a detection only in a single-filter (only H-band) for $z<14$ and in EDF with single-filter-detection (only H-band) for $z<14$.

For the RST, we find that SMSs could be visible in the surveys as follows: in RST SN medium for $z<10.5$, in RST SN deep for $z<10.5$, in RST SN medium with single-filter-detection (only H-band) for $z<13$ and in RST SN deep with single-filter-detection (only H-band) for $z<13$. We find that the observability is also fundamentally limited by the filter wavelength range ($1.38$--$1.77\,\mu$m), rather than by the observable magnitude depth. The maximum observable redshift range for them is the same. We therefore can use both RST SN medium and SN deep surveys together to get a larger effective sky area of $9+5=14\,\mathrm{deg}^2$. 

The above ranges may seem a bit optimistic because we consider the observability of only the brightest SMS explosions. Also, these brightest cases were the models with $\eta>1$, where our model is not too reliable for quantitative statements. We therefore repeat the same procedure for more moderately bright SMS explosions (they also correspond to cases with $\eta<1$) of $L_\mathrm{bol}\sim10^{46}\,\mathrm{erg\,s^{-1}}$ (e.g. FujH1, FujHe1). We find that SMSs are visible in the surveys as follows: in EWF for $z<5.5$, in EDF for $z<7$, in EWF with single-filter-detection (only H-band) for $z<6.5$, in EDF with single-filter-detection (only H-band) for $z<9$.

For the RST, we find that SMS transients would be visible in the surveys as follows: in RST SN medium for $z<8$, in RST SN deep for $z<9$, in RST SN medium with single-filter-detection (only H-band) for $z<9.5$ and in RST SN deep with single-filter-detection (only H-band) for $z<10.5$. For these cases the visibility is limited by the magnitude depth.

Using these redshift ranges, we compute the comoving volume that lies within the observation range. For the lower redshift-cutoff we choose $z=7$ because we expect most SMSs to form at $z=10\sim20$ \citep{Patrick2023:10.1093/mnras/stad1179}, or until lower redshifts ($z> 6$--$7$) in non-standard formation scenarios \citep{kiyuna:2024}. The redshift range was extended for the less bright cases. Similarly to \citealt{Moriya:2023jfq}, we then use the intrinsic timescale of the SMS explosion light curves to compute the expected SMS explosion rate. We do this in the following way: In this work, we found that the brightest SMS light curves with $\eta>1$ can last around $100\,$yrs in the source frame. Moderately bright light curves have a typical duration of $10\,$yrs in the source frame. We compute the bound on the intrinsic rate per comoving time and per comoving volume from these values. Assuming that no detection is made during the survey, this means that, in the source frame, no explosion has happened within at least one transient timescale ($10\,$yrs or $100\,$yrs, respectively). This works if the survey timescale is shorter than the transient timescale, which is well fulfilled here. The rate will therefore give the observational bound on SMS explosion existence in this redshift range. If the true intrinsic SMS explosion rate is higher, we would expect to observe at least one event during the survey time. The results are summarised in Table \ref{tab:results:telescope-survey-SMS-detection-rates}.

Based on the results from \tabref{tab:results:telescope-survey-SMS-detection-rates}, we found that the EUCLID wide field (EWF) has the possibility to most strongly constrain the SMS explosion rate to an order of $\sim 10^{-14}$Mpc$^{-3}$yr$^{-1}$ in a redshift range of $7$--$14$ for the brightest cases, or $\sim 10^{-12}$Mpc$^{-3}$yr$^{-1}$ in $5$--$6.5$ for less bright SMS explosions. But the wide field has the caveat that it only allows for single-epoch detections. This might complicate the identification based on photometric data alone. The EUCLID deep field (EDF) does not have this issue but the constraining potential is smaller by roughly two orders of magnitude at $\sim 10^{-12}$Mpc$^{-3}$yr$^{-1}$ (bright cases) or $\sim 10^{-10}$Mpc$^{-3}$yr$^{-1}$ (moderately bright cases) due to the smaller sky coverage. From the RST we expect to detect rates of above $\sim 10^{-9}$Mpc$^{-3}$yr$^{-1}$, even for the less bright SMS explosions. 

A similar study to ours has also been done by \citealt{Moriya:2023jfq} for some JWST deep fields. They found an SMS rate bound of $8\e{-7}$Mpc$^{-3}$yr$^{-1}$ in the redshift range of $10$--$15$. This is significantly less constraining compared to our EUCLID and RST bounds. The smaller detectable rate is because they assumed a transient timescale of $2\,$yrs in the source frame and due to the smaller survey volume. When we use their survey volume of $6.2\e{5}$Mpc$^3$ with our transient timescales, we get $1.6\e{-8}$Mpc$^{-3}$yr$^{-1}$ (brightest cases) and $1.6\e{-7}$Mpc$^{-3}$yr$^{-1}$ (moderately bright cases), respectively. We therefore stress that the EUCLID and RST surveys should not be overlooked to observe or constrain SMS explosions. 

The intrinsic rate of SMSs based on cosmological simulations has been predicted by several studies (also see references in \citealt{Moriya:2023jfq}) to be around $10^{-8}$--$10^{-12}$Mpc$^{-3}$yr$^{-1}$ at $z\sim10$--$15$. On the lower redshift range, where EUCLID and RST will be most sensitive, some studies \citep{Agarwal:2012MNRAS.425.2854A,Chiaki:2023MNRAS.521.2845C} predict rates of $6\e{-10}$Mpc$^{-3}$yr$^{-1}$ (at $z=6$) \citep{Agarwal:2012MNRAS.425.2854A} and $5\e{-8}$Mpc$^{-3}$yr$^{-1}$ (at $z=10$ but plateauing for smaller $z$) \citep{Chiaki:2023MNRAS.521.2845C}, both assuming an average SMS lifetime of $2\e{6}\,\mathrm{yr}$. 

\begin{figure}
    \centering
    \includegraphics[width=0.49\textwidth]{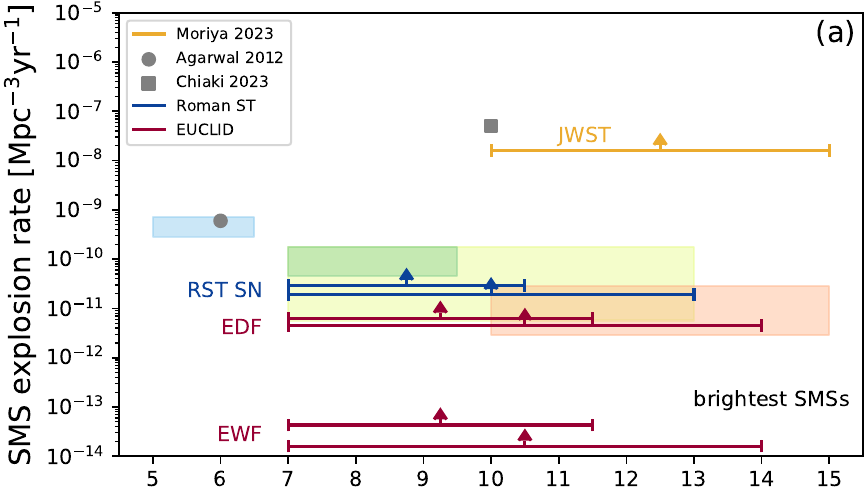}
    \includegraphics[width=0.49\textwidth]{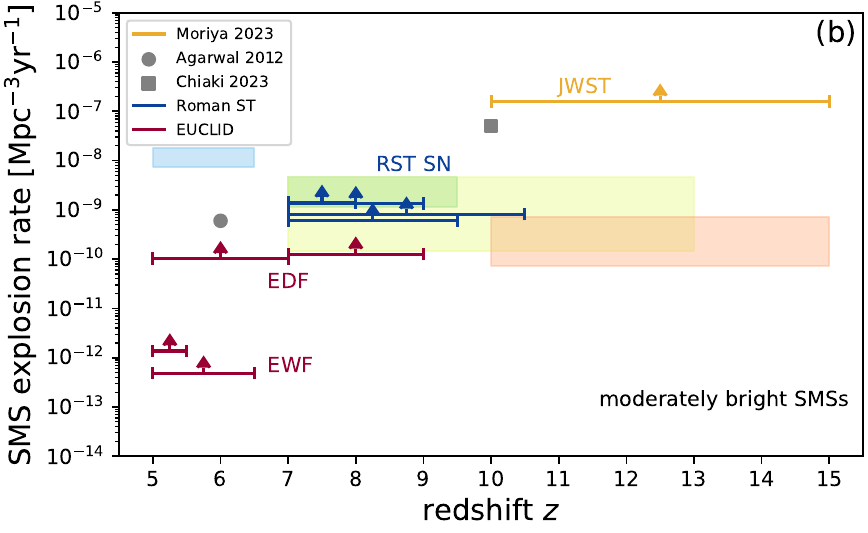}
    \caption{SMS explosion rate estimates and constraints as functions of redshift. The horizontal lines mark the (upper) bounds on the observable rates from the EUCLID and Roman space telescope (RST) surveys listed in \tabref{tab:results:telescope-survey-SMS-detection-rates}. Rates above will be detectable. Panel \textbf{(a)} shows the rates for the brightest SMS cases (with $\eta>1$) and panel \textbf{(b)} for moderately bright SMS explosions (with $\eta<1$). We also show the observable rate for JWST obtained by \protect\citealt{Moriya:2023jfq} and intrinsic SMS explosion rates from cosmological simulations by \protect\citealt{Agarwal:2012MNRAS.425.2854A} and \protect\citealt{Chiaki:2023MNRAS.521.2845C}. The coloured boxes mark estimated rates based on cosmic star formation history \protect\citep{Harikane:2023ApJS..265....5H}, assuming a Salpeter initial mass function and that the fraction of visible exploding SMSs is large ($f_\mathrm{GRISN}\approx 1$). In panel \textbf{(a)} the boxes mark rates for SMSs with $M\sim10^5$--$10^6\,M_\odot$ and in panel \textbf{(b)} for $M\sim10^4$--$10^6\,M_\odot$. The derived rates are for some cases significantly higher than what can be detected. We therefore expect to find multiple SMS explosions using EUCLID and RST, but not JWST.}
    \label{fig:results:SMS-explosion-rate-estimation}
\end{figure}

We additionally follow a calculation similar to \citealt{Moriya:2023jfq} to estimate the expected SMS rate based on the observationally determined cosmic star formation rate (SFR) (see Figure 18 in \citealt{Harikane:2023ApJS..265....5H}). Assuming an initial mass function $\psi$ for SMSs, we can compute the SMS rate $r_\mathrm{SMS}$ from the star formation density rate $\rho_\mathrm{SFR}$ as $r_\mathrm{SMS} = \rho_\mathrm{SFR} f_\mathrm{GRISN} \Psi (\psi)$, with $\Psi(\psi)$ defined below in \eqref{eq:results:salpeter-IMF}. $f_\mathrm{GRISN}$ is the fraction of SMSs that explode via a GRI SN and produce a visible light curve. We here consider an optimistic scenario where $f_\mathrm{GRISN} \approx 1$. But we have to keep in mind that a successful SMS explosion strongly depends on the mass and rotation rate, so the fraction should be lower. However, there are currently no estimates on what value would be appropriate for $f_\mathrm{GRISN}$. With our value, we can therefore get a rough upper estimate for the number of observable SMS explosions. For simplicity, we also assume the Salpeter initial mass function $\psi(x=2.35) \propto M^{-2.35}$, as was done by \citealt{Moriya:2023jfq}, because the true SMS initial mass function is not known. We get
\begin{align}
    \Psi(\psi) = \frac{\int_{m_\mathrm{low}}^{10^6M_\odot} \psi dM}{\int_{0.1M_\odot}^{10^6M_\odot} M \psi dM} \: . \label{eq:results:salpeter-IMF}
\end{align}
To get an estimate for SMS explosions rates, we take $m_\mathrm{low}=10^4\,M_\odot$ for our medium-brightness models (with $L_\mathrm{bol}\sim10^{45-46}\,\mathrm{erg\,s^{-1}}$) and $m_\mathrm{low}=10^5\,M_\odot$ for our brightest models (with $L_\mathrm{bol}\gtrsim10^{47}\,\mathrm{erg\,s^{-1}}$). We thus get $\Psi=4.6\e{-7}\,M_\odot^{-1}$ and $\Psi=1.8\e{-8}\,M_\odot^{-1}$, respectively. Finally, we use the observational information on $\rho_\mathrm{SFR}$ from \citealt{Harikane:2023ApJS..265....5H} (see their Figure 18). We take $\rho_\mathrm{SFR}$ in specific redshift ranges and compute the SMS rate $r_\mathrm{SMS}$ from there. Note that for $z\lesssim7$, taking the total SFR could over-estimate SMS rates because the primordial environment required for SMS formation is more difficult to achieve.

All SMS explosion rates and our calculated constraints are summarised in Figure \ref{fig:results:SMS-explosion-rate-estimation}. The bounds and redshift ranges obtained for the various EUCLID and RST surveys from Table \ref{tab:results:telescope-survey-SMS-detection-rates} are denoted by the horizontal lines. In panel (a) we show the rates for detecting only the brightest SMSs with $L_\mathrm{bol}\gtrsim10^{47}\,\mathrm{erg\,s^{-1}}$ and $\eta>1$. In panel (b) we show the rates for moderately bright cases with $L_\mathrm{bol}\sim10^{45-46}\,\mathrm{erg\,s^{-1}}$ and $\eta<1$. Additionally we show the rates for JWST using the results from \citealt{Moriya:2023jfq}, but using the transient timescales from our work (see discussion above). The intrinsic rates coming from cosmological simulations of \citealt{Agarwal:2012MNRAS.425.2854A} and \citealt{Chiaki:2023MNRAS.521.2845C} are shown as well. The coloured surfaces mark the SMS rate obtained from the observed star formation rate from \citealt{Harikane:2023ApJS..265....5H} with a Salpeter initial mass function (see the discussion around \eqref{eq:results:salpeter-IMF}). 

We find that the derived rates can in some cases be significantly higher than the detectable rate. We therefore expect to find SMS explosions using EUCLID and RST. More specifically, in panel (a), we show the rates and detection bounds for the brightest SMS cases, which correspond to the cases with $\eta>1$. We see that the EUCLID wide field bound lies significantly below the expected SMS explosion rates inferred from the star formation history (coloured surfaces) and from cosmological simulations (results from \citealt{Agarwal:2012MNRAS.425.2854A} and \citealt{Chiaki:2023MNRAS.521.2845C}). With this, for the EUCLID wide field survey with detection in multiple bands, we expect more than $100\times f_\mathrm{GRISN}$ SMS single-epoch detections based on the star formation rate. If we take the numerical results from \citealt{Agarwal:2012MNRAS.425.2854A} and \citealt{Chiaki:2023MNRAS.521.2845C} as a benchmark, then we can expect to detect up to $(10^4$--$10^6)\times f_\mathrm{GRISN}$ SMS explosions. For the EUCLID deep field we would expect to detect on the order of a few up to a few $100\times f_\mathrm{GRISN}$, based on our rate estimate. Based on the cosmic SFR of \citealt{Harikane:2023ApJS..265....5H}, the RST SN surveys on the other hand might detect an order of $10\times f_\mathrm{GRISN}$ massive SMS explosions. But the simulation result from Chiaki is more optimistic and could lead to hundreds of detections in the RST SN survey. For JWST, current bounds are too weak to detect SMSs based on the star formation rate and simulation by \citealt{Agarwal:2012MNRAS.425.2854A}. But based on the result from \citealt{Chiaki:2023MNRAS.521.2845C}, we could expect a few detections. But this is not very clear because the fraction of SMSs that explode via a GRI supernova ($f_\mathrm{GRISN}$), and produce visible signatures, is not known. Also, Chiaki's results show the total rate for all SMSs, not just the brightest cases. The value shown here might thus be an over-estimate for the brightest SMS explosion cases and we do not expect SMS detections in the current data. But with a larger sky area covered, we conclude that JWST could at least provide moderate upper bounds.

For the less massive and less bright SMS cases (with $\eta<1$) one would naively expect less detections due to the lower sensitivity in brightness. But this is not necessarily true, because less massive SMSs are also more likely to form and thus more numerous. We show this in panel (b). The coloured surfaces again mark the expected rates based on the SFR. The rates are higher compared to the more massive stars because less massive SMSs are more likely to form. This is a consequence of the Salpeter initial mass function. The expected numbers of detections in EUCLID and RST are thus higher, even though the redshift detection range is smaller for those stars. For the EWF, we now expect to find on the order of $>10^4 \times f_\mathrm{GRISN}$ SMS explosions at redshifts $5$--$7$ (given they can form at such low redshifts). For the EDF we should find around $100\times f_\mathrm{GRISN}$ SMS explosion transients for $z=7$--$9$. The RST SN surveys detection rates lie within the star formation induced rates, but around two orders of magnitude below the numerically simulated rates of~\cite{Chiaki:2023MNRAS.521.2845C}. Hence, $\gtrsim 10\times f_\mathrm{GRISN}$ detections in the RST SN surveys may be possible. For JWST, the picture is similar as with the more massive SMS cases. But here, the JWST bound lies above the estimates from \citealt{Chiaki:2023MNRAS.521.2845C}. We therefore do not expect to find SMS explosions in the currently available data. 

We should note that generally our estimates are to be understood cautiously as a rough order of magnitude estimate. This is because there is still significant remaining uncertainty in, e.g., the light curve model, properties of the SMS-forming halo, the SMS explosion fraction $f_\mathrm{GRISN}$ and the SMS initial mass function at time of explosion.

As discussed earlier in section \ref{subsec:results:distinguishing-SMS-Explosions-from-alternative-sources}, there is also a possibility to photometrically confuse LRDs with SMS explosion transients. We briefly assess the risk of miss-classification here. According to the COSMOS-Web survey of JWST \citep{Akins:2024:2024arXiv240610341A}, over $434$ LRDs have been observed therein to date and around $500$ more have been found during other surveys as well (see references in \citealt{Akins:2024:2024arXiv240610341A}). The covered survey area of COSMOS-Web was $0.54\,\mathrm{deg}^2$ at $z\sim5$--$9$ and the probed comoving volume ranges over $1.8\e{7}$\,Mpc$^3$ over a timespan of $2.5/(1+z)\,$yrs in the source frame. This roughly leads to a detection rate of LRDs of $7.7\e{-5}$\,Mpc$^{-3}$yr$^{-1}$. The actual rate could be higher because not the whole observation time was devoted to finding LRDs.

Now, if we consider that LRDs can appear in similar redshift ranges and brightness as SMS explosions, we can compare the SMS and LRD rates to estimate the miss-classification risk. Based on the SMS explosion rates inferred from the star formation history and from numerical simulations (see \figref{fig:results:SMS-explosion-rate-estimation}), we find that LRDs are, at least, around $3$--$4$ orders of magnitude more prevalent than SMS explosions. Therefore, around $0.1\,\%$ to $0.01\,\%$ of the detected LRDs have the risk of being photometrically miss-classified SMS explosion transients. Given the currently known total number of LRDs, we find that at most $\sim0.1$--$0.9$ of them could be SMS explosions that are miss-classified as LRDs. When the number of known LRDs rises, we should eventually expect that some of them could be SMS explosions. We therefore conclude that -- even through the miss-classification risk is not too high -- it could be worthwhile to perform follow-up spectral observations of LRD candidates because valuable scientific opportunities to detect SMS explosions might be missed otherwise.

\subsection{Further Investigation Topics and Future Model Improvements} \label{subsec:discussion:further-investigation-and-future-model-improvements}

Another interesting path of study would be less massive large stars of around a few $10^2$--$10^4\,M_\odot$ \citep{Volpato:2023ApJ...944...40V}. They are sometimes called ``very massive stars'' (VMSs), which would collapse to a black hole after the pair-creation instability \citep{1971reas.book.....Z}. These could form in similar environments to SMSs but have evolved under different circumstances. For example, with smaller sustained accretion rates  ($<0.02\,M_\odot \mathrm{yr}^{-1}$), or they have been ejected out of the star-forming disk by N-body interactions during their formation (see e.g. \citealt{kiyuna:2024}), thus cutting off the mass accretion and leading to massive stars of $\sim1000\,M_\odot$. Additionally, these stars might form more frequently due to their smaller mass and because they do not need such ideal/extreme environments as are needed for typical SMSs.

VMSs are of interest because they have a different explosion channel from SMSs. 
Even though the collapse trigger is different from that of SMSs, the outcome could be similar and the mass outflow could be launched by a disk wind if the VMSs are rapidly rotating (e.g., \citealt{Uchida:2018ago}). Because such VMSs would likely also be surrounded by a dense CSM, an interaction powered supernova could happen. A main caveat would be however that VMSs evolve as blue supergiants and the strong UV radiation might ionise or blow away the surrounding CSM. This could significantly impact the dynamics of the ejecta and the observability to distant observers.

We give a rough estimate for the expected brightness and visibility of these VMSs. Assuming that they did not blow away their dense CSM, we can apply our light curve model for this type of supernovae. For the case where the CSM is diluted significantly, the light curve would look more closely to a Type IIp supernova and can be modelled using the model of \citealt{Matsumoto:2015bjg} for an expanding envelope with a recombination phase. 

We take as an example a VMS with mass of $880\,M_\odot$ and a core mass of $800\,M_\odot$, and another one with $6600\,M_\odot$ and a core of $6000\,M_\odot$. We now follow the same arguments as in section \ref{subsec:results:model-selection}. The stars will leave behind BHs of $\sim720\,M_\odot$ and $\sim5400\,M_\odot$, respectively. The remaining mass is ejected or part of an accretion torus. The ejecta kinetic energy and total ejecta mass will be $(2.9\e{53}\,\mathrm{erg}, 88\,M_\odot)$ and $(2.2\e{54}\,\mathrm{erg}, 660\,M_\odot)$. We also assume an initial radius of $1000\,R_\odot$. Using our light curve model, we find peak luminosities of $3\e{45}\,\mathrm{erg\,s^{-1}}$ and $2\e{46}\,\mathrm{erg\,s^{-1}}$, respectively. The light curve time-scales are around $7$\,yrs in the optically thick shock phase for both cases. This is not too different from other more massive SMS explosion cases. But for both VMS cases, the diffusion peak is reached within less than one year after the explosion starts. This will naturally lead to a faster and less persistent transient with higher variability. We find that both VMSs could be visible in the EUCLID wide and deep fields up until redshifts of $8$--$9$. With the higher expected abundance of VMSs compared to SMSs, we expect to find many such explosions in the EUCLID and RST observations. This highlights the opportunity of EUCLID as an ideal tool to detect massive star explosions. 

As a comparison, the light curve evolves as a TypeIIp SN if no dense CSM is present. The light curve can then be modelled in the same ways as in \citealt{Uchida:2018ago} or \citealt{Matsumoto:2015bjg}. \cite{Uchida:2018ago} finds a maximum brightness for $\sim320\,M_\odot$ stars (roughly comparable mass to our VMSs) of $\sim10^{43}$--$10^{44}\,\mathrm{erg\,s^{-1}}$. The lower brightness should make it more difficult to observe such cases. The presence of a dense CSM thus significantly enhances the visibility of massive exploding stars at high redshift. Conversely, the more abundant but less massive VMSs could become less visible due to not having a dense CSM surrounding them. We therefore expect that the visibility of less massive SMSs and VMSs could be inhibited both by smaller intrinsic brightness and a more dilute CSM. 

There are many ways in which the study of SMS explosions can be improved in the future. They can be summarised in three main points: improving the initial data for SMS collapse simulations, improving the collapse simulation and inclusion of more effects (e.g. more detailed nuclear burning networks and SMS atmosphere, see below), and improving the subsequent light curve model. 

For the first point, realistic initial data can be improved by self-consistently coupling SMS formation and stellar evolution models as a standard practice. The collapse of primordial halos, formation of SMSs and their evolution need to be evolved in parallel including back-reactions. This is also necessary to get an idea of the distribution in mass and composition of SMSs when they collapse. 

Regarding the second point, the final evolutionary states of the SMSs then need to be translated accurately as initial conditions for relativistic collapse simulations. For example, \citealt{Fujibayashi:2024vnb} approximates the SMS as a polytropic, chemically homogeneous core and neglects the bloated atmosphere. In the future, we will need simulations that take into account the whole SMS at onset of collapse. More complete nuclear burning networks will also be needed, especially for smaller SMSs and for stars with higher metallicity. The full inclusion of the SMS atmosphere will also be necessary to improve the quality of the ejecta calculation. 

For the third point, the light curve model can be improved by directly using the simulated profiles of the ejecta and then using them in radiative transfer hydrodynamic codes and/or semi-analytic light curve models. The final light curve can be improved in two main ways: with better dynamical modelling and by improving the modelling of the spectrum. Regarding the dynamics, while we approximated the CSM as a simple power law with $\rho\propto r^{-2}$ in this paper, the CSM profile has a great impact on the results as we show in Appendix~\ref{sec:appendix:light-curve-model-validation}. Ideally, one would use the density profile near the SMS directly from a halo collapse simulation. In realistic collapse scenarios, we would expect a protostellar disk close to the SMS \citep{Patrick2023:10.1093/mnras/stad1179}, which could lead to non-spherical shock-interactions and radiation emission. We also modelled the hydrodynamical interaction and radiation transport in the shock region under the one-zone approximation. Another key finding is that the CSM can significantly affect the intrinsic light curve of the shocked shell through ionisation, absorption, and scattering. To quantitatively predict and incorporate the effects of the CSM, future work will need to more holistically account for the various radiation and back-reaction processes occurring within it. This could be improved by using full radiation-transfer-hydrodynamics simulations, taking into account the non-local thermal equilibrium physics. 

Regarding the spectrum, we only discussed it very qualitatively and only took into account explicitly the Lyman-alpha absorption by interstellar gas. In the future, it will be necessary to also model individual emission and absorption lines, including a possible Balmer break in the optically thick phase (also see sec. \ref{subsec:discussion:ionization-of-CSM}), to quantitatively discuss the detectability and distinguishability of SMSs and other sources. The effect of the incomplete thermalisation (i.e., $\eta>1$) in the shock region was also modelled only superficially here. Future models will benefit by including this effect with more detail. The effects of the halo and environment such as possible dust attenuation, radiation self-absorption and scattering by the halo gas, will need to be taken into account in future works as well. A full lifetime simulation that includes all these steps is still some time away and is left for future works.

\section{Conclusions} \label{sec:conclusions}
In this work, we studied the light curves associated with explosions of supermassive stars (SMSs) at high redshift with $z>7$. We explored the scenario where a rapidly accreting SMS is initially surrounded by a dense circumstellar medium (CSM). Massive and energetic ejecta from the SMS are released during the explosion, which subsequently interact with the CSM via shocks. 

The shock region is optically thick and emits electromagnetic radiation. We developed a semi-analytic light curve model that computes its size, velocity, mass and internal energy. Using this, we computed the bolometric luminosity and emission spectrum as a function of time. We used initial values for the ejecta mass, ejecta energy and initial radius that are informed by stellar evolution calculations and general relativistic SMS collapse simulations. 
We found that the luminosity is initially dominated by emissions from an optically thick shock region. For some high-energy models, the radiation can fall out of equilibrium and the spectrum can shift to the blue, compared to a standard blackbody spectrum. This shift happens if the thermal coupling coefficient $\eta$ satisfies the condition $\eta>1$. At late phases, the shock becomes optically thin and emits non-thermal radiation. But it is much smaller than the emission in the optically thick shock phase for all cases considered here.

For moderately energetic cases with $\eta<1$, we found that the thermal emission phase of the light curves lasts $10$--$15$\,yrs in the source frame and can reach peak luminosities of $10^{45}$--$10^{46}\,\mathrm{erg\,s^{-1}}$. This is similarly bright to previously observed LRDs or AGN at high redshifts. The observed light curves can last up to $\sim 250$\,yrs due to cosmic time dilation. During the evolution, the effective surface temperature gradually cools down. This later crosses over to a constant-temperature recombination phase where the surface temperature is around $6000\,$K.

For more energetic cases with $\eta>1$ the evolution is similar, but the optically thick phase lasts longer (up to $200\,$yrs), due to inhibited hydrogen recombination. The maximum luminosity is generally higher, $\sim 10^{47}\,\mathrm{erg/s}$. The spectrum is also much bluer compared to the $\eta<1$ cases. 

We found that for the $\eta>1$ cases, the bluer spectrum produces a large amount of ionising radiation, which could partially ionise the CSM. But due to the special environment of atomically cooled halos, we further find that the CSM could be instead completely ionised by Balmer-continuum photons via ionising hydrogen in the $n=2$ excited state. This leads to a moderate scattering optical depth in the CSM of $\tau\sim4$ around the emission peak. This will smear out details in the light curve but retain the over-all shape. The Balmer-continuum photon ionisation is expected to happen for all cases considered here, except for the pulsating SMS models.

We also found that all sources show small variability. The timescale in which the source varies brightness by similar amounts to LRDs ($20\%$) is $0.5$--$9\,$yrs in the source frame. Thus, light curve variations of SMS explosions would remain undetected for $\sim0.5$--$9\, (1+z)\,$yrs in the observer frame. This is longer than the current operational time of JWST for sources at $z>7$. We therefore presume that some suspected persistent sources could be slowly varying SMS transients instead.

The photometric brightness of SMS explosions in different telescope filters of JWST was computed. SMS explosions can reach $19$--$22$ magnitudes in brightness in multiple long wavelength filters of JWST and can remain visible even up to $z\sim 20$. Pulsating SMSs can be visible in JWST up until $z\sim15$. We discussed and quantified the location and movement of SMS explosion transients at large redshift in colour-magnitude diagrams over time. In JWST NIRCam filters at $z=10\,(20)$, they mostly occupy the range $mag_\mathrm{F444W} > 19\,(21)$ and $mag_\mathrm{F277W}-mag_\mathrm{F444W} > -1\,(0)$. This places them in a similar region as LRDs. The movement in the diagram ranges over $\Delta m_\mathrm{AB}\sim 0.2$--$1$ magnitudes per $(1+z)$ observer years.

The photometric brightness of SMS explosions in EUCLID and the RST was computed as well. We especially focused on the observability of SMSs in different planned and ongoing surveys such as the EUCLID wide and deep fields and the RST supernova surveys. For EUCLID, we find that the brightest SMS explosions will be visible for $z<11.5$ in the wide and deep field surveys. The RST SN surveys can detect SMSs at $z<10.5$ for the brightest cases. For less bright SMS explosions, the redshift ranges are reduced but a detection generally remains viable until $z<5$--$7$ for EUCLID and $z<9$ for the RST. The redshift ranges are extended in all cases by $\Delta z\sim 1$--$3$ if single-filter detections in the longest wavelength filter are possible. 

The SMS formation and explosion rate is currently not well constrained. We suggested to probe it using surveys of EUCLID and RST. We found that they produce surprisingly stringent constraints that are much stronger than current bounds form JWST deep fields. The EUCLID wide field will be able to constrain SMS explosion rates in $z\sim7$--$12$ down to $\sim 4\e{-14}$Mpc$^{-3}$yr$^{-1}$, while the deep field reaches $\sim 10^{-11}$Mpc$^{-3}$yr$^{-1}$. The RST SN surveys could probe rates down to $\sim10^{-11}$Mpc$^{-3}$yr$^{-1}$. This is much deeper than current JWST bounds of $\sim10^{-7}-10^{-8}$Mpc$^{-3}$yr$^{-1}$. Based on the cosmic star formation history, and cosmological simulations, for $7<z<12$, we expect to find up to several hundred SMS explosions in the EUCLID deep field and up to a few dozen in the RST SN surveys. The EUCLID wide field survey could image up to several thousand SMS explosions. But it has the caveat that it cannot directly detect transients because the survey produces single-epoch images. We therefore identify the EUCLID deep field and RST SN survey to be best suited to detect SMS explosions. Both cover relatively large sky areas, have sufficient photometric depth to detect SMSs and provide multi-epoch detections spread over several years. This will make it possible to identify slowly varying transients. 

We discussed spectral properties and the possibility of miss-classification of SMS explosions as either LRDs or AGN. This could most easily happen due to the long variability timescale of the source and if no spectral information is present. Regarding spectral lines, we expect a metal poor spectrum with features of a type IIn supernova explosion. Strong hydrogen and helium lines are to be expected. Spectral lines would be blueshifted asymmetrically by $\sim 3000\,\mathrm{km\,s^{-1}}$ due to the shock shell expansion. Absorption and re-emission of light in the primordial circumstellar medium could give rise to P-Cygni profiles. Dust-attenuation is also possible if there was some prior star formation within the primordial halo. A Balmer break could also be present, depending on the ionisation state of the CSM. Indirect evidence for SMS explosions include metal-poor AGN and a bursty star formation history. 

Finally, we discussed future topics of study and avenues for improving the modelling of SMSs. We identified massive stars in the few $10^2$--$10^4\,M_\odot$ range as interesting future study targets. These stars could, e.g., form as ``failed'' SMSs. If immersed in a dense circumstellar medium, they could reach similar luminosities to SMSs of $\sim10^{45}\,\mathrm{erg\,s^{-1}}$ and be visible to relatively high redshifts of $8$--$9$. They could be easier to detect due to their higher expected number density and faster transient evolution. We also discussed the need to work towards complete lifetime simulations that include SMS formation, evolution and explosion to get a more holistic understanding of SMSs and their intrinsic abundance. We leave more detailed studies for future work.

\section*{Acknowledgements}

We thank Akihiro Suzuki, Samuel Patrick, Piyush Sharda, Yuichi Harikane, Takashi Hosokawa, Kohei Inayoshi, Xihan Ji, Shigeo Kimura, and Kazumi Kashiyama for helpful discussions. Special thank is to Masaomi Tanaka for his careful reading of the manuscript and valuable comments. This work was in part supported by Grant-in-Aid for Scientific Research (grant No.~23H04900) of Japanese MEXT/JSPS.

\section*{Data Availability}

All data used in this work was taken form the cited sources or produced by the authors. The Python code that was used to compute the light curves and generate the figures is available as is, without documentation, on the Git repository: \url{https://github.com/xcex/SMS-lightcurves} [last accessed 2025-07-21].




\bibliographystyle{mnras}
\bibliography{biblio} 




\appendix
\section*{Appendix \label{sec:appendix}}
\section{Analytic Form of the ejecta Energy and Mass Integrals} \label{sec:appendix:analytic-form-of-the-ejecta-energy-and-mass}

The integrals $I_n(\Tilde{\beta})$ and $J_n(\Tilde{\beta})$ that appear in the kinetic energy and mass of the ejecta (\eqref{eq:light-curve-model:initial-conditions-ejecta-kinetic-energy} and \eqref{eq:light-curve-model:initial-conditions-non-rel-ejecta-mass}) can be solved analytically for a range of integers $n$. $n$ corresponds to the power of the density distribution of the ejecta, see \eqref{eq:light-curve-model:ejecta-density-profile}. For selected values of $n$, the integrals are given by:
\begin{align}
    I_0(\Tilde{\beta}) &= \frac{\Tilde{\beta}}{2} \sqrt{1-\Tilde{\beta}^2} - \Tilde{\beta} - \frac{\arcsin(\Tilde{\beta})}{2} + \text{arctanh}(\Tilde{\beta}) \: , \\
    I_1(\Tilde{\beta}) &= \frac{1}{2} \left( \sqrt{1-\Tilde{\beta}^2} - 1 \right)^2 \: , \\
    I_2(\Tilde{\beta}) &= - \frac{\Tilde{\beta}}{2} \sqrt{1-\Tilde{\beta}^2} + \Tilde{\beta} - \frac{\arcsin(\Tilde{\beta})}{2} \: , \\
    J_0(\Tilde{\beta}) &= \frac{1}{2} \left( \arcsin(\Tilde{\beta}) - \Tilde{\beta} \sqrt{1-\Tilde{\beta}^2} \right) \: , \\
    J_1(\Tilde{\beta}) &= \frac{\Tilde{\beta}^2}{2} \: , \\
    J_2(\Tilde{\beta}) &= \frac{1}{2} \left( \arcsin(\Tilde{\beta}) + \Tilde{\beta} \sqrt{1-\Tilde{\beta}^2}\right) \: .
\end{align}
We note that the integrals diverge for $n\geq 3$. This is the reason that we only consider the values for $n=0,1,2$ in this work. To avoid this divergence, and model also density distributions with higher exponents, one could, e.g., use a broken power law distribution where the low-velocity part of the ejecta has, e.g., a uniform distribution and only the high-velocity part follows a power law, as done by, e.g., \cite{Suzuki:2016bmq,Suzuki:2018lvv}. Another option is to assume that there is a lowest velocity in the ejecta $\beta_\mathrm{min} > 0$, instead of using zero as we did in this work.

\section{Light Curve Model Validation} \label{sec:appendix:light-curve-model-validation}

We systematically investigate the general behaviour of our light curve model (see Sec. \ref{sec:light-curve-model}) when changing the model parameters $E_\mathrm{kin,eje}$ (kinetic energy of the ejecta), $M_\mathrm{eje}$ (total ejecta mass) and $R_0$ (radius of the SMS at the onset of collapse). We also investigate the impact on our model when changing the ejecta density distribution and the CSM density distribution. We here suppress the non-equilibrium radiation and manually set $\eta < 1$.

\begin{figure}
    \centering
    \includegraphics[width=0.498\textwidth]{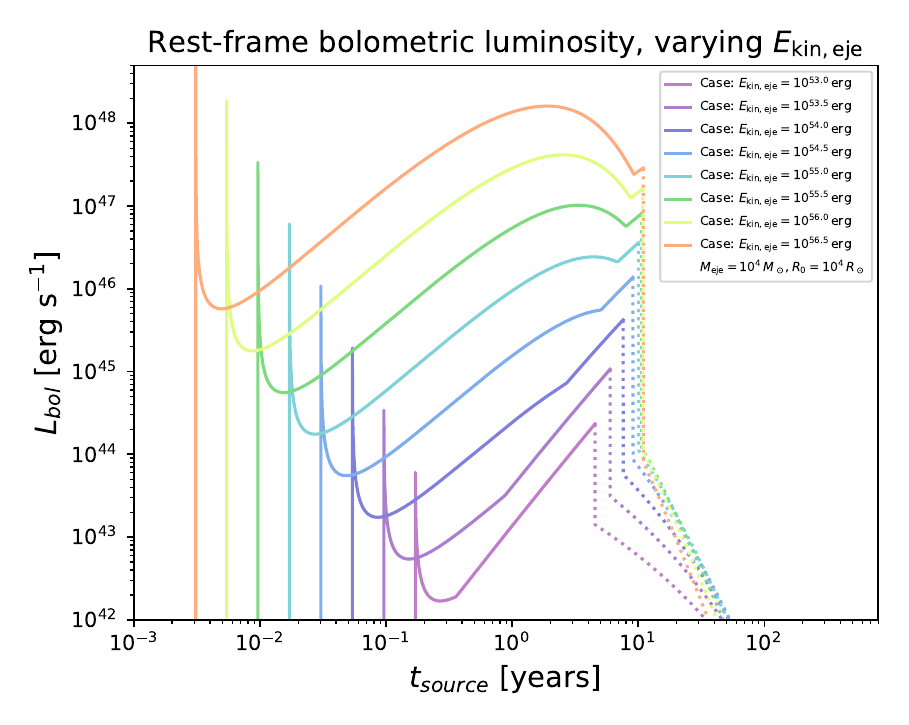}
    \caption{Bolometric Luminosity $L_\mathrm{bol}$ of an SMS explosion as a function of time, computed using our light curve model (see Sec. \ref{sec:light-curve-model}). The ejecta mass $M_\mathrm{eje}$ and the SMS radius $R_0$ are fixed and the ejecta kinetic energy $E_\mathrm{kin,eje}$ is varied.}
    \label{fig:appendix:varying-Ekin_eje}
\end{figure}

In Fig. \ref{fig:appendix:varying-Ekin_eje}, we show the bolometric luminosity as a function of time for fixed values of $M_\mathrm{eje}$ and $R_0$ while varying the kinetic energy $E_\mathrm{kin,eje}$: When it increases, the bolometric luminosity also increases. At the same time, the length of the optically thick shock phase (solid lines) increases as well. At larger energy, the position of the light curve peak moves to earlier times. For lower energies, the light curve peak is in the recombination phase when the surface temperature of the shock is frozen at the ionisation temperature of hydrogen ($T_\mathrm{ion}\sim 6000\,$K). For higher energies, the peak shifts to earlier times and coincides with the diffusion peak of the material at $t_\mathrm{diff} \propto M_\mathrm{eje}^{0.75} E_\mathrm{kin,eje}^{-0.25}$ \citep[][Eq. (24)]{Uchida:2018ago}.

\begin{figure}
    \centering
    \includegraphics[width=0.498\textwidth]{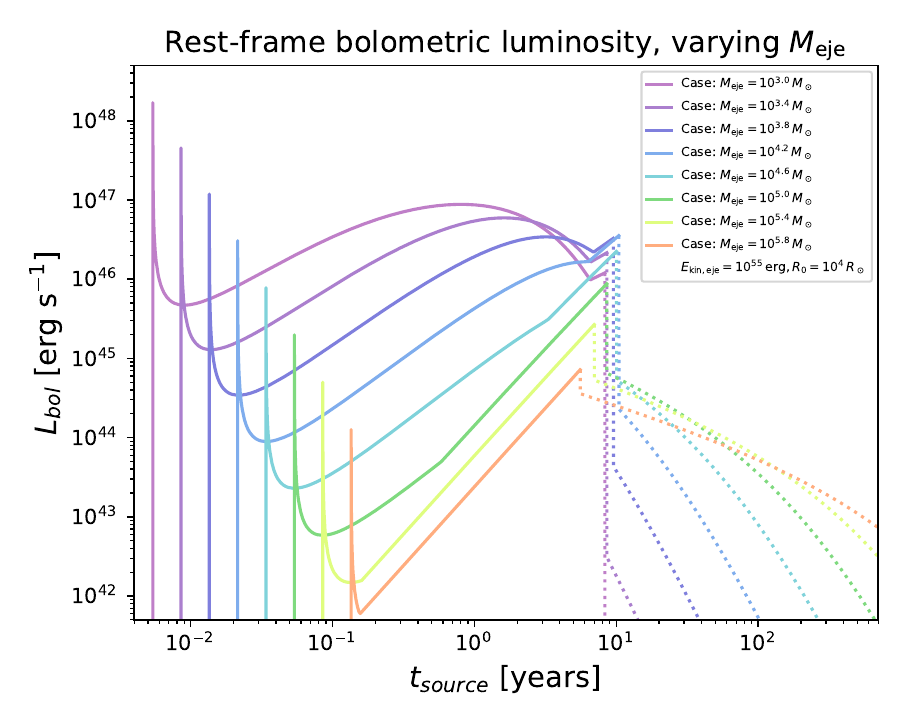}
    \caption{Bolometric Luminosity $L_\mathrm{bol}$ of an SMS explosion as a function of time, computed using our light curve model (see Sec. \ref{sec:light-curve-model}). The ejecta kinetic energy $E_\mathrm{kin,eje}$ and the SMS radius $R_0$ are fixed and the ejecta mass $M_\mathrm{eje}$ is varied.}
    \label{fig:appendix:varying-M_eje}
\end{figure}

In Fig. \ref{fig:appendix:varying-M_eje}, we show the bolometric luminosity as a function of time for fixed values of $E_\mathrm{kin,eje}$ and $R_0$ while varying the ejecta mass. The general impact of changing $M_\mathrm{eje}$ is as follows: Increasing the ejecta mass leads to a lower bolometric luminosity and a later peak-time of $L_\mathrm{bol}$. Increasing $M_\mathrm{eje}$ therefore has a roughly inverse effect to increasing the ejecta kinetic energy. The same reasoning regarding the light curve peak position also applies here. Increasing the ejecta mass also leads to a longer optically thick shock phase up until a point. Increasing the mass further then decreases the optically thick phase again. The point when the optically thick phase is maximised is the point where the position of the diffusion peak coincides with the start of the recombination phase of the light curve. This makes sense because the diffusion peak is the point after which the internal energy stored in the shock region is mostly radiated and is thus able to power the increasing bolometric luminosity in the recombination phase.

\begin{figure}
    \centering
    \includegraphics[width=0.498\textwidth]{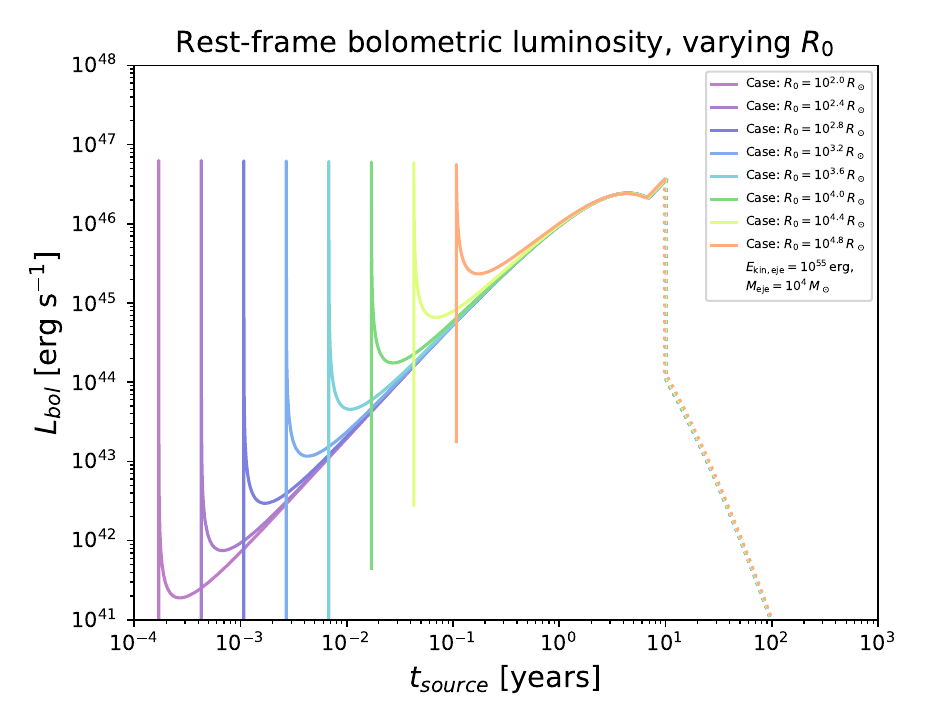}
    \caption{Bolometric Luminosity $L_\mathrm{bol}$ of an SMS explosion as a function of time, computed using our light curve model (see Sec. \ref{sec:light-curve-model}). The ejecta kinetic energy $E_\mathrm{kin,eje}$ and the ejecta mass $M_\mathrm{eje}$ are fixed and the SMS radius $R_0$ is varied.}
    \label{fig:appendix:varying-R0}
\end{figure}

In Fig. \ref{fig:appendix:varying-R0}, we show the bolometric luminosity as a function of time for fixed values of $E_\mathrm{kin,eje}$ and $M_\mathrm{eje}$ while varying the SMS radius. The general impact of changing $R_0$ are as follows: Changing the initial radius leads to a different initial time $t_0$, which corresponds to the dynamical time of the system. This modifies the early time evolution of the light curve. The late-time behaviour however is not modified significantly. The light curve peak time and magnitude are not changed significantly. Also the length of the optically thick shock phase is essentially unaffected by different initial radii. This can be understood, because a different initial radius does not modify the absolute ejecta velocity. This also means that the diffusion time is not affected because it depends only on $E_\mathrm{kin,eje}$ and $M_\mathrm{eje}$. It is therefore justified to simply choose a fiducial value for $R_0$, because the behaviour of the light curve at later times will not be significantly affected, even when varying $R_0$ by several orders of magnitude. 

\begin{figure}
    \centering
    \includegraphics[width=0.498\textwidth]{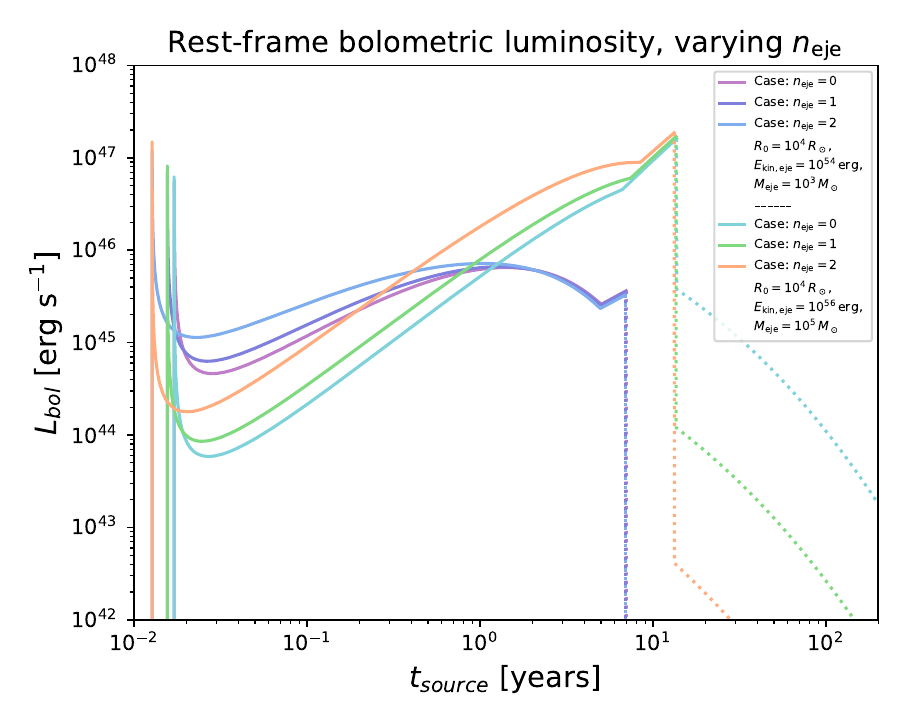}
    \caption{Bolometric Luminosity $L_\mathrm{bol}$ of an SMS explosion as a function of time, computed using our light curve model (see Sec. \ref{sec:light-curve-model}). We vary the density distribution of the ejecta $\rho_\mathrm{eje} \propto (\beta\Gamma)^{-n_\mathrm{eje}}$.}
    \label{fig:appendix:varying-ejecta-density}
\end{figure}
\begin{figure}
    \centering
    \includegraphics[width=0.498\textwidth]{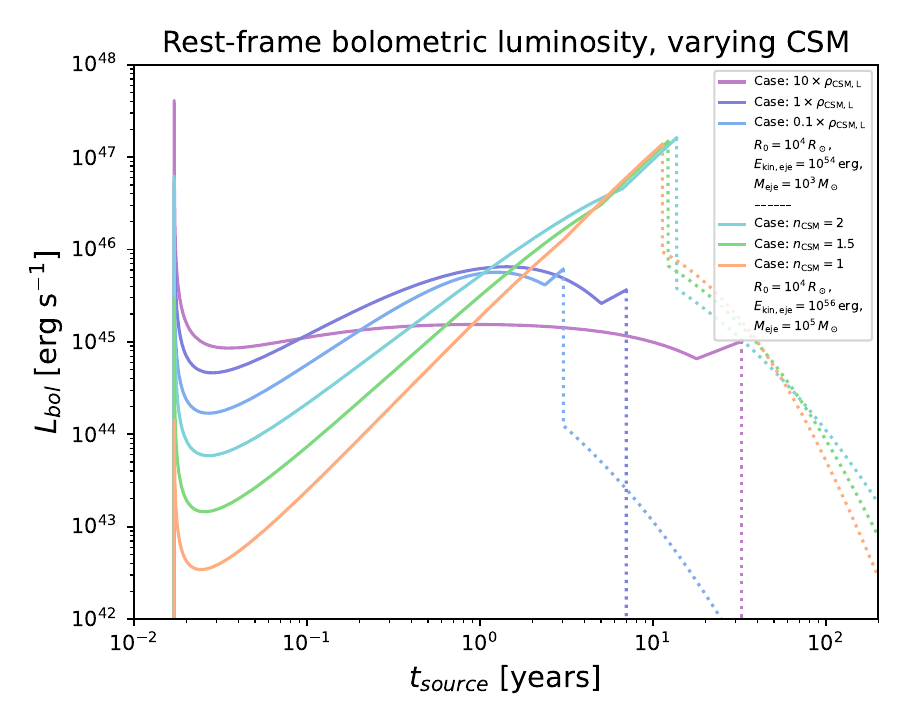}
    \caption{Bolometric Luminosity $L_\mathrm{bol}$ of an SMS explosion as a function of time, computed using our light curve model (see Sec. \ref{sec:light-curve-model}). We vary the density and the distribution of the CSM matter. $\rho_\mathrm{CSM,L}$ stands for the density of the Larson solution, see \eqref{eq:light-curve-model:Larson-CSM-Model}. We vary the distribution as $\rho_\mathrm{CSM} = \rho_{n_\mathrm{CSM}} r^{-n_\mathrm{CSM}}$, where we choose $\rho_{n_\mathrm{CSM}}$ so that the density matches $\rho_\mathrm{CSM,L}$ at $r=0.1\,$pc.}
    \label{fig:appendix:varying-CSM-density}
\end{figure}
\begin{figure}
    \centering
    \includegraphics[width=0.498\textwidth]{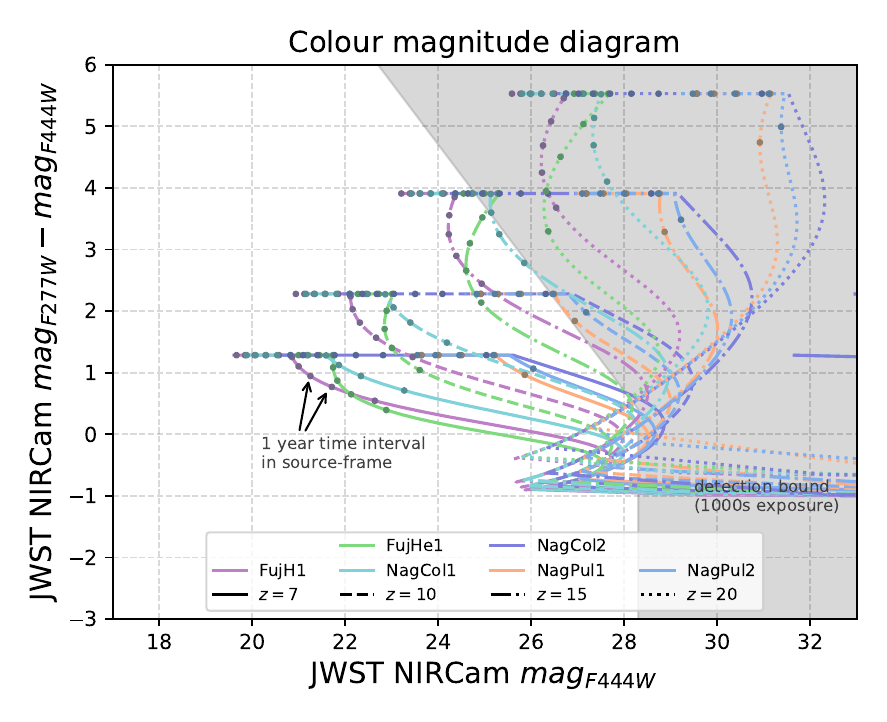}
    \caption{Same as Fig.~\ref{fig:results:colour-magnitude-diagrams}, but the early-time luminosity peak is included.}
    \label{fig:appendix:colour-magnitude-diagrams-early-time-peak}
\end{figure}

We also investigate the model dependence on the ejecta density distribution in \figref{fig:appendix:varying-ejecta-density}. We do this by changing the exponent $n_\mathrm{eje}$ of the power law for the ejecta, see \eqref{eq:light-curve-model:ejecta-density-profile}. We find that changing the ejecta density distribution has only a minor effect on the light curve over all. The maximum brightness changes by at most a factor of two and the time when the shock becomes transparent is essentially unchanged. The main difference is that for larger $n_\mathrm{eje}$ the early-time emissions are slightly enhanced and the time of the peak is slightly earlier. This can be understood because for higher $n_\mathrm{eje}$ the outer layers of the ejecta are less dense but have comparatively higher kinetic energy. This leads to a less massive but more energetic shock region with smaller optical depth, which decreases the diffusion time and thus increases the luminosity.

Because the light curve is powered by shock interaction between the ejecta and the CSM, we also briefly discuss the impact of changing the CSM properties in \figref{fig:appendix:varying-CSM-density}. For this, we modify the CSM density by a constant factor and we also change the slope of the power law. Changing the slope of the CSM leads to small changes in the luminosity in the early light curve evolution. A shallower slope leads to a shorter optically thick shock phase. But the changes are well within the physical uncertainty regarding initial conditions and observability. Changing the CSM density can have significant effect on the light curve. Changing the density by a factor of $10$ changes the length of the optically thick light curve phase by a factor of $3$--$4$ and can also change the maximum brightness by a factor of up to a few. This is to be expected because the shock is powered by the CSM interaction. The CSM is therefore a main source of uncertainty in our model. This highlights the importance of taking the CSM into account self-consistently in future simulations.

\section{Colour Magnitude plot including early-time peak} \label{sec:appendix:colour-magnitude-plot-with-early-time-peak}

In the main text in Fig. \ref{fig:results:colour-magnitude-diagrams}, we show SMS explosions in a colour-magnitude diagram for JWST filters. We here show in \figref{fig:appendix:colour-magnitude-diagrams-early-time-peak} a version of \figref{fig:results:colour-magnitude-diagrams} that included the early emission peak. The main difference here is that the initial peak produces a very blue initial peak of around $m_\mathrm{F444W}\geq26$ and $m_\mathrm{F277W}-m_\mathrm{F444W}\approx -1$. 

We originally cut out the first initial light curve peak because of the following reasons. Firstly, the initial peak will likely be obscured due to self-ionisation of the surrounding CSM by UV radiation from the shock. Second, because the figures are easier to read. Third, the initial peak is very short compared to the light curve evolution (roughly two times $t_0\sim \mathcal{O}(3-30\,\mathrm{days})$) and would be visible only very briefly. The probability that an SMS light curve would be observed in the early phase is therefore quite low.


\bsp	
\label{lastpage}
\end{document}